\documentclass[%
 reprint,
 superscriptaddress,
preprintnumbers,
 amsmath,amssymb,
 aps,
 pra,
floatfix,
longbibliography]{revtex4-2}

\usepackage[colorlinks=true, pdfstartview=FitV, linkcolor=blue, citecolor=blue, urlcolor=blue]{hyperref}

\usepackage{amsfonts}
\usepackage{graphicx}
\usepackage{physics}
\usepackage{color}
\usepackage{relsize}
\usepackage{hyperref}
\usepackage{framed}

\usepackage{tcolorbox}
\tcbset{colframe=white}

\usepackage{ulem}

\begin{document}

\title{Quantum simulation of lattice gauge theories via deterministic duality transformations assisted by measurements}

\author{Hiroki Sukeno}
\author{Tzu-Chieh Wei}

\affiliation{C.N. Yang Institute for Theoretical Physics and Department of Physics and Astronomy, State University of New York at Stony Brook, Stony Brook, NY 11794-3840
}

\date{\today}

\begin{abstract}
Quantum simulation is one of the major applications of quantum devices.
In the noisy intermediate-scale quantum era, however, the general quantum simulation  is not yet feasible, such as that  of lattice gauge theories, which is likely limited due to the violation of the Gauss law constraint and the complexity of the real-time dynamics, especially in the deconfined phase. 
Inspired by the recent works of S. Ashkenazi and E. Zohar [Phys. Rev. A {\bf 105}, 022431 (2022)] and of N. Tantivasadakarn, R. Thorngren, A. Vishwanath, and R. Verresen [arXiv: 2112.01519], we propose to 
simulate dynamics of lattice gauge theories by using
the Kramers-Wannier transformation via cluster-state-like entanglers, mid-circuit measurements, and feedforwarded corrections, which altogether is a constant-depth deterministic operation. 
In our scheme, specifically, we first quantum simulate the time evolution under a corresponding symmetric Hamiltonian from an initial symmetric state, and then apply the Kramers-Wannier procedure. This results in a  wave function that has time evolved under the corresponding lattice gauge theory from a corresponding initial, gauged wave function. 
In the presence of noises in time evolution, the procedure succeeds when we are able to pair up magnetic monopoles represented by non-trivial measurement outcomes. Further, given a noise-free Kramers-Wannier transformation, the resulting wave function from a noisy time evolution satisfies the Gauss law constraint. 
We give explicit examples with the low dimensional pure gauge theories and gauge theories coupled to bosonic/fermionic matters such as the Fradkin-Shenker model.
\end{abstract}

\maketitle

\section{Introduction}

Lattice gauge theory is a cardinal formulation in modern many-body physics.
The Euclidean formulation of the lattice gauge theories~\cite{Kogut, Wegner, PhysRevD.10.2445,Rothe:1992nt} has provided numerous results, such as predicting phase diagrams ~\cite{Wegner, PhysRevD.10.2445, PhysRevD.19.3682, PhysRevD.17.2637, PhysRevD.20.1915, PhysRevD.21.2308, mclerran1981monte} in high-energy physics, while the low-energy properties of lattice gauge theories have been useful guideposts in condensed matter physics \cite{2005PhRvB..71d5110L,2016PhRvB..94w5157V,fradkin_2013, beekman2017dual} and quantum information science~\cite{2002JMP....43.4452D, 2010PhRvB..82h5114T} as well.
On the other hand, the Monte-Carlo simulation of lattice gauge theories often suffers from the sign problem~\cite{2021arXiv210812423N, PhysRevB.41.9301, 2008arXiv0808.2987H, 2017PTEP.2017c1D01G}, and the search for an efficient method to simulate the dynamics of gauge theories in real time or with finite chemical potentials has been a central subject of the field.
In the Hamiltonian lattice gauge theory~\cite{Kogut, PhysRevD.17.2637, Horn:1979fy}, the dimension of the Hilbert space grows exponentially with the number of degrees of freedom, and 
quantum simulation~\cite{feynman2018simulating,lloyd1996universal} is expected to be one of the solutions to these issues, which can potentially enable us to perform the simulation of quantum many-body dynamics with resources linear in the system size and in the time duration.
The quantum simulation of lattice gauge theories is now one of the major subjects of study~\cite{2016RPPh...79a4401Z,2013AnP...525..777W,2016ConPh..57..388D,2020EPJD...74..165B,2006PhRvA..73b2328B,homeier2021z} in the Noisy Intermediate Scale Quantum (NISQ) era~\cite{Preskill:2018}. 

One  challenge in simulating the dynamics of lattice gauge theories using quantum computers is the complexity of the wave function, such as in the toric code~\cite{2003AnPhy.303....2K}.
The ground state of the toric code is identical to those representing the deconfining phase limit of the $\mathbb{Z}_2$ gauge theory in $(2+1)$-dimensions~\cite{Kogut}. 
As the toric code possesses long-range entanglement, preparing its ground state requires a quantum circuit of the size of the system~\cite{2006PhRvL..97e0401B, 2010ChenGuWen}, which is challenging with unitary gates provided in current NISQ devices~\cite{satzinger2021realizing}. 
In particular, the Hamiltonian of the gauge theory involves four-body interaction terms on the square lattice. Thus the quantum simulation of the model requires a large number of quantum gates in general.
Another challenge in simulating the dynamics of lattice gauge theories is that a noisy quantum simulation over a large depth induces significant errors, leading to a wave function with unphysical contributions violating the Gauss law constraint. 
Therefore, enforcing the gauge invariance is one of the primary areas of interest in studies on quantum simulation of gauge theories \cite{2020Natur.587..392Y, 2020PhRvL.125c0503H,2014PhRvA..90d2305K,2013PhRvL.110e5302Z,2012PhRvL.109l5302Z,2011PhRvL.107A5301Z,2022arXiv220413709H,PhysRevD.107.014506,2012PhRvL.109q5302B,2014PhRvL.112l0406S,sukeno2023mbqs, halimeh2021gauge, zhou2022thermalization,  halimeh2022gauge,halimeh2022stabilizing, homeier2022quantum, halimeh2023spin}.

The topologically ordered states cannot be reached by short-depth unitaries alone from a product state~\cite{2010ChenGuWen}. 
However, there is a well-known loophole in the context of the fault-tolerant measurement-based quantum computation \cite{2005PhRvA..71f2313R,2007PhRvL..98s0504R,2007NJPh....9..199R}.
Namely, a combination of the short-range entangler, such as the controlled gate used to create the so-called cluster state~\cite{PhysRevA.68.022312, 2005PhRvA..71f2313R}, and on-site measurements give rise to a constant-depth operation to prepare the ground state of the toric code.
Recently, this idea was extended to further  cases, and it is now recognized that short-range entangled states such as symmetry-protected topologically (SPT) ordered states~\cite{PhysRevB.85.075125, 2010PhRvB..81f4439P,2011PhRvB..84p5139S, 2011PhRvB..83c5107C,PhysRevB.84.235128, PhysRevB.87.155114,PhysRevB.84.235141} can be transformed to topologically ordered states by constant-depth operations including measurements and feedforward~\cite{Hierarchy, SchrodingersCat, ShortestRoute, 2021arXiv211201519T, Bravyi2022a, 2021arXiv211104765A} (see also~\cite{PhysRevLett.127.220503,Lu2022}).
In Refs.~\cite{2021arXiv211201519T,Hierarchy}, such operations were interpreted as the celebrated {\it Kramers-Wannier transformation} \cite{PhysRev.60.252, Wegner, Kogut,2015PhRvX...5a1024H, 2018arXiv180907757R}, implying versatility of the method to many other contexts.
It is indeed a method to promote a global symmetry to the corresponding gauge symmetry (or gauge redundancy); thus, it is also called {\it gauging}.
However, its application to quantum simulation of time evolution in lattice gauge theories
has not been much investigated to date.

\subsection*{Summary of results}

In this work, we use the idea of the Kramers-Wannier transformation based on measurements \cite{2021arXiv211201519T} to real- and imaginary-time quantum simulation of lattice gauge theories.
In $(2+1)$-dimensions, the transverse-field Ising model is transformed into the $\mathbb{Z}_2$ lattice gauge theory.
We show that the time evolution under the transverse-field Ising model on a symmetric initial state is transformed into that of the $\mathbb{Z}_2$ lattice gauge theory on a long-range entangled state with a {\it single-shot} operation~\cite{2021arXiv211201519T,Hierarchy}, which is just an entangling operation followed by single-qubit measurements.

Using the Kramers-Wannier transformation on the time-evolved wave function of the transverse-field Ising model has an advantage over directly simulating the gauge theory.
First, since the transverse-field Ising model Hamiltonian is simpler, the required connectivity of the quantum circuit for this model will be much simpler than that required for simulating the corresponding gauge model.
Second, since the product state maps to the toric code ground state via the Kramers-Wannier transformation, the time-evolved wave function
is changed to a long-range entangled state, which is far from the product states if restricted to unitaries alone. 
Finally, with a generic unitary error channel during the time evolution with {\it e.g.,} the transverse-field Ising model, the resulting state after the duality procedure is gauge symmetric, given a perfect (constant-depth) dualization operation. Due to noises, the correction part of the dualization operation may fail when we find an odd number of $X=-1$ outcomes (instead of $X=+1$) out of the measurement, in which case we would have an isolated magnetic monopole,  explained below. In that case,  we may simply throw away the result and restart the procedure.

The randomness of measurement outcomes  induces unwanted phase factors on the resulting wave functions. 
Such phase factors can be expressed using the so-called  byproduct operators.
For pure gauge theories, the byproduct operators are Pauli $Z$ operators supported on a set of strings whose endpoints correspond to non-trivial measurement outcomes.
For gauge theories coupled to matters, the byproduct operators are $Z$ operators acting on the gauge degrees of freedom whose location corresponds to non-trivial measurement outcomes.
In either case, byproduct operators are correctable.

We generalize our method to a non-trivial SPT Hamiltonian in $(2+1)$-dimensions, which results in a quantum simulation of an Abelian twisted gauge theory \cite{2012PhRvB..86k5109L, 2013PhRvB..87l5114H}, akin to the Dijkgraaf-Witten topological field theory on triangulated lattices \cite{dijkgraaf1990topological}.
We also show that our method works for the gauge group $\mathbb{Z}_N$ \cite{Horn:1979fy}.
Then we generalize the method to obtain matter theories covariantly coupled to $\mathbb{Z}_2$ gauge fields in $(1+1)$-dimensions. 
We discuss cases with a bosonic matter and a fermionic matter separately, and in the latter, the fermion parity in the Kitaev Majorana chain model~\cite{Kitaev:2000nmw} is the global symmetry to be gauged~\cite{borla2021gauging}.
In $(2+1)$-dimensions, we also give an example with an Ising model covariantly coupled to gauge fields, whose phase diagram was studied by Fradkin and Shenker~\cite{PhysRevD.19.3682}.
The Kramers-Wannier transformation we will use for theories with matter is a generalization of the standard one given in, {\it e.g.}, Ref.~\cite{2018arXiv180907757R}, and we embody a physical realization of such mathematical transformations.
We concentrate on the real-time evolution in the main text, but there is no obstruction to generalizing our method to the imaginary-time evolution \cite{lin2021real, 2020NatPh..16..205M, 2022arXiv220209100M}, simply by replacing $t \mapsto -i \tau$ with $\tau$ being the imaginary time.
Post-selection is generically required in implementing the imaginary-time evolution on quantum devices, but the part on the duality transformation in our method is always deterministic.

\begin{figure*}
\includegraphics[width=0.7\linewidth]{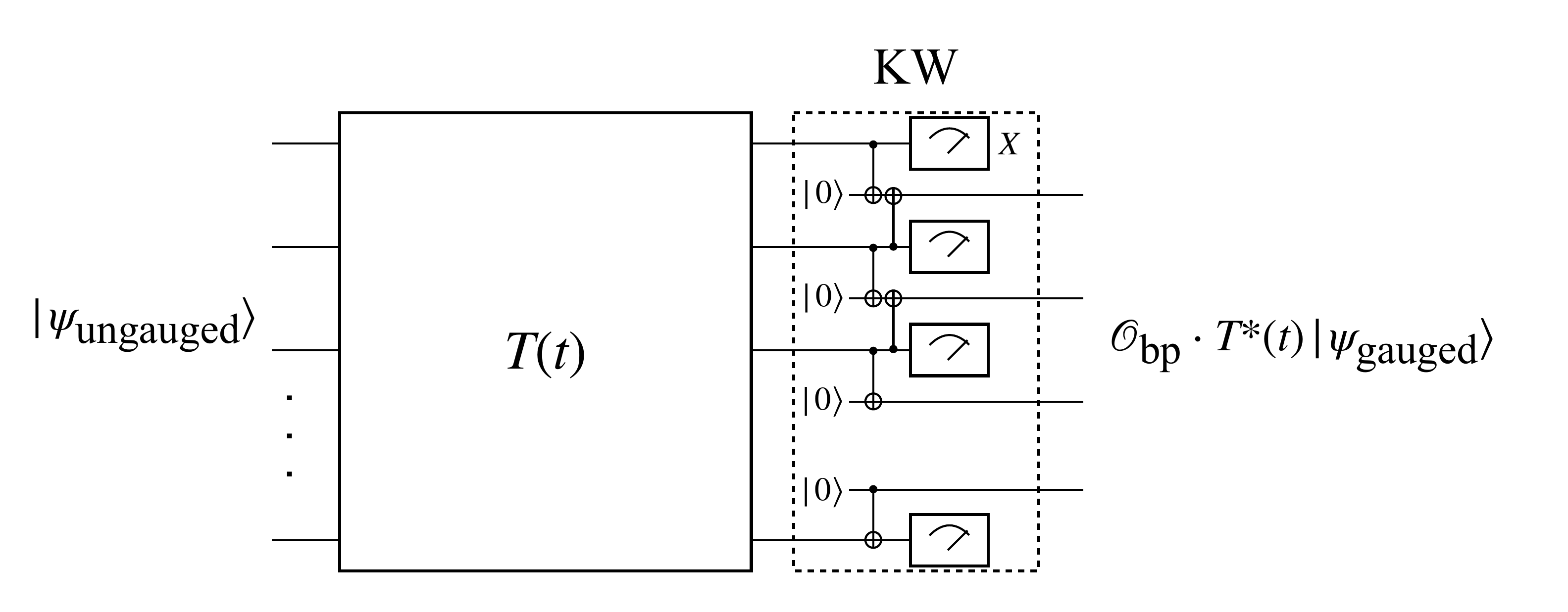} 
\caption{The procedure of the KW transformation of time evolution. 
(Top) A quantum circuit expression of the result. Starting from an ungauged wave function, we perform a time evolution $T(t)$ via a quantum circuit. 
The Kramers-Wannier transformation procedure consists of entangling the original degrees of freedom with additional degrees of freedom by entanglers and measuring the original degrees of freedom.
We obtain a gauged wave function evolved with a time evolution $T^*(t)$ with a Hamiltonian obtained by the Kramers-Wannier duality, up to a byproduct operator $\mathcal{O}_{\text{bp}}$, which depends on the measurement outcomes. 
}
\label{fig:concept}
\end{figure*}

Our results are demonstrated with several examples  and can be compactly summarized in a formula as follows:
\begin{tcolorbox}
[width=\linewidth, sharp corners=all, colback=white!95!black]
\begin{align}
\mathcal{O}^{M^\star}_{\text{bp}} \cdot T^{M^\star}(t) |\psi_\text{gauged} \rangle 
= \widehat{\text{Map}} \cdot T^{M}(t) |\psi_\text{ungauged} \rangle .  \end{align}
\end{tcolorbox}
\noindent Here $T^M(t)$ and $T^{M^\star}(t)$ represent the Trotterized time evolution in the original and dual theories, respectively.
The map $\widehat{\text{Map}}$ is a physical operation that implements the duality map, which consists of applying entanglers and measurements.
The wave function $|\psi\rangle$ is the initial state, whose form  will be explained in detail below, and the duality map relates the gauged and ungauged ones.
For example, in $(2+1)$-dimensions, when $|\psi_{\text{ungauged}}\rangle$ is a product $|+\rangle$ state, $|\psi_{\text{gauged}}\rangle$ is the ground state of the toric code.
The operator $\mathcal{O}^{M^\star}_{\text{bp}}$ is the byproduct operator, and its specific form depends on the dualization.
 Figure~\ref{fig:concept} has a schematic diagram illustrating our result.
For convenience, we also summarize our results from different examples in Table~\ref{tab:summary}.
(In each case, the equality is up to a normalization constant but is omitted  for convenience.)

\begin{table*}
\begin{center}
{\renewcommand{\arraystretch}{2.0}
\begin{tabular}{|c||c|c||c|c||c|c| c| }
\hline
     dim &$M$ & d.o.f. & $M^\star$  & d.o.f. & $\widehat{\text{Map}}$ &$\mathcal{O}^{M^\star}_{\text{bp}}$  & Section \\ \hline
     2& Ising$/\mathbb{Z}_2$ (TFI) & $V$ & pure $\mathbb{Z}_2$ gauge theory (GT) & $E^*$ & $\widehat{\text{KW}}$ \eqref{eq:kw-map}&  \eqref{eq:bp-Z2gauge} & \ref{sec:GT}\\ 
    2 & twisted Ising$/\mathbb{Z}_2$ (tTFI) & $V$ & twisted $\mathbb{Z}_2$ gauge theory (tGT) & $E^*$& $\widehat{\text{KW}}$ \eqref{eq:kw-map} &  \eqref{eq:bp-Z2gauge} & \ref{sec:tGT} \\
    2 &  clock$/\mathbb{Z}_N$ ($\mathbb{Z}_N$ clock) & $V$ &  $\mathbb{Z}_N$ gauge theory ($\mathbb{Z}_N$ GT) & $E^*$& $\widehat{\text{KW}}^{\mathbb{Z}_N}$ \eqref{eq:kw-map-ZN}&  \eqref{eq:bp-ZN} & \ref{sec:ZN-gauge-theory}\\
    1 &  Ising (TL-Ising) & $V$ & gauge theory with Ising matter (GM) & $V^* \cup E^*$ & $\widehat{\text{KW}}^{\text{GM}}$ \eqref{eq:kw-map-gm} & \eqref{eq:bp-gm} & \ref{sec:GM}\\
     1 & Ising (TL-Ising) & $V$ & gauged Majorana chain (QED) & $V^* \cup E^*$ & $\widehat{\text{JW}}$ \eqref{eq:jw-map} & \eqref{eq:bp-gm} & \ref{sec:QED} \\
    2 &  star-plaquette (SP) & $E$ & gauge theory with Ising matter (FS) & $V^* \cup E^*$ & $\widehat{\text{FS}}$ \eqref{eq:fs-map} & \eqref{eq:bp-fs} & \ref{sec:FS} \\\hline
\end{tabular}
}
\caption{Summary of results in our examples.
Here, ``$/\mathbb{Z}_K$" ($K=2,N$) indicates that the Hamiltonian and the initial wave function we consider is symmetric under the transformations that form such groups.
The names next to the models inside the parenthesis are used for labels of the models, $M$ or $M^\star$.
}
\label{tab:summary}
\end{center}
\end{table*}

\subsection*{Related works}

In Refs.~\cite{2020arXiv200811395Y,Gustafson:2020yfe} the Kramers-Wannier duality was used as a mathematical dictionary for the quantum simulation of the $(2+1)$d $\mathbb{Z}_2$ lattice gauge theory. 
On the other hand, our method allows us to physically prepare the simulated wave function of the gauge theory. 
In Ref.~\cite{2021arXiv211104765A}, a physical implementation of the Kramers-Wannier transformation using measurements and post-selection was studied in the context of quantum simulation of gauge theories. 
Here we explicitly formulate a procedure for dualizing time evolutions, including a method to handle the randomness of the measurement outcomes. 
We apply the deterministic measurement-based gauging method, first elucidated in Ref.~\cite{2021arXiv211201519T} for obtaining topologically ordered ground states.
In particular, we reveal that our method does not require post-selections even with time evolution unitaries, just as in the case of transforming ground states.
Finally, although our models of consideration and procedures are distinct from theirs, we mention an interesting work, Ref.~\cite{lee2022measurement}, which utilized the measurement-based Kramers-Wannier transformation to obtain imaginary-time evolution under some Ising/gauge quantum Hamiltonian assisted by classical processing.

\subsection*{Organization of the paper}

This paper is organized as follows.
In Section \ref{sec:KW-by-measurements}, we explain our method with the example of the duality between the transverse-field Ising model and the pure gauge theory in $(2+1)$-dimensions.
We define the map $\widehat{\text{KW}}$ and demonstrate our main result.
In Section \ref{sec:KW-generalization-to-pure}, we generalize our result in two directions. One is to twist the models on both sides of the duality, in the sense of twists in (symmetry-protected) topological orders.
The other is to extend the result to cyclic groups $\mathbb{Z}_N$. 
In Section \ref{sec:KW-gauge-matter}, we generalize the idea to incorporate matter fields after the dualization. 
Section \ref{sec:conclusions-discussion} is devoted to conclusions and discussion.

\section{Time evolution of gauge theory via measurement-based Kramers-Wannier} \label{sec:KW-by-measurements}

\subsection{Prelude: Measurement-based Kramers-Wannier transformation in 1D }

Let us begin with a simple example to demonstrate the core ideas in our method.
We consider a spin model defined on vertices in a one-dimensional periodic lattice. 

We take the model to be symmetric under the $\mathbb{Z}_2$ global symmetries generated by
\begin{align}
U^{(0)} := \prod_{v \in V} X_v ,
\end{align}
where $X$ is the Pauli X operator.
An example of such theory is described by the transverse-field Ising model Hamiltonian,
\begin{align}
H = - \lambda \sum_{v \in V} X_v -  \sum_{e \in E} \Big( \prod_{v \subset e} Z_v \Big) , \label{eq:1dtfi-ham}
\end{align} 
where the second term is the ordinary Ising interaction $Z_v Z_{v'}$ for $e=\langle v,v'\rangle$, with  $Z$ being the Pauli  $Z$ operator. 
The relation $v \subset e$ under the product means that we take a product over those vertices contained in an edge $e$.

In the present work, we consider the (quenched) time evolution starting from an initial wave function, and we assume that it is symmetric under global symmetry.
In the current example with the transverse-field Ising model, the initial wave function is written using the Hilbert space spanned by the basis which is a tensor product of $Z$ eigenvectors over vertices. 
We write it as $\bigotimes_{v \in V} |a_v\rangle_v$, or more simply $| \{ a_v \}\rangle$ with $a_v = 0,1$.
The symmetric initial wave function is then written as
\begin{align}
&|\psi_\text{in} \rangle = \sum_{ \substack{ a_v \in \{0,1\} \,  \\ \, v \in V } } C(\{a_v\}) \bigotimes_{v \in V} |a_v\rangle_v , \label{eq:sym-ini-state-1d} \nonumber \\
&\text{such that  } U^{(0)} |\psi_\text{in} \rangle = |\psi_\text{in} \rangle ,
\end{align}
where $C(\{a_v\}): (\mathbb{Z}_2)^{\otimes |V|} \rightarrow \mathbb{C}$ is a suitable complex coefficient.
The time-evolved wave function with the first-order Trotter decomposition is written as 
\begin{align}
|\psi (t) \rangle = 
\Big( \prod_{e \in E } e^{- i \Delta t \prod_{v \subset e} Z_v }  
\prod_{v \in V } e^{- i \Delta t \lambda X_v } 
\Big)^k | \psi_\text{in}\rangle .
\end{align}

\begin{figure*}
   \includegraphics[width=0.9\linewidth]{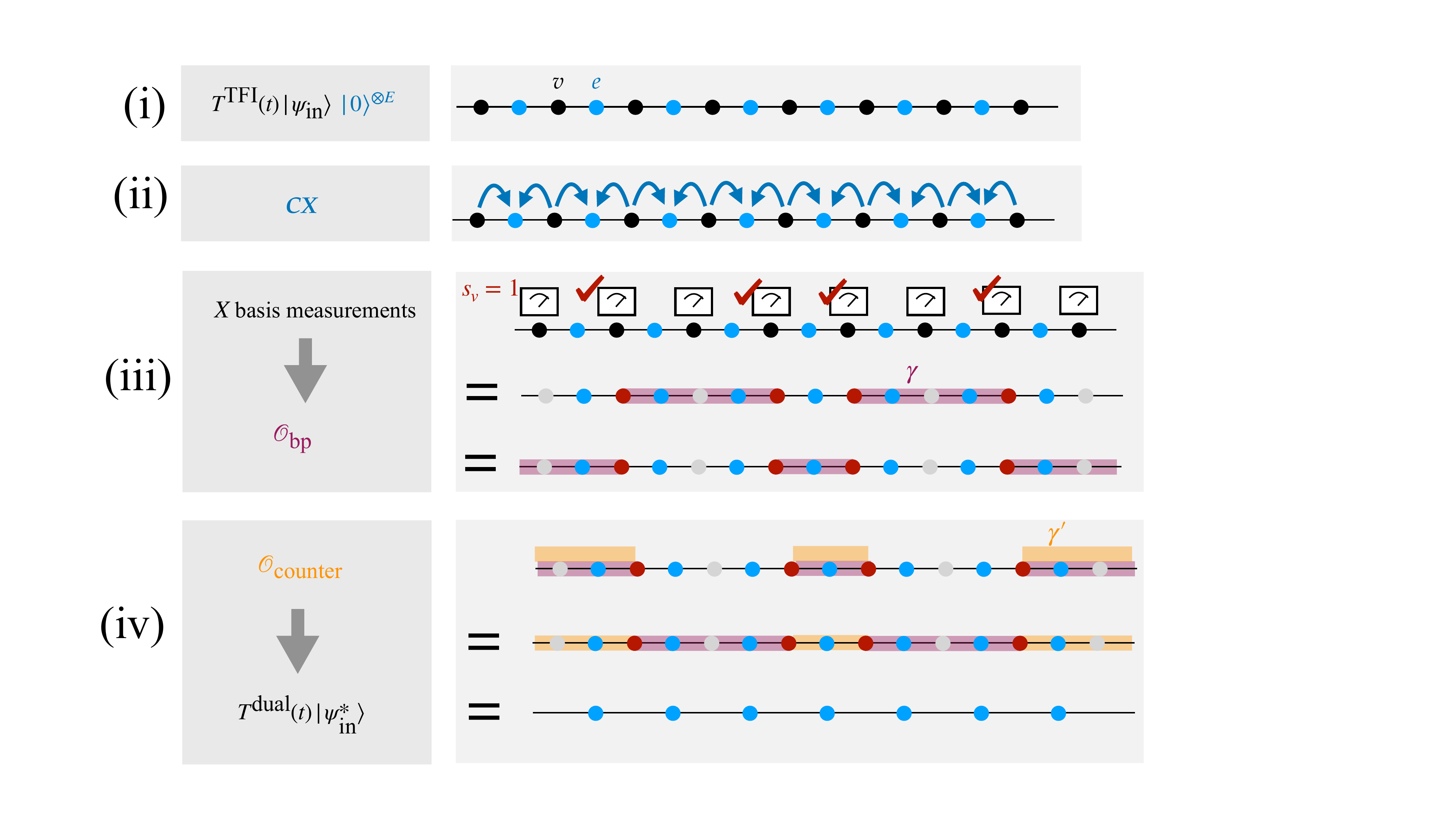}
   \caption{The measurement-assisted Kramers-Wannier transformation in a one-dimensional periodic lattice. (i) A state $|\psi_\text{in}\rangle$ is initialized on vertices (black dots) of the 1d chain, and is evolved under the Hamiltonian of the TFI model. Ancillary qubits $|0\rangle^{\otimes E}$ are placed on edges (blue dots) separately. (ii) We apply the controlled-$X$ gate, where the vertex qubits control $X$ gates on edge qubits, as indicated by the arrows in the figure. (iii) We measure the vertex qubits in the $X$ basis. When the measurement outcome is $X=-1$ (occuring at check marks and red dots below), phase operators effectively act on the desired dual time evolution. It can be expressed as $\mathcal{O}_\text{bp}$, the product of $Z$ operators on edge qubits that overlap with purple lines which connect red dots. (iv) One can negate the byproduct operator by applying the counter operator $\mathcal{O}_\text{counter}$, the product of  $Z$ operators on edge qubits marked by orange lines that connect red dots. As $\mathcal{O}_\text{bp}\times \mathcal{O}_\text{counter}$ is trivial on the state after the map, we obtain the dual time evolution which is not affected by randomness of measurement outcomes.}
   \label{fig:1D-KW}
\end{figure*}

We often use the controlled-NOT gate:
\begin{align}
CX_{c,t} = |0\rangle \langle 0|_c \otimes I_t + |1\rangle \langle 1|_c \otimes X_t ,
\end{align}
where $c$ is the controlling qubit and $t$ is the target qubit.
Now, we describe our transformation procedure, which is adopted from Ref.~\cite{2021arXiv211201519T}, for example:
\begin{enumerate}
\item[(i)] Introduce ancillary degrees of freedom $|0\rangle^{\otimes E}$ on edges.
\item[(ii)] Apply entanglers 
\begin{align}
\mathcal{U}_{\text{1D KW}} = \prod_{v \in V} \Big( \prod_{e \supset v} CX_{v,e}\Big) 
\end{align}
to the product state $|\psi(t)\rangle \otimes  |0\rangle^{\otimes E}$. 
The relation $e \subset v$ under the product indicates that we take a product over those edges that contain the vertex $v$.
\item[(iii)] Measure all the vertex degrees of freedom in the Pauli $X$ basis. 
\item[(iv)] Construct a counter operator $\mathcal{O}_{\text{counter}}$, which we describe later, based on the measurement outcomes in (iii). Apply it to the post-measurement state.
\end{enumerate}
We illustrate the procedure in Fig.~\ref{fig:1D-KW} and
we claim that the resulting state is 
\begin{align}
\Big( \prod_{e \in E } e^{- i \Delta t Z_e }  
\prod_{v \in V } e^{- i \Delta t \lambda \prod_{e \supset v} X_e } 
\Big)^k | \psi^*_\text{in}\rangle, \label{eq:example-tfi-1}
\end{align}
with 
\begin{align}
| \psi^*_\text{in}\rangle
=  
\sum_{ \substack{ a_v \in \{0,1\} \,  \\ \, v \in V } } C(\{a_v\}) \bigotimes_{e \in E} \big| \sum_{v \subset e} a_v \big\rangle_e .
\end{align}
Note the basis associated with edges is now a sum of bits that were associated with the original wave function on vertices.
It is important to observe that this wave function also obeys the (dual) global symmetry condition:
\begin{align}
& | \psi^*_\text{in}\rangle = U^{(0,\text{dual})}| \psi^*_\text{in}\rangle , \\
&U^{(0,\text{dual})} := \prod_{e \in E} Z_e .
\end{align}
The wave function \eqref{eq:example-tfi-1} is now a quenched time evolution of the transverse-field Ising model on the dual lattice with the Hamiltonian being
\begin{align}
H_{\text{dual}} 
&= - \sum_{e \in E} Z_e
- \lambda \sum_{v \in V} \prod_{e \supset v} X_e \\
&= - \sum_{v^* \in V^*} Z_{v^*}
- \lambda \sum_{e^* \in E^*} \prod_{v^*  \subset e^*} X_{v^*}, \label{eq:1dtfi-ham-dual}
\end{align}
where $V^*$ and $E^*$ are the sets of dual vertices and  dual edges, respectively.
Note that the Hamiltonian is symmetric, i.e., $[H_{\text{dual}} ,U^{(0,\text{dual})}]=0 $.

\subsubsection{Preparation}

Before proving the above claim, we collect some facts to facilitate our demonstration.
First, the global symmetry in the original theory gives rise to a constraint on the measurement outcomes.
Writing the measurement basis in the $X$ basis denoted as $|s_v\rangle^{(X)}_v$ with $s_v=0$ or $1$ (depending on the outcome), we have
\begin{align}
\Big( \bigotimes_{v \in V} \langle s_v |^{(X)}_v \Big) |\psi_{\text{in}}\rangle 
&=
\Big( \bigotimes_{v \in V} \langle s_v |^{(X)}_v \Big)\prod_{v \in V} X_v  |\psi_{\text{in}}\rangle \nonumber \\ 
&= \prod_{v \in V }(-1)^{s_v} \Big( \bigotimes_{v \in V} \langle s |^{(X)}_v \Big) |\psi_{\text{in}}\rangle , \label{eq:contractMeaurement}
\end{align}
which tells us that 
\begin{align}
\sum_{v \in V} s_v = 0 \quad \text{mod }2,
\end{align}
where we have used the global symmetry in the first equality and the fact that $X|0/1\rangle^{(X)}=\pm |0/1\rangle^{(X)}$ in the second equality.

As the number of vertices with $s_v=1$ is even, one can construct a set of edges $\gamma \subset E$ that pair them up.
Namely, 
\begin{align}
\text{endpoints of } \gamma = \{v \text{ such that } s_v = 1\}. 
\end{align}
The choice of $\gamma$ is not unique; there are two possibilities in a 1D periodic chain.
We will use the following relation later that applies to states on the dual lattice (or edges on the original lattice) to give an equivalent account of the phase resulting from measuring vertex degrees of freedom: 
\begin{align}
\prod_{v \in V} (-1)^{a_v s_v} \bigotimes_{e \in E} |\sum_{v \subset e} a_v \rangle_e
=
\prod_{e \in \gamma} Z_e \bigotimes_{e \in E} |\sum_{v \subset e} a_v \rangle_e . \label{eq:rewriting-phase-1d}
\end{align}
This relation holds with either of two choices for $\gamma$: one that uses the minimum distance for pairing and the other that involves a path wrapping around in the other direction.
Then we define the byproduct operator as 
\begin{align}
 \mathcal{O}_{\text{bp}} (\gamma) = \prod_{e \in \gamma} Z_e .  
\end{align}
Although the precise definition depends on specific examples, throughout this paper, we call this type of operator --- a product of Pauli operators that expresses phase factors associated with the measurement outcomes --- the byproduct operators, adopting the colloquial terminology in Measurement-Based Quantum Computation \cite{PhysRevLett.86.5188, PhysRevA.68.022312,wei2018quantum}.

We note that the operators that appear in the original Hamiltonian get conjugated by the entangler $\mathcal{U}_{\text{1D KW}}$ as follows:
\begin{align}
X_v &\mapsto X_v \prod_{e \supset v} X_e , \\
\prod_{v \subset e} Z_v  &\mapsto  \prod_{v \subset e} Z_v . 
\end{align}
With this, we are ready to give our first example of deterministic duality transformation assisted by measurements and feedforwarded corrections.

\subsubsection{Demonstration}

\noindent{\bf Step (i-ii):} The wave function after applying the entanglers can be written as
\begin{align}
&\left\{ \prod_{e \in E } \exp(- i \Delta t \prod_{v \subset e} Z_v ) 
\prod_{v \in V } \exp(- i \Delta t \lambda X_v \prod_{e \supset v} X_e ) 
\right\}^k \nonumber \\
&\times \sum_{ \substack{ a_v \in \{0,1\} \,  \\ \, v \in V } } 
C(\{a_v\}) \bigotimes_{v \in V} |a_v\rangle_v
\bigotimes_{e \in E} \big| \sum_{v \subset e} a_v \big\rangle_e. 
\end{align}
We further rewrite this expression by noticing the following  simple fact. 
When the Ising term $\prod_{v \subset e} Z_v$ acts on the basis $|a_v\rangle_v$, it gives the phase factor $\prod_{v \subset e}(-1)^{a_v}$. 
This phase factor is precisely reproduced by the phase operator $Z_e$, which acts on $|\sum_{v \subset e} a_v \rangle_e$. 
Namely, the Ising term can be replaced by a $Z_e$ operator. 
One can indeed show that this replacement is allowed even in the presence of the other operator in the Hamiltonian $X_v \prod_{e \supset v} X_e$.
For example, it is simple to check that the following holds, 
\begin{align}
\Big(\prod_{v' \subset d}  Z_{v'}\Big)  \Big( X_u \prod_{e' \supset u} X_{e'}\Big)\bigotimes_{v \in V} |a_v\rangle_v
\bigotimes_{e \in E} \big| \sum_{v \subset e} a_v \big\rangle_e\nonumber\\
=\Big( Z_{d}\Big)  \Big( X_u \prod_{e' \supset u} X_{e'}\Big)\bigotimes_{v \in V} |a_v\rangle_v
\bigotimes_{e \in E} \big| \sum_{v \subset e} a_v \big\rangle_e,
\end{align} 
for any $d \in E$ and $u \in V$.
Hence we have 
\begin{align}
&\left\{ \prod_{e \in E } \exp(- i \Delta t \textcolor{red}{Z_e} ) 
\prod_{v \in V } \exp(- i \Delta t \lambda X_v \prod_{e \supset v} X_e ) 
\right\}^k \nonumber \\
&\times \sum_{ \substack{ a_v \in \{0,1\} \,  \\ \, v \in V } } 
C(\{a_v\}) \bigotimes_{v \in V} |a_v\rangle_v
\bigotimes_{e \in E} \big| \sum_{v \subset e} a_v \big\rangle_e, \label{eq:preMeasuement}
\end{align}
which is highlighted by the red color in the $Z_e$ operator.

\noindent{\bf Step (iii):} We write the measurement outcomes as $s_{v}=0,1$ in the $X$ basis. 
After contracting eq.~(\ref{eq:preMeasuement}) with   $\bigotimes_{v \in V} \langle s_v |_V^{(X)}$ and using eq.~(\ref{eq:contractMeaurement}), the post-measurement wave function is written as (up to an unimportant normalization constant)
\begin{align}
&\left\{ \prod_{e \in E } \exp(- i \Delta t Z_e ) 
\prod_{v \in V } \exp(- i \Delta t \lambda \textcolor{red}{(-1)^{s_v} } \prod_{e \supset v} X_e ) 
\right\}^k \nonumber \\
&\times \sum_{ \substack{ a_v \in \{0,1\} \,  \\ \, v \in V } } 
C(\{a_v\}) \textcolor{red}{(-1)^{a_v s_v} }
\bigotimes_{e \in E} \big| \sum_{v \subset e} a_v \big\rangle_e  \nonumber \\
 &= \!\left\{ \prod_{e \in E } \exp(- i \Delta t Z_e ) 
\prod_{v \in V } \exp(- i \Delta t \lambda \textcolor{red}{(-1)^{s_v} } \prod_{e \supset v} X_e ) \!
\right\}^k \nonumber \\
&\times \mathcal{O}_{\rm{bp}}(\gamma) \sum_{ \substack{ a_v \in \{0,1\} \,  \\ \, v \in V } } 
C(\{a_v\}) 
\bigotimes_{e \in E} \big| \sum_{v \subset e} a_v \big\rangle_e,
\end{align}
where we have used the relation \eqref{eq:rewriting-phase-1d} to replace the phase factor in the summand with the byproduct operator $\mathcal{O}_{\text{bp} }(\gamma)$. 
We note that the commutation relation 
\begin{align}
\prod_{e \supset v} X_e  \times \mathcal{O}_{\text{bp} }(\gamma)
= (-1)^{s_v } \mathcal{O}_{\text{bp} }  (\gamma)\times \prod_{e \supset v} X_e   
\end{align}
can be used to obtain 
\begin{align}
& \mathcal{O}_{\text{bp}}(\gamma)
\left\{ \prod_{e \in E } \exp(- i \Delta t Z_e ) 
\prod_{v \in V } \exp(- i \Delta t \lambda  \prod_{e \supset v} X_e ) 
\right\}^k \nonumber \\
&\times |\psi^{*}_\text{in} \rangle  .
\end{align}
Next, we ask: how can we handle the byproduct operator?

\smallskip
\noindent{\bf Step (iv):} With the measurement outcomes, we choose a set of edges $\gamma' \subset E$ just as we defined $\gamma$. 
We construct a counter operator 
\begin{align}
\mathcal{O}_{\text{counter}}(\gamma') = \prod_{e \in \gamma'} Z_{e} .
\end{align}
We note that due to the ambiguity (or freedom) in constructing $\gamma$ for the byproduct operators, we have the following relation:
\begin{align}
\mathcal{O}_{\text{bp}}(\gamma) \times 
\mathcal{O}_{\text{counter}} (\gamma') = 1 \text{ or } U^{(0,\text{dual})} .
\end{align}
Either operator on the right-hand side acts trivially on the post-measurement wave function due to $[H_{\text{dual}}, U^{(0,\text{dual})}] = 0$ and $U^{(0,\text{dual})}  |\psi^{*}_\text{in} \rangle  =  |\psi^{*}_\text{in} \rangle $.
Therefore, after the feedforwarded correction, we obtain the dualized time-evolved wave function,
\begin{align}
& 
\left\{ \prod_{e \in E } \exp(- i \Delta t Z_e ) 
\prod_{v \in V } \exp(- i \Delta t \lambda  \prod_{e \supset v} X_e ) 
\right\}^k |\psi^{*}_\text{in} \rangle  .
\end{align}

We have encountered a bit of clutter in writing some of the above equations. 
To suppress it, we will introduce a set of useful notations in the following subsection.
Then the rest of this paper is to apply the idea we just explained above to many other examples --- gauge theories with or without matter fields in (1+1)- or (2+1)-dimensions. Different from the one-spatial dimension here, in higher dimensions, we will be able to explore phases that contain intrinsic topological order.

\subsection{Setting up machinery}

Algebraic topology~\cite{MR1867354} is a useful tool to study topological quantum codes, which allows for straightforward generalization; see Ref.~\cite{2013arXiv1311.0277B} for an introduction. 
In this section, we first provide homological terminology, which will be useful as a shorthand notation and to understand the mechanism of the duality transformation in a unified manner.
Readers who are only interested in the twisted gauge theory and Majorana fermion QED can skip this subsection as we will not use this language there.

\begin{figure}
\includegraphics[width=0.7\linewidth]{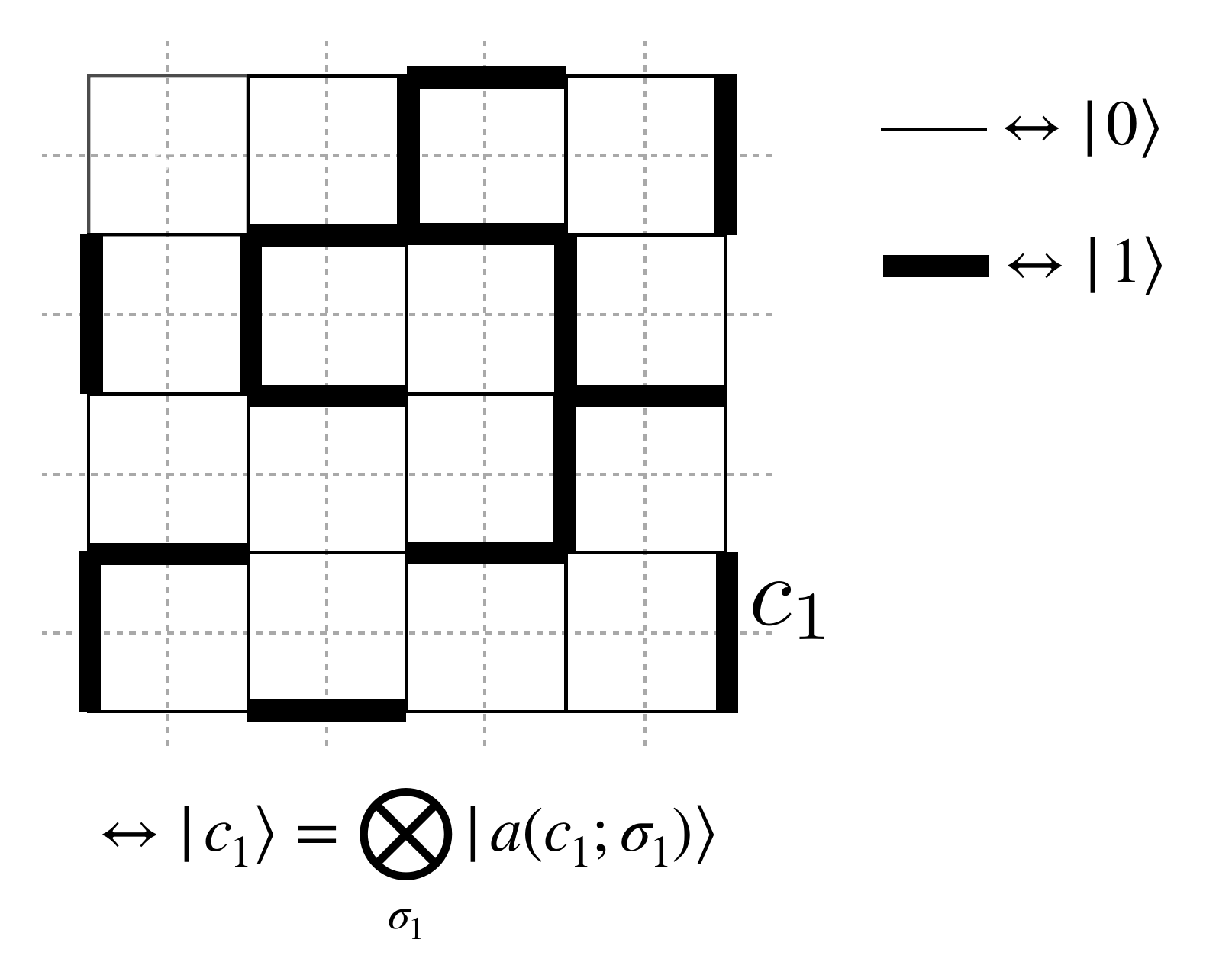}
\caption{An example of 1-chain $c_1$, which defines a basis $|c_1\rangle$ in the Hilbert space. The solid lines correspond to $1$, while ordinary thin lines to $0$. }
\label{fig:chain}
\end{figure}

In $d$-dimensional lattices, let $V$ (also denoted as $\Delta_0$) be the set of vertices, $E$ (or $\Delta_1$) the set of edges, and $P$ (or $\Delta_2$) the set of plaquettes, and so on.
We also write the elements $v$ as $\sigma_0$, $e$ as $\sigma_1$, and $p$ as $\sigma_2$, and
they are called 0-, 1-, and 2-cells, respectively.
For the cyclic group $\mathbb{Z}_2$, we introduce an $i$-chain as a formal linear combination with $\mathbb{Z}_2$ coefficients, $a(c_i ; \sigma_i ) \in \{0,1 ~\text{mod}~2\}$; see Fig.~\ref{fig:chain}:
\begin{align}
c_i & = \sum_{\sigma_i  \in \Delta_i} a(c_i ; \sigma_i) \sigma_i. 
\end{align} 
The boundary operator $\partial$ is a map such that $\partial \sigma_i$ is a sum of $(i-1)$-cells that appear in the boundary of $\sigma_i$. 
We also make use of the dual lattice, and denote the dual vertices as $v^* \in V^*$ (or $\sigma^*_0 \in \Delta^*_0$), dual edges as $e^* \in E^*$ (or $\sigma^*_1 \in \Delta^*_1$) and dual plaquettes as $p^* \in P^*$ (or $\sigma^*_2 \in\Delta^*_2$).
Both the dual chains $c^*_i$ ($i=0,1,2$) and the dual boundary $\partial^*$ are defined in the same manner as for the primal ones.
Note that there is a natural identification between the primal cells and the dual cells:
\begin{align} \label{eq:dual-primal-identification}
\sigma^*_{i}  \simeq \sigma_{d-i} \ \quad (d\text{-dimensions}) . 
\end{align}
We write the set that consists of all possible $i$-chains as $C_i$.

\begin{figure}
\includegraphics[width=0.6\linewidth]{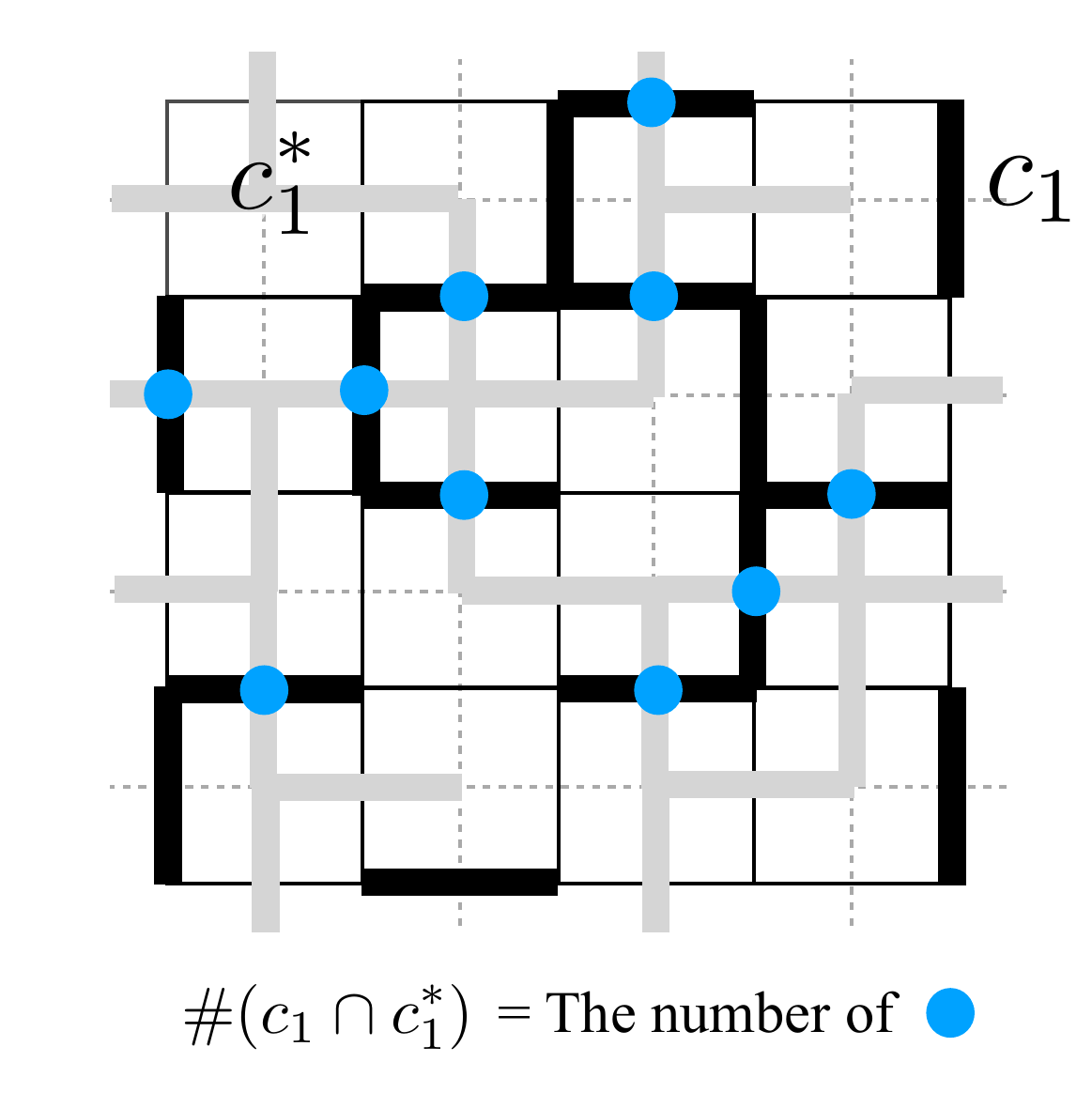}
\caption{An example of the intersection pairing number $\#(c_1\cap c^*_1)$ between a 1-chain $c_1$ and a dual 1-chain $c^*_1$.}
\label{fig:intersection}
\end{figure}

For a pair of chains $c_i$ and $c^*_{d-i}$ in $d$-dimensions we define the intersection pairing 
\begin{align}
\#(c_i \cap c^*_{d-i}) = \sum_{\sigma_i \in \Delta_i} a(c_i; \sigma_i) a(c^*_{d-i}; \sigma^*_{d-i}) \quad {\rm mod} ~ 2.
\end{align}
Intuitively, $\#(c_i \cap c^*_{d-i})$ counts the number of overlaps between $c_i$ and $c^*_{d-i}$; see Fig.~\ref{fig:intersection}.
The duality relation,
\begin{align} \label{eq:duality}
\#(\partial c_{i+1} \cap c^*_{d-i} ) = \#( c_{i+1} \cap \partial^* c^*_{d-i} ),
\end{align}
for $i=0,...,d-1$ and the nilpotency of the boundary operators, {\it i.e.},
\begin{align}
\partial^2 = 0 , \qquad (\partial^*)^2 = 0, 
\end{align}
will become useful later.

We will have many-body interaction terms in Hamiltonians.
As a convention, we express a product of operators $A$ supported on multiple $i$-cells, with which we associate an $i$-chain $c_i$, as
\begin{align}
A(c_i) := \prod_{\sigma_i \in \Delta_i} A^{a(c_i;\sigma_i)}_{\sigma_i}. 
\end{align}

Note the operators to be considered in this paper are mostly the Pauli operators, such as $X$ and $Z$.
We denote the eigenvectors of the Pauli operators as
\begin{align}
Z|s\rangle = (-1)^s|s\rangle \qquad (s=0,1), \\
X|\tilde{s}\rangle = (-1)^s | \tilde{s}\rangle \qquad (s=0,1).
\end{align}
Namely,  $|{0} \rangle $ and $|{1} \rangle $ are the basis states in the standard $Z$ basis whereas $|\tilde{0} \rangle = |+\rangle=(|0\rangle+|1\rangle)/\sqrt{2}$ and $|\tilde{1} \rangle = |-\rangle=(|0\rangle-|1\rangle)/\sqrt{2}$ are in the dual ($X$) basis.

We define a basis of a wave function on $i$-cells with the following product states; see Fig.~\ref{fig:chain}:
\begin{align} \label{eq:Z2-bases-Z}
&| c_i \rangle := \bigotimes_{\sigma_i \in \Delta_i} | a(c_i ; \sigma_i) \rangle^{(Z)}_{\sigma_i}, \\ 
&| \widetilde{c_i} \rangle := \bigotimes_{\sigma_i \in \Delta_i} | a(c_i ; \sigma_i) \rangle^{(X)}_{\sigma_i}, 
 \label{eq:Z2-bases-X}
\end{align}
where the superscripts $(Z)$ and $(X)$ denote that they are eigenvectors of respective operators~\footnote{
In figures etc, we often abbreviate $a(c_i ; \sigma_i)$ as $a_{k} \in \{0,1\}$ where $k$ is a label for $i$-cells $\sigma_i$.
For a 0-chain $c_0$, for example, the basis $|c_0 \rangle = | a_1, a_2, ... \rangle$ consists of a bit string in the computational basis for qubits defined on 0-cells (vertices).
The $Z$ eigenvalue 0/1 corresponds to the coefficient in the linear expansion of the 0-chain with the 0-cells.  
}.
A useful example of this notation is the product state of the $Z$ eigenvectors $|0\rangle$ living on $i$-cells, where the associated $i$-chain is simply zero, $0_i$:
\begin{align}
|0_i \rangle =  
\bigotimes_{\sigma_i \in \Delta_i} | 0 \rangle^{(Z)}_{\sigma_i} .
\end{align}
We also use a notation such as
\begin{align}
|c_i \rangle \otimes | c_j \rangle 
= |c_i, c_j \rangle 
\end{align}
to indicate a tensor product of two (or more) Hilbert spaces.

Our notation allows us to write multi-qubit operations compactly, {\it e.g.}, 
\begin{align}
Z(c_i) |c'_i\rangle &= (-1)^{\#(c_i \cap c'_i)} |c'_i\rangle ,  \label{eq:phase-chain}\\
X(c_i) |c'_i\rangle &= |c'_i+c_i\rangle  , \label{eq:shift-chain}\\
 X(c_i) Z(c'_i)&= (-1)^{\#(c_i \cap c'_i)}  Z(c'_i) X(c_i)  , \label{eq:commutation-chain}\\
\langle \widetilde{c_i} |c'_i\rangle &= \frac{1}{2^{|\Delta_i|/2}}(-1)^{\#(c_i \cap c'_i)}  . \label{eq:inner-product-chain}
\end{align}
In the previous example with the 1D transverse-field Ising models, the equations would be written as, {\it e.g.,}
\begin{align}
\text{Eq.~\eqref{eq:1dtfi-ham} :} \quad 
&H = - \lambda \sum_{\sigma_0 \in \Delta_0} X(\sigma_0) 
- \sum_{\sigma_1 \in \Delta_1} Z(\partial \sigma_1) , \nonumber \\
\text{Eq.~\eqref{eq:1dtfi-ham-dual}:} \quad 
&H_{\text{dual} }
=  -  \lambda\sum_{\sigma^*_1 \in \Delta^*_1} X(\partial^*\sigma^*_1) 
- \sum_{\sigma^*_0 \in \Delta^*_0} Z(\sigma^*_0) , \nonumber  \\
\text{Eq.~\eqref{eq:sym-ini-state-1d}:} \quad 
&
|\psi_{\text{ini} } \rangle 
= \sum_{c_0 \in C_0} C(c_0) |c_0 \rangle ,
\nonumber
\end{align}
and we will see the usefulness of this language in the following sections. This language is less suitable with some examples in which the homological structure is not prominent, {\it e.g.}, twisted gauge theories with cocycle factors, or Majorana fermion QED. In these cases, we shall not stick to the notation we have just presented here.

\subsection{Lattice models in 2D}
\label{sec:Lattice-models-in-2D}

\begin{figure}
\includegraphics[width=0.9\linewidth]{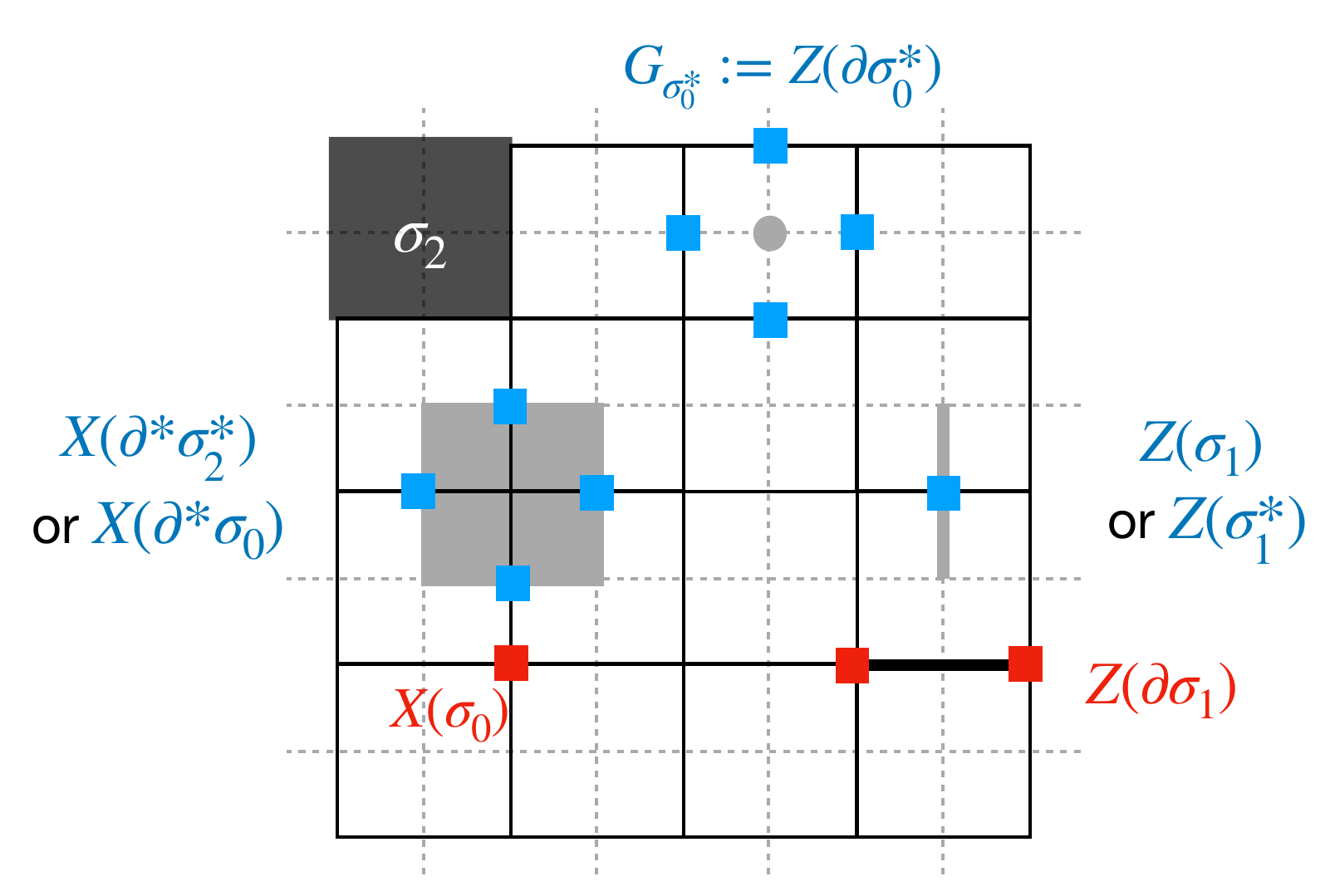}
\caption{The 2d square lattice and its dual. The red boxes represent the operators that appear in the Hamiltonian $H^{\text{TFI}}$ and the blue ones in the Hamiltonian $H^{\text{GT}}$ and the generator of the gauge transformation.}
\label{fig:2dlattice}
\end{figure}

Consider two models in (2+1)-dimensions related by the Kramers-Wannier duality.
Namely, the transverse-field Ising model (TFI),
\begin{align}
H_{\text{TFI}} =  - \sum_{\sigma_1 \in \Delta_1} Z(\partial \sigma_1) - \lambda \sum_{\sigma_0 \in \Delta_0} X({\sigma_0}),
\end{align}
and the gauge theory (GT),
\begin{align}
H_{\text{GT}}= - \sum_{\sigma^*_1 \in \Delta^*_1} Z({\sigma^*_1}) - \lambda \sum_{\sigma^*_2 \in \Delta^*_2} X(\partial^* \sigma^*_2),
\end{align}
where the degrees of freedom of TFI are defined on the primary lattice, while those of GT live on the dual lattice;
see Fig.~\ref{fig:2dlattice}.
The first-order Trotter decomposition for the real-time evolution of the respective model is given by
\begin{align}
T^{\text{TFI}}(t) &= \Bigl( \prod_{\sigma_0 \in \Delta_0} e^{ i \Delta t \lambda X_{\sigma_0}}
\prod_{\sigma_1 \in \Delta_1} e^{ i \Delta t Z(\partial \sigma_1)} \Bigr)^{k},  \\
T^{\text{GT}}(t) &= \Bigl( \prod_{\sigma^*_2 \in \Delta^*_2} e^{ i \Delta t \lambda X(\partial^* \sigma^*_2) }
\prod_{\sigma^*_1 \in \Delta^*_1} e^{ i \Delta t Z_{\sigma^*_1}} \Bigr)^{k}, 
\end{align}
with $t=k \Delta t$. 
We note that in the above for single vertices or edges, we still use the notations, $X_{\sigma_0}$ and $Z_{\sigma_1^*}$.

\subsection{Gauging and measuring symmetric wave functions}
\label{sec:Gauging-and-measuring-symmetric-wave functions}

\begin{figure}
\includegraphics[width=\linewidth]{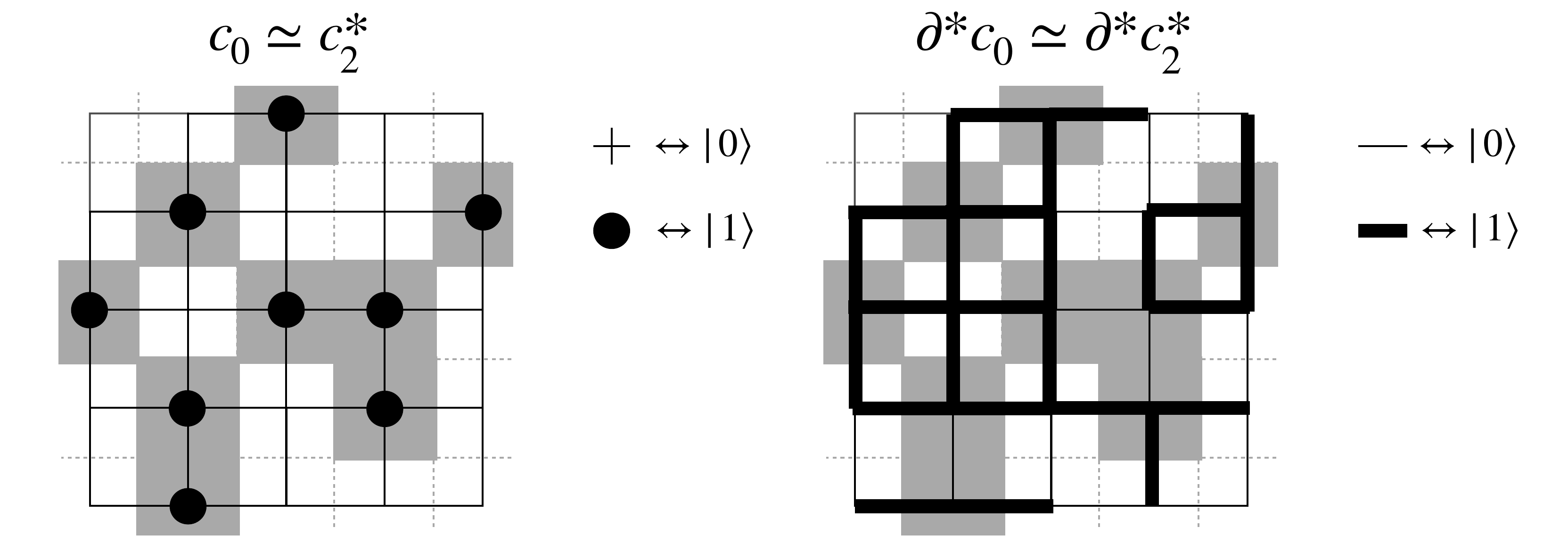} 
\caption{(Left) The ungauged basis $|c_0\rangle$ is a tensor product of $Z$ basis vectors whose eigenvalues are coefficients in the expansion of 0-chain $c_0$ with the 0-cell basis. A 0-chain is naturally identified with a dual 2-chain. (Right) The gauged basis $|\partial^* c_0\rangle$ is a tensor product of bits living on edges, each of which is a sum of bits on vertices surrounding it.}
\label{fig:gauging-basis}
\end{figure}

Now we define the (un)gauged wave function \cite{2012PhRvB..86k5109L,2018arXiv180501836K} as
\begin{align} 
&|\psi^{(0)}_{\text{ungauged}}\rangle
= \sum_{c_0 \in C_0} C(c_0) | c_0 \rangle, \label{eq:Z2-ungauged}  \\
&|\psi^{(1)}_{\text{gauged}}\rangle
= \sum_{c_0 \in C_0} C(c_0) | \partial^* c_0 \rangle , \label{eq:Z2-gauged}
\end{align}
where $C(c_0)$ is a complex coefficient, and $\partial^* c_0$ represents the $\mathbb{Z}_2$ sum of the bits between adjacent vertices connected by the corresponding edge as illustrated in Fig.~\ref{fig:gauging-basis}.
$|\psi^{(0)}_{\text{ungauged}}\rangle$ is assumed to be a $\mathbb{Z}_2$ symmetric wave function, namely,
\begin{align} \label{eq:Z2-global-symmetry-ungauged}
\prod_{\sigma_0 \in \Delta_0} X_{\sigma_0} |\psi^{(0)}_{\text{ungauged}}\rangle= |\psi^{(0)}_{\text{ungauged}}\rangle,
\end{align}
or equivalently
\begin{align}
C(c_0)=C(c_0+\sum_{\sigma_0 \in \Delta_0} \sigma_0). 
\end{align}

Consider measurements in the $X$-basis.
We write the measurement outcome at $\sigma_0$ as $s(\sigma_0) \in \{0,1\}$ to construct an associated chain
\begin{align} \label{eq:Z2-measurement-chain}
 s_0  := \sum_{\sigma_0 \in \Delta_0} s(\sigma_0) \sigma_0 \ . 
\end{align}
After measurements, we have a product state associated with a set of measurement outcomes, $|\widetilde{s_0}\rangle$ (we remind readers that the symbol~\,\,$\widetilde{}$\,\, indicates the $X$ basis).
The $\mathbb{Z}_2$ symmetry implies
\begin{align}
\langle \widetilde{s_0}| \prod_{\sigma_0 \in \Delta_0} X_{\sigma_0}|\psi^{(0)}_{\text{ungauged}}\rangle
&= \prod_{\sigma_0 \in \Delta_0} (-1)^{s(\sigma_0)} \langle \widetilde{s_0}   |\psi^{(0)}_{\text{ungauged}}\rangle \nonumber  \\
&= \langle \widetilde{s_0} |  \psi^{(0)}_{\text{ungauged}}\rangle.
\end{align}
Therefore, we obtain the constraint,
\begin{align}
\prod_{\sigma_0 \in \Delta_0} (-1)^{s(\sigma_0)} = 1,
\end{align}
meaning that the number of the non-trivial outcomes at 0-cells $\sigma_0$'s with $s(\sigma_0)=1$ is even.

On the other hand, the gauged wave function lives on 1-cells, and the global symmetry is promoted to a local symmetry generated by
the gauge transformation associated with each dual vertex $\sigma^*_0 \in \Delta^*_0$:
\begin{align}
&G_{\sigma^*_0} |\psi^{(1)}_{\text{gauged}} \rangle = |\psi^{(1)}_{\text{gauged}} \rangle, \label{eqn:Gsigma} \\
&G_{\sigma^*_0}:= Z(\partial \sigma^*_0). 
\end{align}
The operator $Z(\partial \sigma^{*}_0)$ is a ``divergence operator" in $\mathbb{Z}_2$ electromagnetism as the product of $Z$ is taken over dual edges wrapping around a dual vertex. (Note the identification $\partial \sigma^{*}_0 = \partial \sigma_2$ with $ \sigma^{*}_0 \simeq \sigma_2$.)
The gauge invariance is a consequence of the homological structure:
for each basis of the gauged wave function in eq.~\eqref{eq:Z2-gauged}, we see that
\begin{align}
Z(\partial \sigma^*_0) | \partial^* c_0 \rangle &\overset{\text{eq.~\eqref{eq:phase-chain}}}{=} (-1)^{\# ( \partial \sigma^*_0 \cap \partial^* c_0)} | \partial^* c_0 \rangle \nonumber \\
&\overset{\text{eq.~\eqref{eq:duality}}}{=}
| \partial^* c_0 \rangle . 
\end{align}

The symmetry above is a manifestation of the Gauss law constraint without charges in the present case.
The Hamiltonian of the gauge theory is indeed invariant under the symmetry transformation: $[H^{\text{GT}}, G_{\sigma^*_0} ]=0$.

\subsection{Kramers-Wannier transformation of time evolution} \label{sec:GT}

We now move on to investigate the Kramers-Wannier transformation of the unitary evolution and present our results below.
\subsubsection{Results}

\begin{figure*}
(a)
\includegraphics[width=0.20\linewidth]{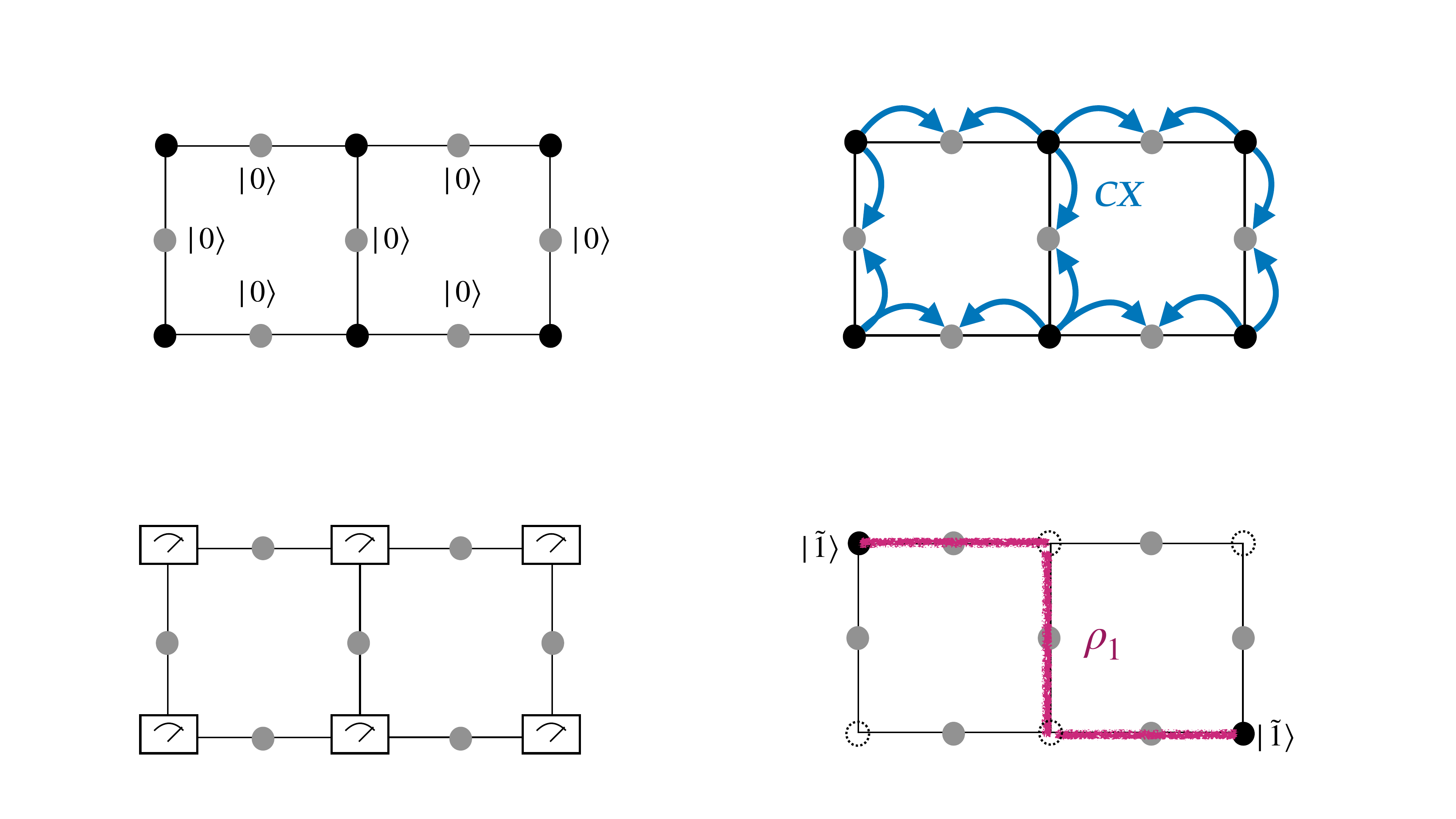}
(b)
\includegraphics[width=0.20\linewidth]{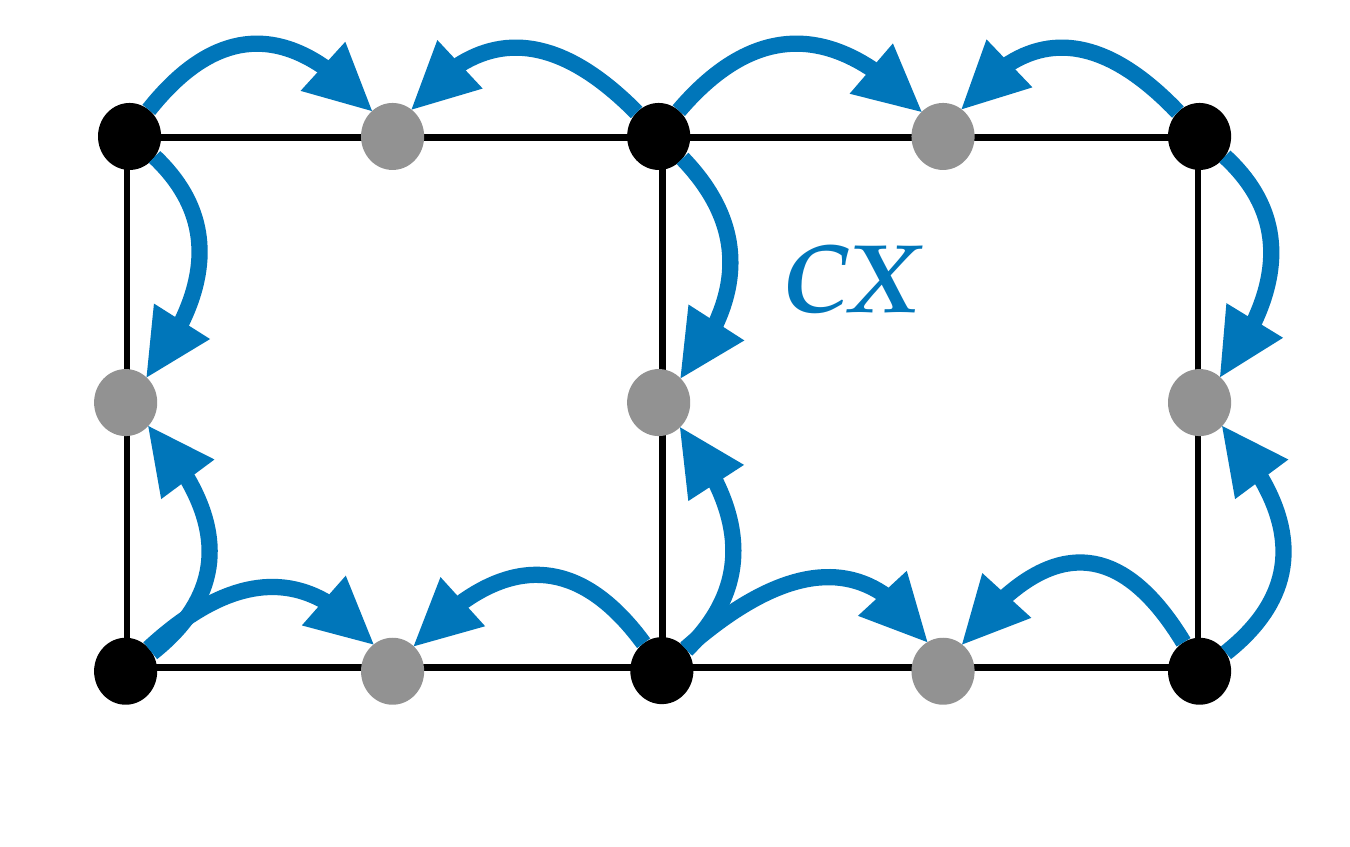}
(c)
\includegraphics[width=0.20\linewidth]{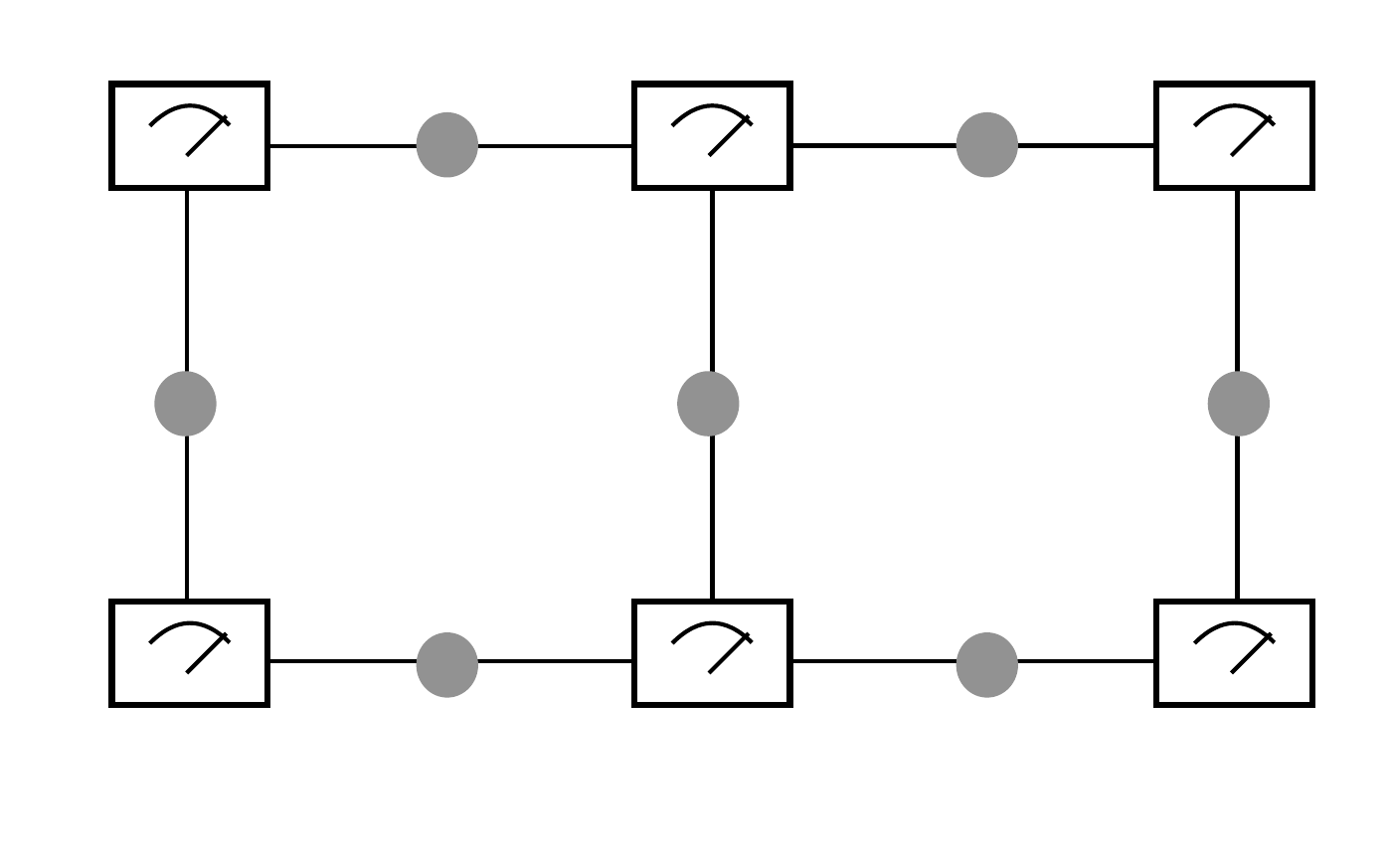}
(d)
\includegraphics[width=0.24\linewidth]{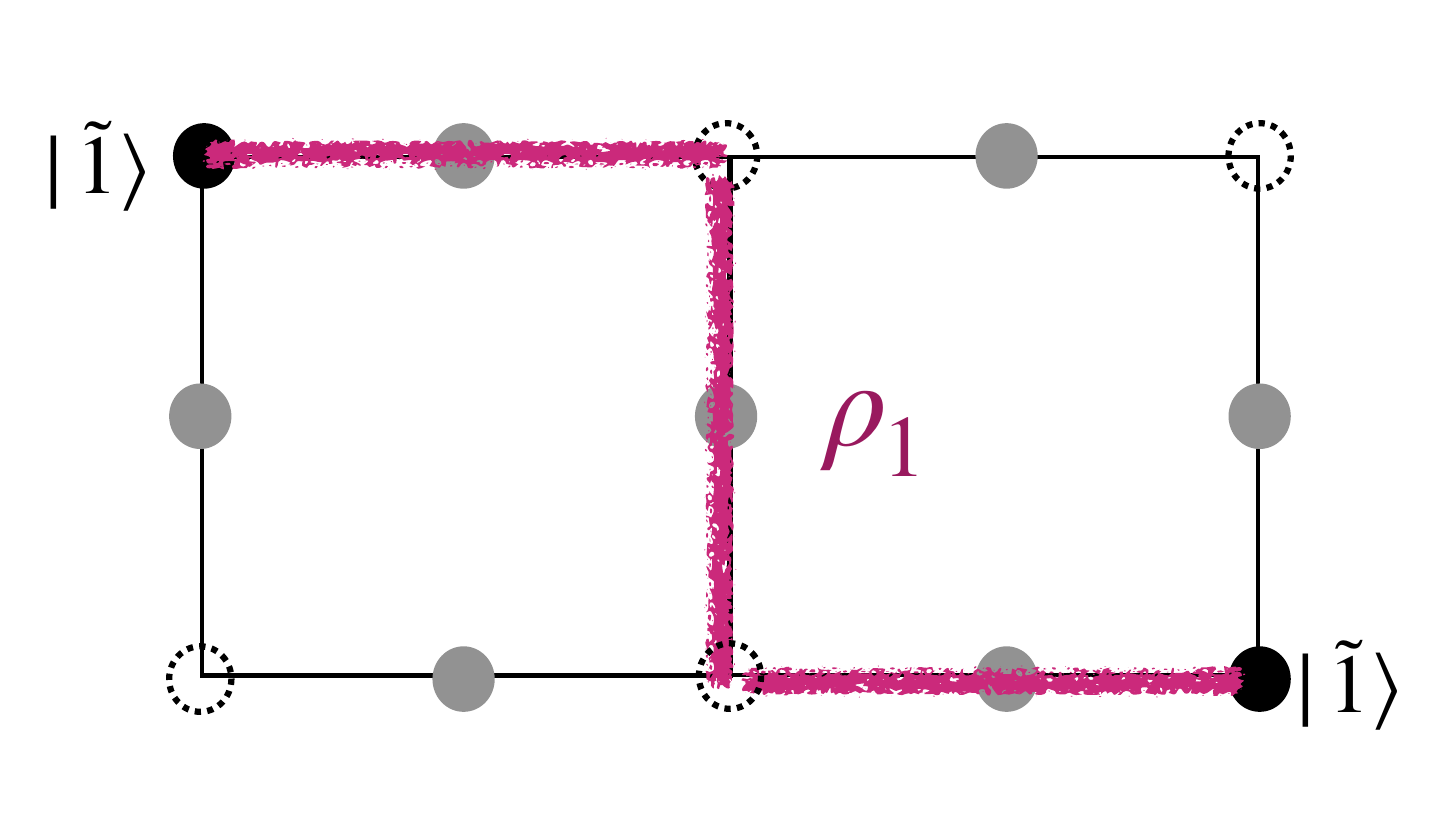}
\caption{Graphic explanation of eq.~\eqref{eq:3-2-duality}. (a) On vertices, we have the time-evolved wave function, $T^{\rm TFI}(t) |\psi^{(0)}_{\rm ungauged} \rangle$. 
We introduce the ancillary product state $|0\rangle$ to the edges of the lattice.
(b) We apply the entangler $\mathcal{U}_{\rm KW}$ which consists of the $CX$ gates whose controlling qubits are on vertices and target qubits are on edges.
(c) We measure the vertex degrees of freedom with the $X$ basis. 
(d) The byproduct operator $\mathcal{O}_{\rm bp}$ is a product of Pauli $Z$ operators supported on a set of paths $\rho_1$, whose endpoints are at the vertices with $s_v =1$. }
\label{fig:2dKW-Ising-gauge}
\end{figure*}

We use the Kramers-Wannier map introduced in~\cite{2021arXiv211201519T}, which is given by
\begin{align} \label{eq:kw-map}
\widehat{\text{KW}}|\psi\rangle
&= \langle
\widetilde{s_0} | \,
\mathcal{U}_{\text{KW}} \, 
|0_1\rangle \otimes |\psi\rangle , \\
\mathcal{U}_{\text{KW}} 
&= \prod_{
\substack{ \sigma_0 \in \Delta_0 \\ \sigma_1 \in \Delta_1}  } \big[CX_{\sigma_0 \sigma_1}\big]^{a(\partial^* \sigma_0 ; \sigma_1)},
\end{align}
where $CX_{ct}$ is the controlled-$X$ gate defined as
\begin{align}
CX_{ct} = |0\rangle \langle 0 |_c \otimes I_t +  |1\rangle \langle 1 |_c \otimes X_t ,
\end{align}
and $|\psi\rangle$ is a wave function defined on 0-cells.
As indicated by the power $a(\partial^* \sigma_0;\sigma_1)$, which is 1 when an edge and a vertex are adjacent to each other and 0 otherwise, the entangler $\mathcal{U}_{\text{KW}}$ consists of $CX$ gates, each of which is controlled by a vertex qubit and applies $X$ on qubits on edges that surround the vertex qubit as illustrated in Fig.~\ref{fig:2dKW-Ising-gauge}~(b).

\begin{figure*}
\includegraphics[width=0.6\linewidth]{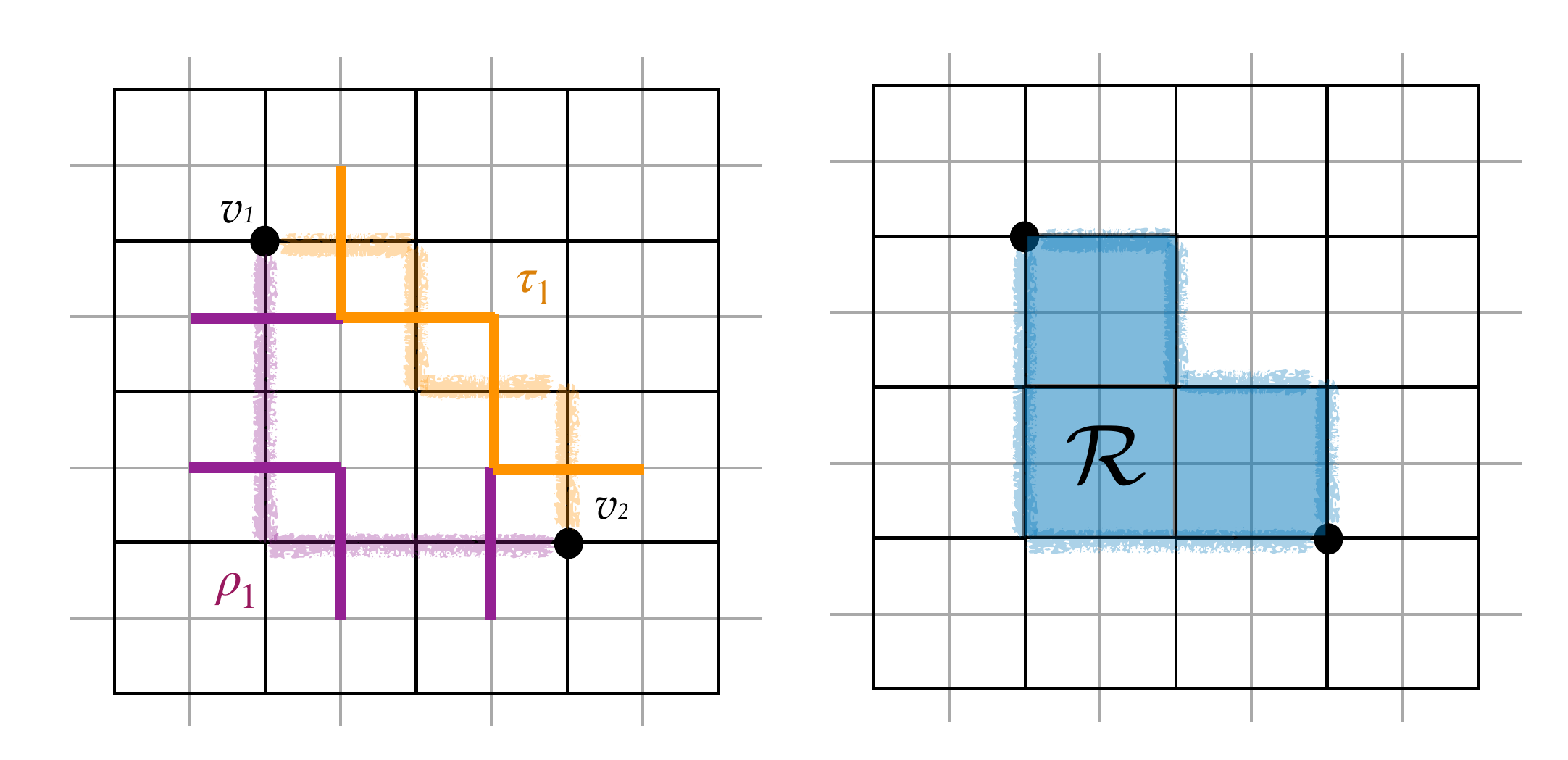}
\caption{An example of the byproduct operator $\mathcal{O}_{\text{bp}}$ resulting of the randomness of measurements.
The vertices $v_1$ and $v_2$ has nontrivial measurement outcomes: $s_0 = v_1 + v_2$.
The Pauli $Z$ operator is applied along the path $\rho_1$ such that $\partial \rho_1 = s_0$, depicted with purple lines. 
Magnetic monopoles are induced at these two vertices (dual plaquettes).
We apply a counter phase operator $\mathcal{O}_{\text{counter}}$ along the path $\tau_1$ (orange) such that it ends at $v_1$ and $v_2$; $\partial \tau_1 = s_0$.
The resulting phase operator $\mathcal{O}_{\text{bp}} \times \mathcal{O}_{\text{counter}} $ is the product of $G_{\sigma^*_0}$ over dual vertices inside $\mathcal{R}$. 
}
\label{fig:Diracstring}
\end{figure*}

Since the number of nontrivial outcomes $s(\sigma_0)=1$ is even, one can construct pairs of 0-cells with nontrivial outcomes and connect them with a set of paths that consist of 1-cells.
For a general set of measurement outcomes $s(\sigma_0) \in \{0,1\}$ and the associated 0-chain $s_0$, we define a 1-chain $\rho_1$ as
\begin{align}
\rho_1 = \sum_{\sigma_1 \in \Delta_1} a(\rho_1; \sigma_1) \sigma_1  \quad \text{such that} \quad \partial \rho_1 = s_0  
\end{align}
with $a(\rho_1; \sigma_1)\in \{0,1\}$.
Then the byproduct operator $\mathcal{O}_{\text{bp}}$ for this duality map is given by
\begin{align} \label{eq:bp-Z2gauge}
\mathcal{O}_{\text{bp}}(\rho_1) = Z(\rho_1)  ,
\end{align}
which is illustrated in Fig.~\ref{fig:Diracstring}.

Now, our claim is summarized by the following expression:
\begin{tcolorbox}
[width=\linewidth, sharp corners=all, colback=white!95!black]
\vspace{-11pt}
\begin{align} \label{eq:3-2-duality}
&\mathcal{O}_{\text{bp}} (\rho_1)\cdot T^{\text{GT}}(t)  | \psi^{(1)}_{\text{gauged}} \rangle  \nonumber \\ 
& = \widehat{\text{KW}} \cdot T^{\text{TFI}} (t) | \psi^{(0)}_{\text{ungauged}} \rangle   \ . 
\end{align}
\end{tcolorbox}
\noindent We refer readers to Fig.~\ref{fig:2dKW-Ising-gauge} for a graphic illustration of the procedure.
Furthermore, we note that the equality holds up to an unimportant normalization constant $ 2^{-|\Delta_0|/2} $.

By taking the limit $\Delta t \rightarrow 0$ with $t = k \Delta t$ fixed, we also establish the following corollary:
\begin{tcolorbox}
[width=\linewidth, sharp corners=all, colback=white!95!black]
\noindent \text{{\bf Corollary }}
\begin{align}
&\mathcal{O}_{\text{bp}}(\rho_1) \cdot e^{-it H_{\text{GT}}}  | \psi^{(1)}_{\text{gauged}} \rangle  \nonumber \\ 
&= \widehat{\text{KW}}\cdot  e^{-it H_{\text{TFI}}}  | \psi^{(0)}_{\text{ungauged}} \rangle \ .
\end{align}
\end{tcolorbox}

The byproduct operator $\mathcal{O}_{\text{bp}}(\rho_1)$ is written with a sum of paths whose endpoints are at 0-cells with non-trivial outcomes.
It can be removed in the following way. 
Since after the dualization procedure we have the data of measurement outcomes, we can construct a 1-chain $\tau_1$ whose endpoints are again at 0-cells with non-trivial outcomes, $\partial \tau_1 = s_0$, as illustrated in Fig.~\ref{fig:Diracstring}.
Then we apply the counter phase operator given by
\begin{align}
\mathcal{O}_{\text{counter}}(\tau_1)
= Z(\tau_1) . 
\end{align}
The net effect of the byproduct operator will be
\begin{align}
\mathcal{O}_{\text{counter}}(\tau_1) \times \mathcal{O}_{\text{bp}} (\rho_1)
= Z(\tau_1 + \rho_1) ,
\end{align}
which satisfies $\partial(\tau_1 + \rho_1)=0$, {\it i.e.}, it is a closed loop $Z$ operator; see Fig.~\ref{fig:Diracstring}.
It is a product of the Gauss law operators
\begin{align}
Z(\tau_1 + \rho_1) = \prod_{\sigma^*_0 \in \mathcal{R}} G(\sigma^*_0),
\end{align}
with $\partial \mathcal{R}= \tau_1 + \rho_1$  (see Fig.~\ref{fig:Diracstring}), and thus the action of such operator on the physical Hilbert space is always trivial: $Z(\tau_1 + \rho_1)=1$.
We emphasize that because of this mechanism, the dualization is indeed deterministic.

\subsubsection{Observation}

We move on to investigate the Kramers-Wannier transformation of the unitary evolution.
To be pedagogical, we first provide some elementary calculations in a more transparent notation, and later we will switch to the notation with algebraic topology to be more efficient. 

First, for a pair of adjacent vertices $u,u' \in V$, $\langle u,u' \rangle \in E$, we have
\begin{align}
& \langle \{\tilde{s}_v\}|_V \prod CX_{v,e} e^{i \xi Z_u Z_{u'}} |0\rangle^{\otimes E} 
|\psi^{(0)}_{\text{ungauged}}\rangle_{V} \nonumber \\
&= \Big( \prod_{e \in \text{string}} Z_e \Bigr) \, 
e^{i \xi Z_e} | \psi^{(1)}_{\text{gauged}} \rangle_{E}  ,
\end{align}
where $\langle \{\tilde{s}_v\}|_V \equiv \langle \widetilde{s_0 }|$ is the $X$ eigenvectors on vertices with eigenvalues $s_v=0,1$, and we have suppressed an overall normalization constant on the right-hand side of the equality.
Since the number of the non-trivial outcomes $s_v=1$ is even, one can construct a set of paths on edges, which we call ``string", to pair up vertices with non-trivial outcomes, see Fig.~\ref{fig:Diracstring}.
Taking the inner product between $\langle \{\tilde{s}_v\}|_V$ and $|\psi^{(0)}_{\text{ungauded}}\rangle_V$, the basis $|c_0\rangle \equiv |\{a_v\}\rangle_V$ gives us a phase $\prod_{v \in V} (-1)^{s_v a_v}$.
For a simple example, let us say we have $s_w=1$, $s_{w'}=1$ ($w,w' \in V$) and $s_v=0$ otherwise.
The phase is then $(-1)^{a_w}(-1)^{a_{w'}}$.
The same phase is obtained by acting $Z$ operators on {\it edge} degrees of freedom along the path connecting $w$ and $w'$, since 
\begin{align}
&\Big(\prod_{e \in \text{string}} Z_e \Big) |\{a_v+a_{v'} \} \rangle_E  \nonumber \\
&= \Big(\prod_{ \langle v , v' \rangle  \in  \text{string}} (-1)^{a_v + a_{v'}} \Big) |\{a_v+a_{v'} \} \rangle_E  \nonumber \\
&= (-1)^{a_w}(-1)^{a_{w'}}  |\{a_v+a_{v'} \} \rangle_E   ,
\end{align}
where $|\partial^* c_0\rangle \equiv |\{a_v+a_{v'} \} \rangle_E$ is the basis for the gauged wave function.

As the next stepping stone, we consider the Kramers-Wannier transformation of the other unitary evolution, which is calculated as follows:
\begin{align}
& \langle \{\tilde{s}_v\}|_V \prod CX_{v,e} e^{i \xi X_u} |0\rangle^{\otimes E} 
|\psi^{(0)}_{\text{ungauged}}\rangle_{V} \nonumber \\
&= 
e^{i \xi (-1)^{s_u} X(\partial^* u)} \nonumber \\
& \qquad \times
\sum_{a_v = 0,1} 
\Big( \prod_{v \in V} (-1)^{s_v a_v}\Bigr) 
C(\{a_v\}) | \{a_v+a_{v'}\} \rangle_E \nonumber \\
&= 
e^{i \xi (-1)^{s_u} X(\partial^* u)} \nonumber \\
&\times \qquad \Big( \prod_{e \in \text{string}}   Z_e \Bigr)  
\sum_{a_v = 0,1} 
C(\{a_v\}) | \{a_v+a_{v'}\} \rangle_E \nonumber \\
&=  \Big( \prod_{e \in \text{string}}   Z_e \Bigr)  \, e^{i \xi  X(\partial^* u)} 
\sum_{a_v = 0,1} 
 C(\{a_v\}) | \{a_v+a_{v'}\} \rangle_E \nonumber \\
&=  \Big( \prod_{e \in \text{string}}   Z_e \Bigr) \, e^{i \xi  X(\partial^* u)}  | \psi^{(1)}_{\text{gauged}}\rangle_{E} \ ,
\end{align}
where $C(c_0) \equiv C(\{a_v\})$; see also eq.~(\ref{eq:Z2-ungauged}).
The short-hand notation $\partial^* u$ denotes the product over edges surrounding the vertex $u$ in the primary lattice.
In other words, $\prod_{e \in \text{nb}(u)} X_e =: X(\partial^* u)$.
In the 2d dual lattice picture, it is a product over edges surrounding a plaquette.
The second equality follows from the argument given above.
In the third equality, we handled the extra sign in $\exp(i \xi (-1)^{s_u} X(\partial^* u))$ (which is $-1$ only at the endpoints $u$'s of strings, which have $s_u=1$) using the commutation relation with $\prod_{e \in \text{string}} Z_e$.
The operators $ X(\partial^* u)$ and $\prod_{e \in \text{string}} Z_e$ anti-commute only at the endpoints of the strings, and that is precisely where the extra sign appears.

\subsubsection{Demonstration}

Now we switch to the notation with algebraic topology and provide the full proof of our statement.
Consider the state before the measurements,
\begin{align}
&\mathcal{U}_{\text{KW}} \, 
|0_1\rangle \otimes T^{\text{TFI}}(t)|\psi^{(0)}_{\text{ungauge}}\rangle= \nonumber \\
&
\prod_{
\substack{ \sigma_0 \in \Delta_0 \\ \sigma_1 \in \Delta_1}  } 
\!\! [CX_{\sigma_0 \sigma_1}]^{a(\partial^* \sigma_0 ; \sigma_1)}
\Bigl( \prod_{\sigma_0 \in \Delta_0} 
\! e^{ i \Delta t \lambda X_{\sigma_0}}
\prod_{\sigma_1 \in \Delta_1} \! e^{ i \Delta t Z(\partial \sigma_1)} \Bigr)^{k} \nonumber\\
& \qquad \times \Big(
\sum_{c_0 \in C_0} C(c_0)  |c_0, 0_1\rangle
\Big).
\end{align}
By propagating the controlled-$X$ operators, we have $X$ operators in the exponent conjugated as $X(\sigma_0) \mapsto X(\sigma_0) X(\partial^* \sigma_0) $ due to the commutation relation between the controlled-$X$ and $X$ gates, 
\begin{align}
\includegraphics[width=0.8\linewidth]{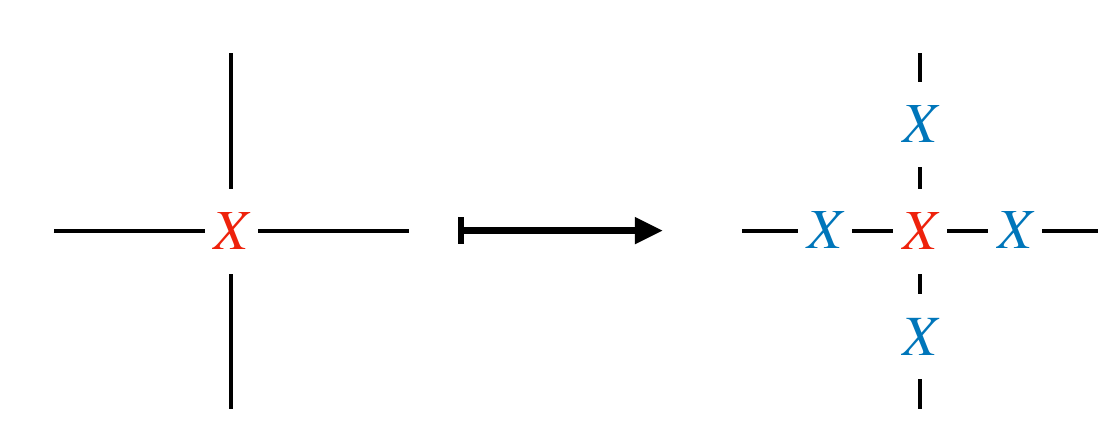} 
\nonumber
\end{align}
and the basis is mapped as $|c_0,0_1\rangle\mapsto |c_0,\partial^* c_0\rangle $:
\begin{align}
\includegraphics[width=\linewidth]{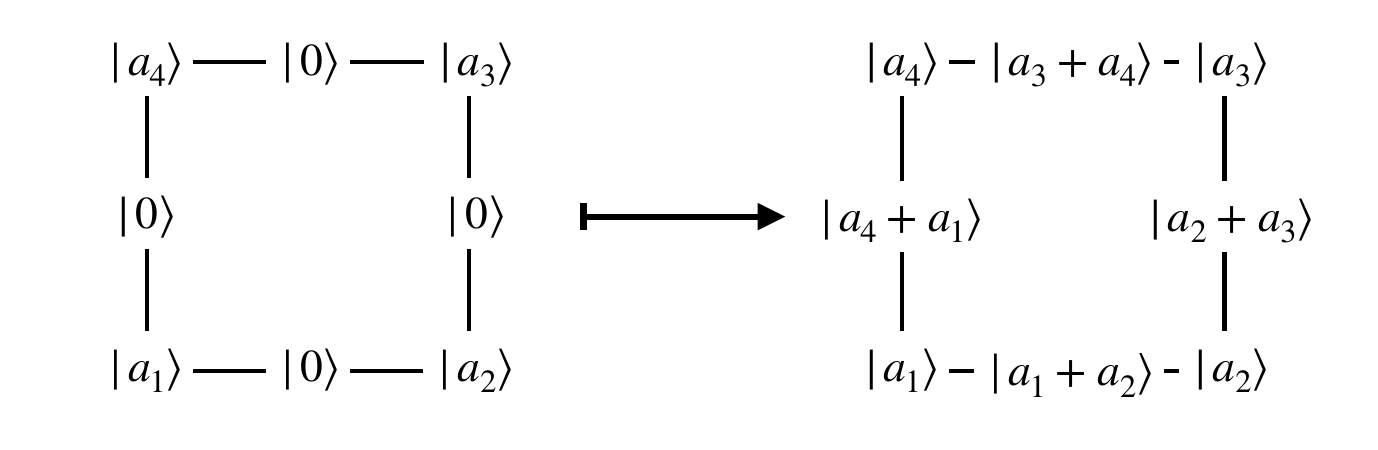} 
\nonumber
\end{align}
We also note that as the controls are on $\Delta_0$, the $Z$ operators remain the same under the controlled-$X$ operators: $Z(\sigma_0) \mapsto Z(\sigma_0)$. Thus, we find the pre-measurement wave function is equal to
\begin{align}
|\psi_{\text{pre}} \rangle = 
& 
\Big( \prod_{\sigma_0 \in \Delta_0 }
e^{i \Delta t \lambda X_{\sigma_0} X(\partial^* \sigma_0)}
\prod_{\sigma_1 \in \Delta_1} e^{ i \Delta t Z(\partial \sigma_1)} \Big)^k \nonumber \\ 
& \times \sum_{c_0 \in C_0}
C(c_0) | c_0, \partial^* c_0 \rangle . 
\end{align}
The term in the first exponent $X_{\sigma_0} X(\partial^* \sigma_0)$ can also be expressed as $X (\sigma_0 + \partial^* \sigma_0)$.
Note that the last (i.e., right-most) Trotter unitary term $e^{ i \Delta t Z(\partial \sigma_1)}$ can be written as $e^{i \Delta t Z_{\sigma_1}}$ because
\begin{align} \label{eq:Z-replacement}
   | \partial^* c_0 \rangle
   \otimes Z(\partial \sigma_1) |c_0\rangle
   = Z(\sigma_1)  | \partial^* c_0 \rangle
   \otimes |c_0\rangle, 
\end{align}
or graphically:
\begin{align}
\includegraphics[width=0.8\linewidth]{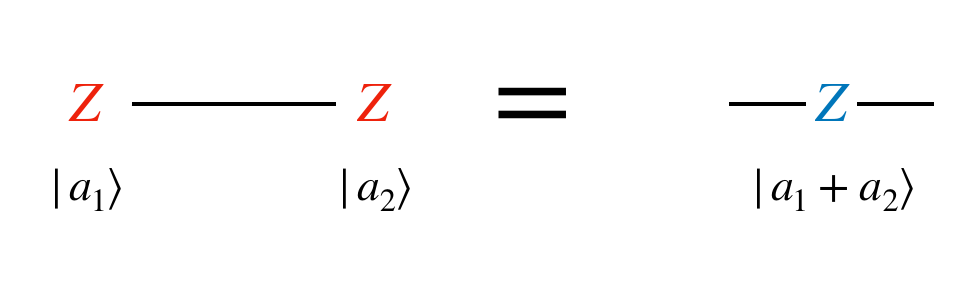} \nonumber
\end{align}
The phase from the $Z$ operator is $(-1)^{\#(\partial \sigma_1 \cap c_0)}$ on the left-hand side and $(-1)^{\#(\sigma_1 \cap \partial^* c_0)}$ on the right-hand side (due to eq.~\eqref{eq:phase-chain}), and they are equal due to the duality in eq.~\eqref{eq:duality}.

Second, the second to last unitary $e^{ i \lambda \Delta t X_{\sigma_0} X(\partial^* \sigma_0)}$ can be expanded in powers of $X_{\sigma_0} X(\partial^* \sigma_0)$, giving a product of $\cos(\Delta t \lambda) +i\sin(\Delta t \lambda)X_{\sigma_0} X(\partial^* \sigma_0)$.  
Crucially, the action of this operator on the basis preserves the structure with which the phase of $Z(\partial \sigma_1)$ is equal to that of $Z_{\sigma_1}$. 
For example, 
\begin{align}
\includegraphics[width=0.9\linewidth]{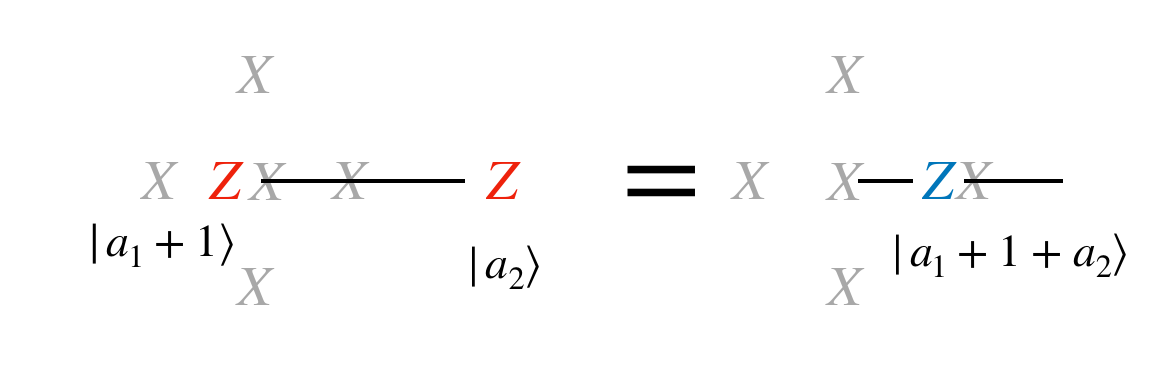}\nonumber 
\end{align}
Indeed, this replacement can be done for all the phase operators in the Trotter unitary; see Appendix~\ref{sec:replacing}. 
Thus the pre-measurement state is equal to
\begin{align}
|\psi_{\text{pre}} \rangle = & 
\Big(
\prod_{\sigma_1 \in \Delta_1 }
 e^{i \Delta t \lambda X_{\sigma_0} X(\partial^* \sigma_0)}
\prod_{\sigma_1  \in \Delta_1} e^{ i \Delta t Z_{\sigma_1}} \Big)^k \nonumber \\
&\times \sum_{c_0 \in C_0}
C(c_0) | c_0, \partial^* c_0\rangle  . 
\end{align}

Now, we consider the post-measurement wave function. 
We take the inner product between $\langle \widetilde{s_0}|$ and $|\psi_{\text{pre}}\rangle $. 
The effect is (1) the operator $X_{\sigma_0}$ collapses to its eigenvalue $(-1)^{s(\sigma_0)}$, and (2) the inner product gives a phase, $\langle \widetilde{s_0}|c_0\rangle = 2^{-|\Delta_0|/2}(-1)^{\#(s_0 \cap c_0)}$ due to eq.~\eqref{eq:inner-product-chain}. Note that the 0-cell $s_0$ that represents the measurement outcomes was previously defined in eq.~(\ref{eq:Z2-measurement-chain}).
Therefore, 
\begin{align}
| \psi_{\text{post}} \rangle = & \Big(
\prod_{\sigma_0 \in \Delta_0 }
e^{i \Delta t \lambda (-1)^{s(\sigma_0)} X(\partial^* \sigma_0)}
\prod_{\sigma_1 \in \Delta_1} e^{ i \Delta t Z_{\sigma_1}} \Big)^k \nonumber \\
& \times   2^{-|\Delta_0|/2}  \sum_{c_0}
C(c_0) 
(-1)^{\#(s_0 \cap c_0)} |\partial^* c_0 \rangle  . 
\end{align}

To deal with the signs originating from the measurement outcome, we first note the following relation:  
\begin{align}
\mathcal{O}_{\text{bp}}(\rho_1)  |\partial^* c_0 \rangle 
&
\overset{\text{eq.~\eqref{eq:phase-chain}}}{=}
(-1)^{\#(\rho_1 \cap \partial^* c_0)} | \partial^* c_0 \rangle \nonumber \\
&
\overset{\text{eq.~\eqref{eq:duality}}}{=}
(-1)^{\#(\partial \rho_1 \cap  c_0)} | \partial^* c_0 \rangle \nonumber\\
&= (-1)^{\#(s_0 \cap  c_0)} | \partial^* c_0 \rangle . 
\end{align}
Let us give a simple example to be more concrete. 
If we have nontrivial outcomes at vertices $1$ and $3$, (i.e., $s(\sigma_0)=1$ at $\sigma_0=v_1,v_3$) the phase can be written as $(-1)^{\#(s \cap c_0)} = (-1)^{a_1} (-1)^{a_3}$. 
Graphically, it can be rewritten with $Z$ operators supported on $\rho_1$ with $\partial \rho_1 = v_1 + v_3$:
\begin{align}
\includegraphics[width=0.9\linewidth]{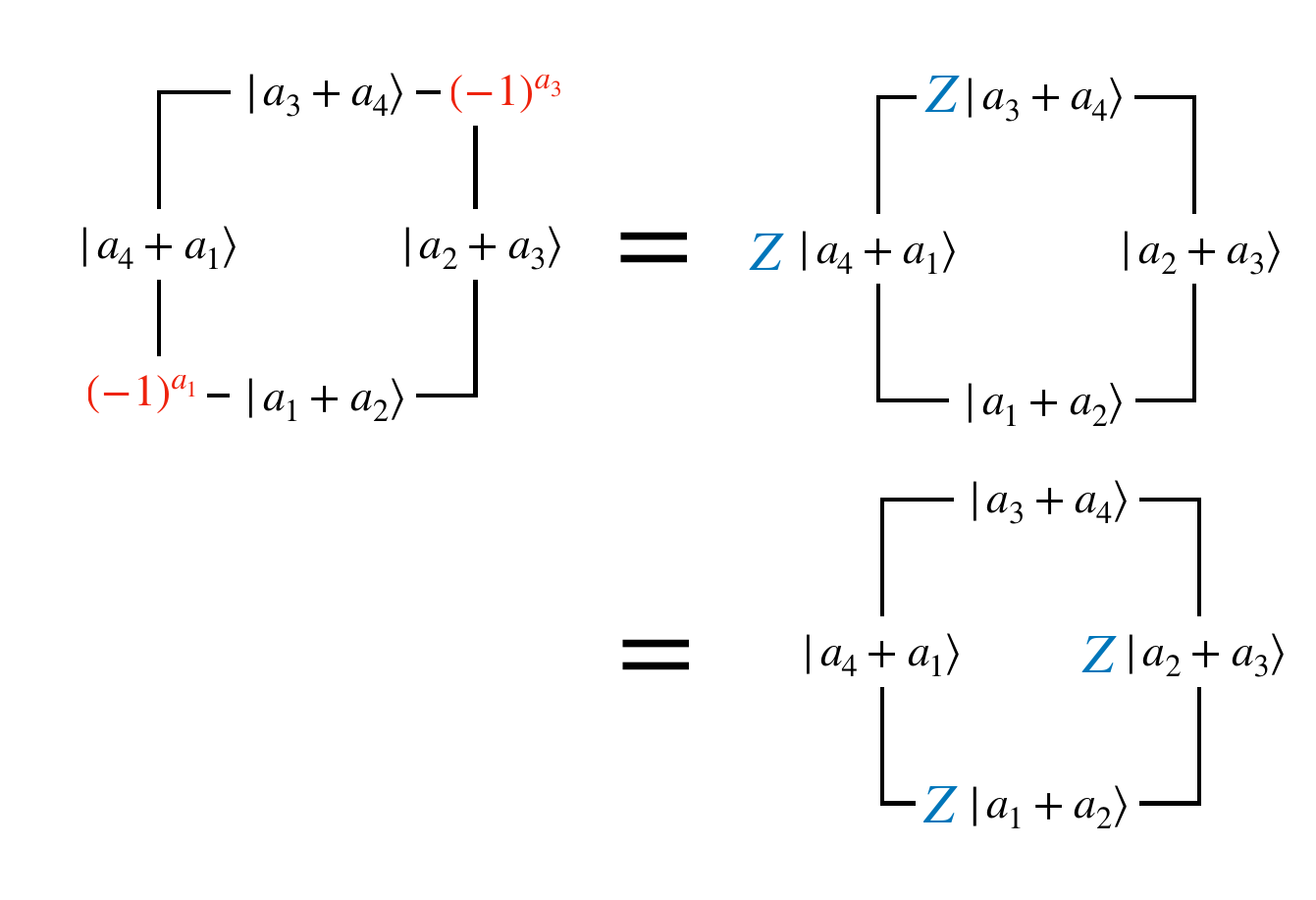} \nonumber
\end{align}
Different choices of $\rho_1$ give the same phase.

The inner product is thus equal to 
\begin{align}
| \psi_{\text{post}}\rangle  
=
&\Big( \prod_{\sigma_0 \in \Delta_0 }
e^{i \Delta t \lambda (-1)^{s(\sigma_0)} X(\partial^* \sigma_0)}
\prod_{\sigma_1 \in \Delta_1} e^{ i \Delta t Z_{\sigma_1}} \Big)^k \nonumber \\
&\times  2^{-|\Delta_0|/2}  \mathcal{O}_{\text{bp}}(\rho_1)
\sum_{c_0 \in C_0}
C(c_0)  |\partial^* c_0 \rangle
\ . 
\end{align}
The sum is now equal to $|\psi^{(1)}_{\text{gauged}}\rangle$.
We use the relation
\begin{align}
(-1)^{s(\sigma_0)} X(\partial^* \sigma_0) \mathcal{O}_{\text{bp}} (\rho_1)
= \mathcal{O}_{\text{bp}}(\rho_1) X(\partial^* \sigma_0), 
\end{align}
to move the byproduct operator to the left, canceling the unwanted signs $(-1)^{s(\sigma_0)}$. 
Again, this relation is due to eq.~\eqref{eq:commutation-chain} and the duality $\#(\partial^* \sigma_0 \cap \rho_1) = \#( \sigma_0 \cap \partial \rho_1) =\#(\sigma_0 \cap s_0)=s(\sigma_0)$ given in eq.~\eqref{eq:duality}.

Finally, we formally dualize the lattice.
The operator $ X(\partial^* \sigma_0 )$ defined in the primal lattice is equal to $X(\partial^* \sigma^*_2)$ in the dual lattice.
Also, $Z_{\sigma_1}$ in the primal lattice is equal to $ Z_{\sigma^*_1}$ in the dual lattice.
(The edges along a path $\rho_1$ are also now interpreted as edges in the dual lattice.)
Hence we obtain
\begin{align}
| \psi_{\text{post}} \rangle 
=
 2^{-|\Delta_0|/2}  \cdot \mathcal{O}_{\text{bp}} (\rho_1)\cdot  T^{\text{GT}}(t) |\psi^{(1)}_{\text{gauged}} \rangle,
\end{align} 
which completes the demonstration.

\subsection{Effect of noise}

Let us briefly analyze the effect of the noise.
In order to focus on the effect of the noise before the dualization, here we treat the entangler and the measurements as noise-free. 
Note that the dualization operation is constant-depth so that the ``volume" of such quantum operation is typically much smaller than that of the time evolution before the dualization. In other words, the errors will be dominated by the time evolution rather than the dualization operation itself.

We write a generic wave function right before the dualization as
\begin{align}
\label{eq:pre-KW-with-errors}
|\psi_{\text{pre-KW}} \rangle 
= \sum_{c_0 \in C_0} D(c_0) |c_0\rangle, 
\end{align}
with a complex coefficient $D(c_0)$.
Due to noise, we no longer assume that it is $\mathbb{Z}_2$ symmetric. 
It implies that the number of non-trivial measurement outcomes $s(\sigma_0)=1$ may be even or odd; in the latter case we cannot find a pair of paths and thus the correction fails.
When it is even, however, we still can construct a sum of paths $\tau_1$. After the dualization and correction, we have 
\begin{align}
|\psi_{\text{post-KW}} \rangle 
=  Z(\tau_1 + \rho_1) \times \sum_{c_0 \in C_0} D(c_0) |\partial^* c_0\rangle . 
\end{align}
It is indeed gauge symmetric:
\begin{align} \label{eq:gauss-law-with-noise}
&G_{\sigma^*_0}|\psi_{\text{post-KW}} \rangle \nonumber \\ 
& =  Z(\tau_1 + \rho_1) \times \sum_{c_0 \in C_0} D(c_0)  Z(\partial \sigma^*_0)|\partial^* c_0\rangle \nonumber \\
& 
\overset{\text{eq.~\eqref{eq:phase-chain}}}{=}
Z(\tau_1 + \rho_1) \times \sum_{c_0 \in C_0} D(c_0)  (-1)^{\#(\partial \sigma^*_0 \cap \partial^* c_0 )}|\partial^* c_0\rangle \nonumber \\
&
\overset{\text{eq.~\eqref{eq:duality}}}{=}
|\psi_{\text{post-KW}} \rangle .   
\end{align}
We also remark that $Z(\tau_1 + \rho_1)=1$ holds since 
\begin{align} 
Z(\tau_1 + \rho_1)|\partial^* c_0 \rangle
& \overset{\text{eq.~\eqref{eq:phase-chain}}}{=} (-1)^{\#(\tau_1 + \rho_1 \cap  \partial^* c_0 )} |\partial^* c_0 \rangle \nonumber \\
& \overset{\text{eq.~\eqref{eq:duality}}}{=}   (-1)^{\#(\partial ( \tau_1 + \rho_1 )\cap   c_0 )}  |\partial^* c_0 \rangle \nonumber \\
&= |\partial^* c_0 \rangle.
\end{align}

If we observe that the number of non-trivial measurement outcomes of $s(\sigma_0)=1$ is odd, then there is an isolated magnetic monopole (see also the next subsection) which we would fail to pair up with the counter operator. Despite this, the post-measurement wave function is still invariant under $G_{\sigma_0^*}$, if we assume the dualization does not induce any error.
(With strong unbiased noises, this would occur with probability $1/2$.)
We know by measurement outcomes that errors have occurred and can drop the result and repeat the procedure until we observe the number of $s(\sigma_0)=1$ being even. 
There is some error detection capability in our scheme, but it is not fault-tolerant, however. 
Then, when the correction succeeds, the gauge symmetry is guaranteed for the resulting wave function by the structure of Kramers-Wannier entanglers represented by $|\partial^* c_0 \rangle$, as elucidated in eq.~\eqref{eq:gauss-law-with-noise}; the ``fidelity" of gauge symmetry is unaffected by the form of time evolution or noises represented by $D(c_0)$.

We also note that the number of quantum gates and degrees of freedom involved in the quantum simulation of the transverse-field Ising model is smaller than those in the gauge theory; the number of edges is roughly twice as large as the number of vertices on the square lattice, and the gauge theory involves a four-body interaction term. It is thus likely that we obtain a time evolution with a higher fidelity through the dualization than directly simulating the gauge theory itself. Related to this point, we remark that a direct simulation of the gauge theory on quantum devices typically involves contributions that violate the Gauss law constraint, whose amount would be proportional to the spacetime volume of the simulation. 
One can perform a syndrome measurement for the Gauss law constraint to perform error correction, just as we do for the star operators in the toric code~\cite{2003AnPhy.303....2K}.
In our dualization, on the other hand, as long as the Kramers-Wannier operation is error-free and the protocol succeeds ({\it i.e.,} when the number of vertices with $s_v=1$ is even), the Gauss law constraint is satisfied. 
Of course, the complete error-free assumption is not realistic, but we emphasize that the Kramers-Wannier operator is constant depth.

\subsection{Meaning of byproduct operators}

The byproduct operator $\mathcal{O}_{\text{bp}}(\rho_1)$ is supported on paths $\rho$ connecting two dual plaquettes (primal vertices).
Physically interpreted, it is a set of {\it Dirac strings} which possess a magnetic flux.  
In the picture of the transverse-field Ising model, a monopole is generated by Pauli $Z$ at primal vertices (see Ref.~\cite{fradkin_2013}, for example), and in our language, it arises from the $|\tilde{1}\rangle=Z|\tilde{0}\rangle$ outcome of the Kramers-Wannier measurements.

\begin{figure}
\includegraphics[width=0.9\linewidth]{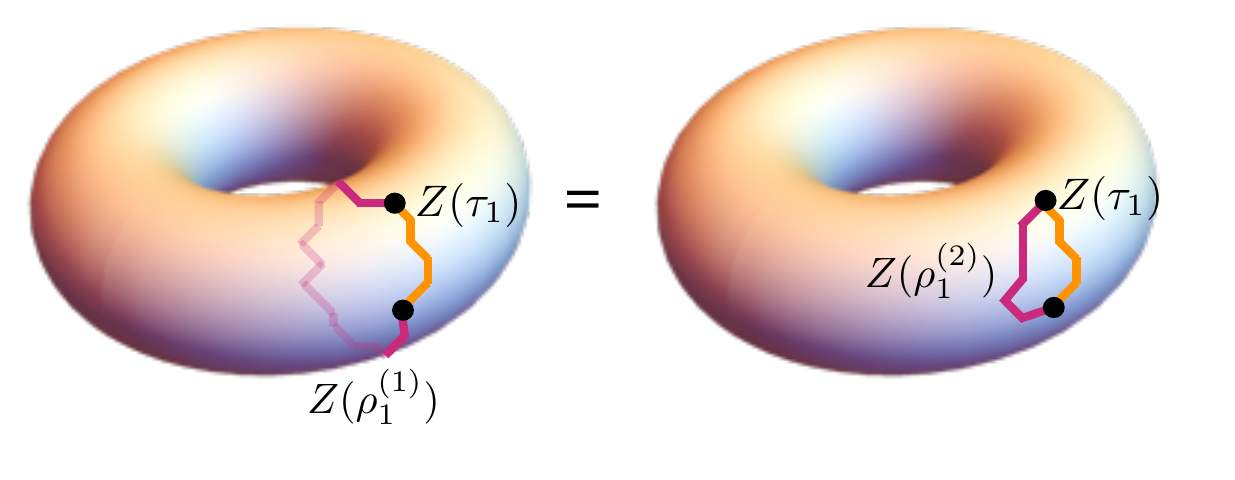}
\caption{The byproduct operator $Z(\rho^{(i)}_1)$  ($i=1,2$) and the counter operator $Z(\tau_1)$ on a torus.}
\label{fig:torus}
\end{figure}

Consider a string $\rho^{(i)}_1$ ($i=1,2$) where $\rho^{(i)}_1$ ends at $\sigma_0 = v_1$ and $\sigma_0=v_2$ individually.
The wave function after the measurements is the same whether we regard the byproduct operator as $Z(\rho^{(1)}_1)$ or $Z(\rho^{(2)}_1)$.
We apply the counter phase operator $Z(\tau_1)$.
Now one can imagine a situation where $\rho^{(2)}_1+\tau_1$ is contractible on a torus, but $\rho^{(1)}_1+\tau_1$ is not as depicted in Fig.~\ref{fig:torus}.
Since both scenarios should give the same physical wave function, it follows that $Z(\rho^{(1)}_1+\tau_1)= Z(\rho^{(2)}_1+\tau_1)=1$. 
Therefore the phase operator supported on a non-contractible loop has to be trivial. 

On the other hand, the $Z$ operator on a non-contractible loop around, say, a torus is called a 't Hooft loop~\cite{1979NuPhB.153..141T}, and it is one of the non-trivial gauge invariant operators in the $\mathbb{Z}_2$ gauge theory~\cite{fradkin_2013}. 
As noted in Refs.~\cite{2014JHEP...04..001K,2018arXiv180907757R}, the $\mathbb{Z}_2$ symmetric spin models are dual to the $\mathbb{Z}_2$ gauge theory with an additional topological constraint, which is a restriction of values of loop operators on non-contractible cycles, and thus the dualization of the transverse-field Ising model leads to a subsector of the full $\mathbb{Z}_2$ gauge theory on a non-trivial manifold. 
This is in harmony with the trivialness of the loop operators that arise from measurement-based dualization.

\section{Generalization to other pure gauge theories}
\label{sec:KW-generalization-to-pure}

In this section, we first consider obtaining the quantum simulation in the $(2+1)$-dimensional double-semion order \cite{2012PhRvB..86k5109L} by dualizing a time evolution in the Levin-Gu $\mathbb{Z}_2$ SPT order.
Then, we discuss the generalization to broader Abelian groups, taking the $\mathbb{Z}_N$ gauge theory as an example.

\subsection{Twisted gauge theory from twisted transverse-field Ising model}
\label{sec:tGT}
\begin{figure*}
(a) 
\includegraphics[width=0.30\linewidth]{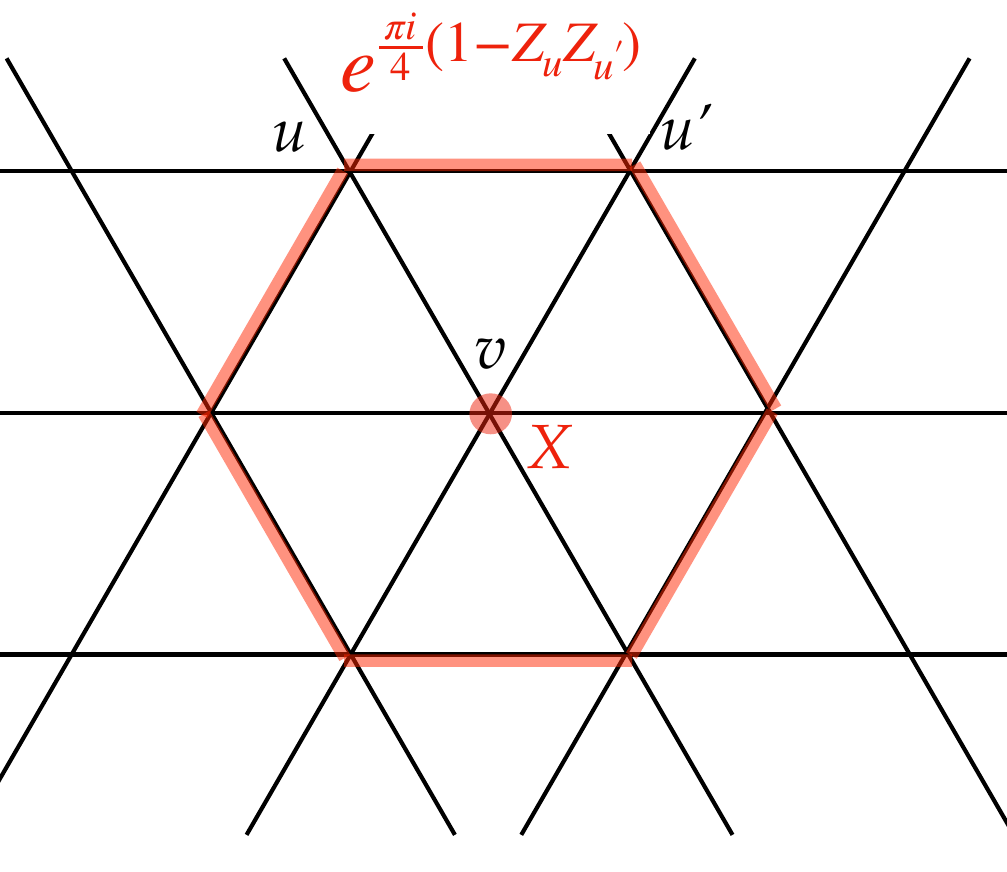}
(b)
\includegraphics[width=0.35\linewidth]{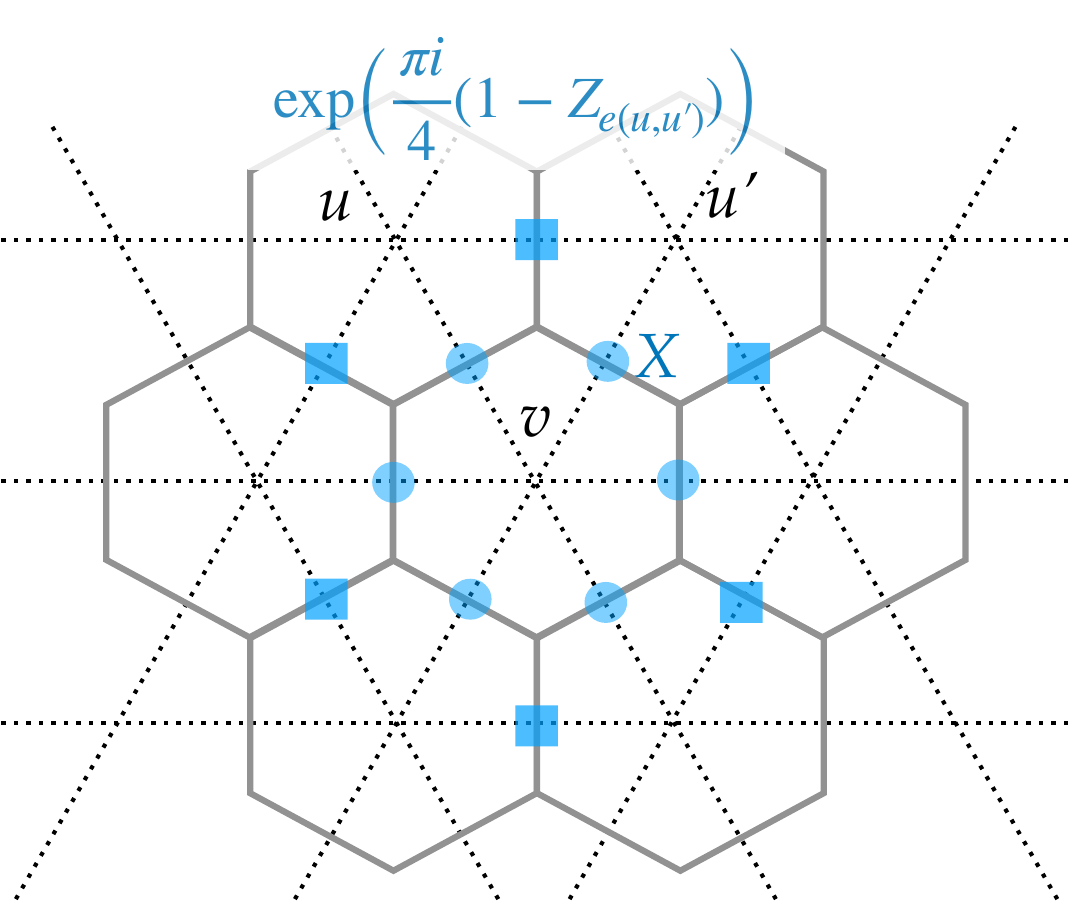} 
(c)
\includegraphics[width=0.20\linewidth]{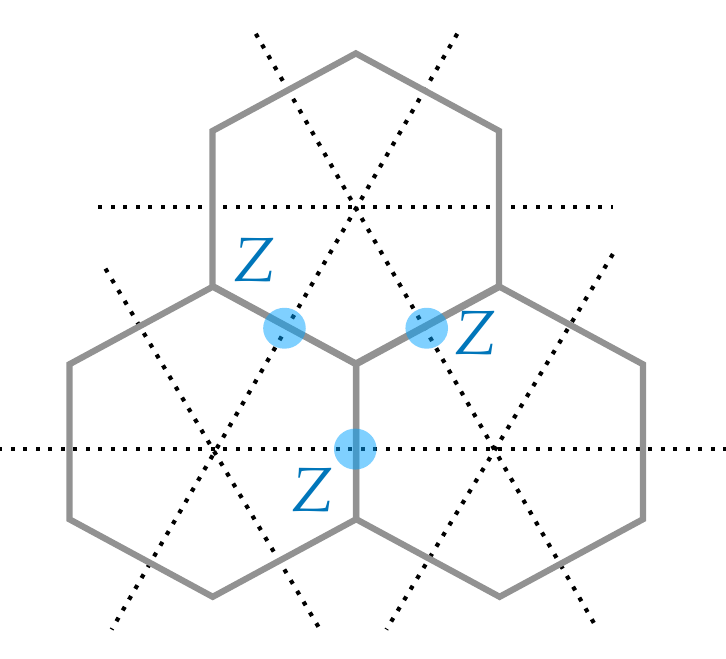}
\caption{(a) The $\mathcal{O}_v$ term in the Hamiltonian \eqref{eq:ttfi-hamiltonian}.
(b) The first term (plaquette term) in the twisted gauge theory Hamiltonian \eqref{eq:tgt-hamiltonian}. 
(c) The Gauss law operator in the twisted gauge theory.  }
\end{figure*}

Throughout this subsection, we use $v \in V$, $e \in E$, etc. to denote cells for convenience.
Consider the following Hamiltonian defined on vertices on a triangular lattice, which we call {\it twisted transverse-field Ising model} (tTFI):
\begin{align} \label{eq:ttfi-hamiltonian}
H_{\text{tTFI}} = -\sum_{v \in V} \mathcal{O}_v - g \sum_{\langle u,u'\rangle \in E} Z_u Z_{u'} \ 
\end{align}
with
\begin{align}
\mathcal{O}_v &= X_v \prod_{\langle  v u u' \rangle } e^{\frac{\pi i }{4} (1-Z_u Z_{u'})},
\end{align}  
where $\langle  v u u' \rangle$ is a triangle that consists of $v$, $u$, and $u'$, 
$\langle u,u'\rangle$ is an edge connecting vertices $u$
and $u'$. 
In this section, we use the notation of vertices, edges, and plaquettes, instead of cells. 
This Hamiltonian is symmetric under the $\mathbb{Z}_2$ symmetry generated by $\prod_{v \in V} X_v$.
When $g = 0$, the ground state is described by the Levin-Gu SPT state~\cite{2012PhRvB..86k5109L}~\footnote{Note that we have chosen an opposite sign in the Hamiltonian since $X=+1$ corresponds to the ground state when $g=0$ and if we remove the twist factor.}.
Our model is a $\mathbb{Z}_2$ symmetric deformation of the Levin-Gu SPT Hamiltonian.
The first-order Trotter decomposition of the time evolution is
\begin{align}
T_{\text{tTFI}}(t) = \Big( \prod_{v \in V} e^{i\Delta t \mathcal{O}_v } \prod_{\langle u , u' \rangle \in E} e^{ i \Delta t g Z_u Z_{u'}} \Big)^k. 
\end{align}
As before, we consider the ungauged wave function $|\psi_{\text{ungauged}}\rangle$ and assume it is $\mathbb{Z}_2$ symmetric.
One can, for example, load the Levin-Gu SPT state as $|\psi_{\text{ungauged}}\rangle$. 
It can be prepared with a finite depth circuit, {\it i.e.}, Hadamard, $Z$, controlled-$Z$, and controlled-controlled-$Z$ gates~\cite{wei2018quantum}.

For the dualized model, consider a deformed version of the double-semion model (a twisted gauge theory) \cite{2012PhRvB..86k5109L}, whose Hamiltonian is given by
\begin{align} \label{eq:tgt-hamiltonian}
H_{\text{tGT}} = - \sum_{v \in V } 
\tilde{\mathcal{O}}_v 
- g \sum_{e \in E} Z_e  .
\end{align}
with 
\begin{align}
\tilde{\mathcal{O}}_v = \prod_{e \supset v} X_e  \prod_{\langle v u u' \rangle } e^{\frac{\pi i}{4} (1-Z_{\langle u,u'\rangle} )}  . 
\end{align}
When $g=0$, it is one of the stabilizers of the double-semion model. 
The $Z_e$ term is the electric term, and the other term is the (twisted) magnetic (plaquette) term.
It is symmetric under the gauge transformation generated by 
\begin{align}
G_\triangle = \prod_{ e \in \triangle } Z_e,
\end{align}
where $\triangle$ denotes a triangle in the primal lattice. 
In the dual lattice picture, this is a divergence operator associated with a dual vertex. 
The gauged wave function $|\psi_{\text{gauged}}\rangle$ is also symmetric under this gauge transformation.
When the ungauged wave function $|\psi_{\text{ungauged}}\rangle$ is the Levin-Gu SPT state $|\psi_{\text{Levin-Gu}} \rangle $, then the gauged wave function $|\psi_{\text{gauged}}\rangle_E$ is the ground state of the double-semion model $|\psi_{\text{DS}} \rangle $ \cite{2021arXiv211201519T}:
\begin{align}
&\tilde{\mathcal{O}}_v   |\psi_{\text{DS}} \rangle
= |\psi_{\text{DS}} \rangle,  \quad \text{for~all~} v \in V,  \\
&G_{\triangle} |\psi_{\text{DS}} \rangle 
= |\psi_{\text{DS}} \rangle,  \quad \text{for~all~}\triangle   . 
\end{align}

The transformation with the Kramers-Wannier map straightforwardly generalizes to this model  between the time evolution of symmetric $|\psi_{\text{ungauged}}\rangle$ under $H_{\text{tTFI}}$ and that of $|\psi_{\text{gauged}}\rangle$ under $H_{\text{tGT}}$:

\begin{tcolorbox}
[width=\linewidth, sharp corners=all, colback=white!95!black]
\vspace{-11pt}
\begin{align}
&\mathcal{O}_{\text{bp}}(\rho_1) \cdot  T^{\text{tGT}}(t) |\psi_{\text{gauged}}\rangle  \nonumber \\
&= \widehat{\text{KW}} \cdot  T^{\text{tTFI}}(t) |\psi_{\text{ungauged}}\rangle \ . 
\end{align}
\end{tcolorbox}
We give detailed proof in Appendix~\ref{sec:replacing}.

\subsection{$\mathbb{Z}_N$ gauge theory from $\mathbb{Z}_N$ transverse-field clock model}
\label{sec:ZN-gauge-theory}
Here we show that the Kramers-Wannier-based gauging extends to the cyclic group $\mathbb{Z}_N = \{0,1,...,N-1 \, \text{mod}~N\}$. 
In this subsection, the Pauli operators are generalized to the qudit version of them.
The bases are generalized as follows:
\begin{align}
Z | s \rangle = \omega ^{s} | s \rangle,  \quad
X | \tilde{s} \rangle = \omega ^{s} | \tilde{s} \rangle,
\end{align}
with $\omega = e^{2 \pi i /N}$ and $s \in \{0,1,...,N-1 \, \text{mod}~N\}$.
The qudit Pauli operators are written using the generalized computational basis as
\begin{align}
Z=  \sum_{a = 0,..,N-1}\omega^{a} |a\rangle \langle a |, \quad 
X = \sum_{a = 0,..,N-1}|a+1\rangle \langle a |. 
\end{align}
They satisfy the commutation relation $ZX= \omega XZ$.
The $X$-basis is expressed explicitly as
\begin{align}
|\tilde{s} \rangle = \frac{1}{\sqrt{N}} \sum_{a=0,..,N-1} \omega^{-as} |a\rangle. 
\end{align}
The controlled-$X$ gate and the controlled-$X^{-1}$ gate are defined as
\begin{align}
CX_{c,t} &= \sum_{a,b=0,...,N-1} |a \rangle \langle a |_c \otimes |b+a \rangle \langle b |_t, \\
CX^{-1}_{c,t} &= \sum_{a,b=0,...,N-1} |a \rangle \langle a |_c \otimes |b-a \rangle \langle b |_t,
\end{align}
which satisfies 
\begin{align}
(CX_{c,t})^{\epsilon_1}\, X^{\epsilon_2}_c = X_c^{\epsilon_2} X_t^{\epsilon_1 \epsilon_2}\, (CX_{c,t})^{\epsilon_1}, \quad (\epsilon_1 , \epsilon_2 = \pm 1) \ .
\end{align}

\begin{figure*}
(a)
\includegraphics[width=0.15\linewidth]{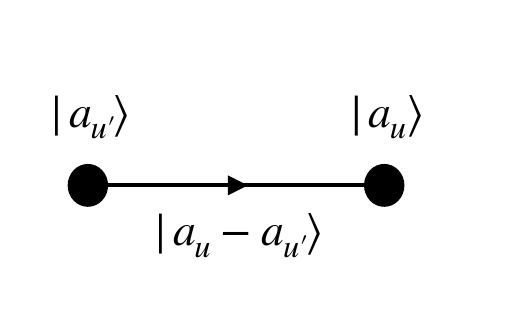}
(b)
\includegraphics[width=0.25\linewidth]{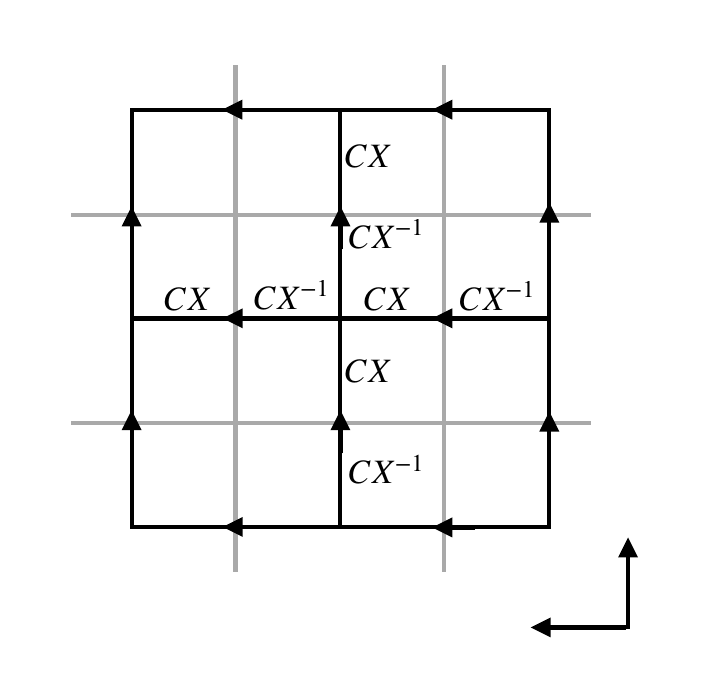} 
(c)
\includegraphics[width=0.25\linewidth]{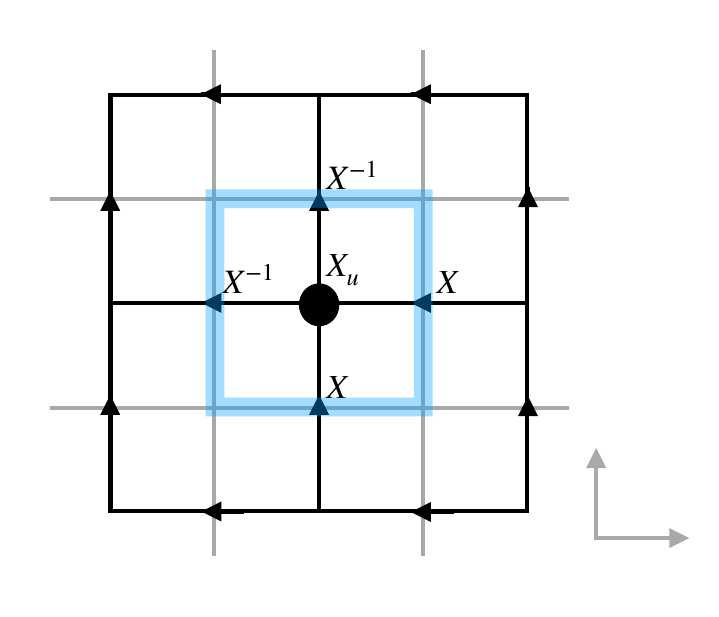}
(d)
\includegraphics[width=0.15\linewidth]{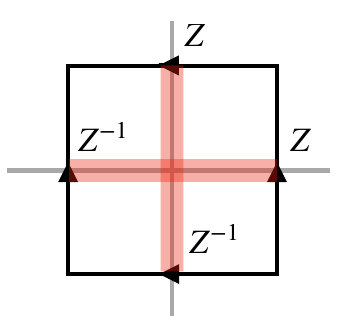}
\caption{
 (a) The action of the entangler in the computational basis.
The symbol $a_u$ is a $\mathbb{Z}_N$ integer.
(b) The pattern of the entangler. 
The controlled-$X$ gate acts as $[CX_{\sigma_0,\sigma_1}]^{a(\partial^* \sigma_0; \sigma_1)}$, where $a(\partial^* \sigma_0; \sigma_1) = a(\partial^* \sigma^*_2; \sigma^*_1)=\pm 1$ (or $0$) is determined according to the orientation of dual edges relative to the boundary of the dual plaquette, as explained in eq.~\eqref{eq:boundary-dual-2-cell}.
The basis $|c_0, 0_1\rangle$ is mapped to $|c_0, \partial^* c_0\rangle$ as before.
(c) 
The entangler transforms the $X_{\sigma_0}$ term in the clock model to the product of the five Pauli $X$'s in the figure. 
The blue square is the plaquette term in the gauge theory, {\it i.e.}, $X(\partial^* \sigma^*_2)$. 
(d) The Gauss law operator in the gauge theory, $Z(\partial \sigma_2)=Z(\partial \sigma^*_0)$. This is a symmetry in the image of the Kramers-Wannier map $\widehat{\text{KW}}^{\mathbb{Z}_N}$:
$Z(\partial \sigma_2)|\partial^* c_0 \rangle = \omega^{\#( \partial \sigma_2 \cap \partial^* c_0) } |\partial^* c_0 \rangle = |\partial^* c_0 \rangle$.
}
\label{fig:ZN-lattice}
\end{figure*}

We will discuss the time-evolved duality between the two following theories. (1) The $\mathbb{Z}_N$ clock model will be defined on the primal lattice and (2) the $\mathbb{Z}_N$ gauge theory will be defined on the edges in the dual lattice.
For $\mathbb{Z}_N$, we generalize the $i$-chains to have $\mathbb{Z}_N$ coefficients, $a(c_i;\sigma_i) \in \{0,...,N-1 ~\text{mod}~N\}$. 

We generalize the boundary operator in the following way. 
Consider a dual plaquette surrounded by four dual edges $\sigma^{*(i)}_{1}$ ($i=1,2,3,4$) with coordinates $\sigma^{*(1)}_1 = \{ \ell \hat{x} | 0\leq \ell \leq 1  \}$, $\sigma^{*(2)}_1 = \{\hat{x} +  \ell  \hat{y}| 0\leq \ell \leq 1  \}$, $\sigma^{*(3)}_1 = \{ \ell  \hat{x} +  \hat{y}| 0\leq \ell \leq 1  \}$, $\sigma^{*(4)}_1 = \{  \ell \hat{y}| 0\leq \ell \leq 1  \}$. 
(We omitted the shift of the dual lattice by $\frac{1}{2}(\hat{x}+\hat{y})$ for simplicity.)
We set the basis so that each dual edge is oriented either towards $+\hat{x}$ or $+\hat{y}$ directions.
Then the boundary of the dual plaquette is expressed as 
\begin{align} \label{eq:boundary-dual-2-cell}
    \partial^* \sigma^{*}_2 = \sigma^{*(1)}_1 + \sigma^{*(2)}_1 - \sigma^{*(3)}_1- \sigma^{*(4)}_1, 
\end{align} 
{\it i.e., } $a(\partial^* \sigma^{*}_2; \sigma^{*(1)}_1)=1$, $a(\partial^* \sigma^{*}_2; \sigma^{*(2)}_1)=1$, $a(\partial^* \sigma^{*}_2; \sigma^{*(3)}_1)=-1=N-1$, $a(\partial^* \sigma^{*}_2; \sigma^{*(4)}_1)=-1=N-1$.

On the other hand, the boundary of the primal 1-cell is defined as follows.
Consider primal edges with coordinates $\sigma^{(x)}_1 = \{ (1-\ell) \hat{x} | 0\leq \ell \leq 1  \}$
and $\sigma^{(y)}_1 = \{ \ell \hat{y} | 0\leq \ell \leq 1  \}$.
The edges on the primal lattice are taken so that they are oriented towards $-\hat{x}$ or $+\hat{y}$ directions.
({\it i.e.,} edges on the primal lattice and those on the dual lattice are identified by local +90 degrees rotation.)
The boundary of these edges is defined as
\begin{align}
\partial \sigma^{(x)}_1 &= \{\vec{0}\} - \{\hat{x}\}, \\ 
\partial \sigma^{(y)}_1 &= \{\hat{y}\} - \{\vec{0}\}, 
\end{align}
where the points on the right-hand side indicate 0-cells.
With this set of definitions, it follows that 
\begin{align}\label{eq:ZN-Poincare}
\#(c_1 \cap \partial^* c^*_2) = \#( \partial c_1 \cap c^*_2) \quad \text{mod}~N. 
\end{align}

We present the Hamiltonian of the clock model, a $\mathbb{Z}_N$ version of the transverse-field Ising model, and the $\mathbb{Z}_N$ gauge theory (GT).
The $\mathbb{Z}_N$ clock model is given by
\begin{align}
H^{\mathbb{Z}_N}_{\text{clock}}
= - \sum_{\sigma_1 \in \Delta_1} (Z(\partial \sigma_1) + \text{h.c.})
- \lambda \sum_{\sigma_0 \in \Delta_0} (X_{\sigma_0} + \text{h.c.}), 
\end{align}
which is defined on the vertices on the primal lattice.
This Hamiltonian is invariant under a global symmetry generated by $\prod_{\sigma_0 \in \Delta_0} X_{\sigma_0}$. 
We assume that the initial ungauged wave function $|\psi^{(1)}_{\text{ungauged}}\rangle$ is also symmetric under this global symmetry. 
The same argument we gave for the case with $\mathbb{Z}_2$ tells us that the set of measurement outcomes in the $X$ basis $|\widetilde{s_0}\rangle$ is constrained as $\prod_{\sigma_0\in \Delta_0} \omega^{s(\sigma_0)} = 1$.
Hence we have
\begin{align} \label{eq:ZN-measure-constraint}
\sum_{\sigma_0 \in \Delta_0} s(\sigma_0) = 0 \quad \text{mod}~N. 
\end{align}

Then the $\mathbb{Z}_N$ gauge theory is defined with the Hamiltonian
\begin{align}
H^{\mathbb{Z}_N}_{\text{GT}}
= - \sum_{\sigma^*_1 \in \Delta^*_1} (Z_{\sigma^*_1} + \text{h.c.})
- \lambda \sum_{\sigma^*_2 \in \Delta^*_2}\big( X(\partial^* \sigma^*_2) + \text{h.c.}
\big).  
\end{align}
See Fig.~\ref{fig:ZN-lattice} for the illustration of the plaquette term (c) and the Gauss law divergence operator (d).
The first-order Trotter decompositions of respective models are given by
\begin{align}
&T^{\mathbb{Z}_N \text{clock}}(t) \nonumber \\
&=
\Big(\prod_{\sigma_1\in \Delta_1} e^{i \Delta t (Z(\partial \sigma_1) + \text{h.c.}) } 
\prod_{\sigma_0 \in \Delta_0} e^{i \lambda \Delta t ( X_{\sigma_0}  + \text{h.c.}) }\Big)^k, 
\\
&T^{\mathbb{Z}_N \text{GT}}(t)\nonumber \\
&=
\Big(\prod_{\sigma^*_1 \in \Delta^*_1} e^{i \Delta t (Z_{\sigma^*_1} + \text{h.c.}) } 
\prod_{\sigma^*_2 \in \Delta^*_2} e^{i \lambda \Delta t ( X(\partial^*\sigma^*_2)+ \text{h.c.}) }\Big)^k. 
\end{align}

The Kramers-Wannier map is implemented as follows.
We prepare the ancillary degrees of freedom on edges $|0_1\rangle$.
The entangler is 
\begin{align} \mathcal{U}^{\mathbb{Z}_N}=\prod_{\substack{\sigma_0 \in \Delta_0 \\ \sigma_1  \in \Delta_1}} [CX_{\sigma_0, \sigma_1}]^{a(\partial^* \sigma_0 ; \sigma_1)}  ,
\end{align}
where $\sigma_0 \simeq \sigma^*_2$ and $\sigma_1 \simeq \sigma^*_1$; see Fig.~\ref{fig:ZN-lattice}~(a)(b). The exponent  $a(\partial^*\sigma_0;\sigma_1) \in \{0,+1,-1\}$ is used to specify the orientation of the edges relative to a vertex and thus whether $CX$, its inverse, or the identity gate is applied.
The Kramers-Wannier map for the $\mathbb{Z}_N$ is given by 
\begin{align} \label{eq:kw-map-ZN}
\widehat{\text{KW}}^{\mathbb{Z}_N} = 
\langle  \widetilde{s_0} | \,\mathcal{U}^{\mathbb{Z}_N} \,|0_1\rangle . 
\end{align}
Our result is as follows:
\begin{tcolorbox}
[width=\linewidth, sharp corners=all, colback=white!95!black]
\vspace{-11pt}
\begin{align}
&\mathcal{O}^{\mathbb{Z}_N}_{\text{bp}} (\rho_1)\cdot  T^{\mathbb{Z}_N \text{GT}}(t)
 |\psi_{\text{gauged}} \rangle \nonumber \\
 &= \widehat{\text{KW}}^{\mathbb{Z}_N} \cdot  T^{\mathbb{Z}_N \text{clock}}(t)
 |\psi_{\text{ungauged}} \rangle.
\end{align}
\end{tcolorbox}

The ungauged and gauged wave function is defined by the same formal expression as eqs.~\eqref{eq:Z2-ungauged} and \eqref{eq:Z2-gauged}, with the bases in eqs.~\eqref{eq:Z2-bases-Z} and \eqref{eq:Z2-bases-X}, where the coefficient $a(c_i;\sigma_i) \in \{0,1 ~\text{mod}~2\}$ is now replaced by $a(c_i;\sigma_i) \in \{0,...,N-1 ~\text{mod}~N\}$.
The byproduct operator takes the form 
\begin{align}\label{eq:bp-ZN}
\mathcal{O}^{\mathbb{Z}_N}_{\text{bp}}(\rho_1)
= \prod_{\sigma_1 \in \Delta_1} 
Z(\sigma_1)^{a(\rho_1;\sigma_1)} = Z(\rho_1),
\end{align}
with $\rho_1 = \sum_{\sigma_1 \in \Delta_1} a(\rho_1;\sigma_1) \sigma_1$ such that $\partial \rho_1 = \sum_{\sigma_0 \in \Delta_0} s(\sigma_0) \sigma_0$. 
Constructing $\rho_1$ as well as finding $\tau_1$ is always possible since $\sum_{\sigma_0 \in \Delta_0} s(\sigma_0) = 0 ~ \text{mod}~N$. 

The dualization can be shown in the same way as in the case with $\mathbb{Z}_2$.
We note that the effect of the entangler on the ungauged basis is formally the same as in $\mathbb{Z}_2$, {\it i.e.,} $|c_0, 0_1\rangle \mapsto | c_0 , \partial^* c_0 \rangle $. 
We have expressions for $\mathbb{Z}_N$ as in eqs.~\eqref{eq:phase-chain}, \eqref{eq:commutation-chain}, \eqref{eq:inner-product-chain}, \eqref{eq:duality} with $-1$ replaced by $\omega$ and 2 by $N$.
Thus the process of rewriting equations is exactly the same throughout the proof except that there are h.c. terms and changing factors associated with $\mathbb{Z}_2$ to those in $\mathbb{Z}_N$.

One can understand the correction of the phase factors in the case with $\mathbb{Z}_N$ in the following way. Suppose we found $n_s$ vertices with $s(\sigma_0)\neq 0$. 
For two such vertices, say $\sigma^{(1)}_0$ and $\sigma^{(2)}_0$ among them, one can find an oriented path $\gamma^{(1,2)}$ connecting these two: $\partial \gamma^{(1,2)} = \sigma^{(1)}_0 - \sigma^{(2)}_0$.
Applying $\prod_{\sigma_1 \in \gamma^{(1,2)}} Z(\sigma_1)^{-s(\sigma^{(1)}_0)}$ cancels the phase factor associated with the vertex $\sigma^{(1)}_0$, leaving a phase factor $\omega^{a(c_0;\sigma^{(2)}_0)( s(\sigma^{(2)}_0) + s(\sigma^{(1)}_0))}$ at the vertex $\sigma^{(2)}_0$.
This procedure reduces the number of vertices with $s(\sigma_0)\neq0$ as $n_s \rightarrow n_s-1$ (or $n_s \rightarrow n_s-2$ if the two precisely cancel each other, {\it i.e.}, $s(\sigma^{(1)}_0) +s(\sigma^{(2)}_0) = 0 $ mod $N$). Repeating it, we eventually arrive at the corrected wave function with $n_s = 0 $. (Given that $Z$ operators commute, one can perform all the necessary corrections in one single step.)
In the language of topological order, the phase factors shall be regarded as Abelian anyons in the $\mathbb{Z}_N$ toric code \cite{2003AnPhy.303....2K}, and the reducing the number of nontrivial phases can be seen as the annihilation of anyons using ribbon operators.

\section{Generalization to matter models covariantly coupled to gauge fields}
\label{sec:KW-gauge-matter}

\subsection{(1+1)d Ising matter coupled to topological gauge fields}
\label{sec:GM}

Consider the following Hamiltonian with transverse and longitudinal terms defined on the vertices in the one-dimensional primal lattice:
\begin{align} \label{eq:hamiltonian-tl-ising}
H^{\text{TL-Ising}}  =& - \sum_{ \sigma_{1} \in \Delta_1} Z(\partial \sigma_1) - g \sum_{\sigma_0 \in \Delta_0} X_{\sigma_0} 
\nonumber\\
& - h \sum_{\sigma_0 \in \Delta_0} Z_{\sigma_0} . 
\end{align}
The standard Kramers-Wannier transformation $\{ Z(\partial \sigma_1), X_{\sigma_0}\} \mapsto \{ Z_{\sigma^*_0}, X(\partial^* \sigma^*_1)\} $ is not well-defined for this Hamiltonian due to the last term. 
We can, however, generalize the duality by introducing a topological gauge field on the edges in the dual lattice, $\Delta^*_1$, and imposing the Gauss law constraint on the degrees of freedom in the dual lattice; see {\it e.g.,}~\cite{2018arXiv180907757R}.
Namely, the alternative Kramers-Wannier transformation is given by 
\begin{align}
X_{\sigma_0} &\mapsto X_{\sigma^*_1} X(\partial^* \sigma^*_1),  \\
Z_{\sigma_0} & \mapsto Z_{\sigma^*_1},
\end{align}
and the Gauss law constraint is given by 
\begin{align} \label{eq:gauss-law-mg}
G_{\sigma^*_0} := Z_{\sigma^*_0} Z(\partial \sigma^*_0) =1.  
\end{align}
The other transformation $Z(\partial \sigma_1) \mapsto Z_{\sigma^*_0}$ is obtained by substituting the second line of the transformation to the Gauss law constraint.
The resulting theory is given by the Hamiltonian describing a gauge-matter theory:
\begin{align}
H^{\text{GM}}= & 
- \sum_{\sigma_0^* \in \Delta^*_0} Z_{\sigma^*_0} - g \sum_{\sigma^*_1 \in \Delta^*_1} X_{\sigma^*_1} X(\partial^* \sigma^*_1)
\nonumber \\ 
& -h \sum_{\sigma^*_1 \in \Delta^*_1} Z_{\sigma^*_1}\,. 
\end{align}
The Hamiltonian is invariant under the gauge transformation, $[H^{\text{GM}},G_{\sigma^*_0}]=0$.
The first term is an ordinary matter term.
The second term is a matter kinetic term covariantly coupled to the gauge field. 
And the last term can be seen as an electric term in the gauge sector.
In this subsection, we show that such alternative Kramers-Wannier transformation for time evolution can be realized by entanglers and measurements, and we denote it by KW$^{GM}$.
This is a generalization of measurement-assisted gauging in the literature \cite{2021arXiv211201519T}.

To distinguish the undualized and dualized degrees of freedom, we write the degrees of freedom on primal 0-cells with the double bracket $| \ \rangle\!\rangle$.
As before, we use the following bases for the wave functions: 
\begin{align} 
&| {c_0} \rangle\!\rangle := \bigotimes_{{\sigma_0} \in \Delta_0} | a(c_0 ; \sigma_0) \rangle\!\rangle^{(Z)}_{{\sigma_0}},  \\
&| \widetilde{ {c_0}} \rangle\!\rangle := \bigotimes_{{\sigma_0} \in \Delta_0} | a(c_0 ; \sigma_0) \rangle\!\rangle^{(X)}_{{\sigma_0}}, 
\end{align}
and
\begin{align}
&| c_0 \rangle := \bigotimes_{\sigma_0 \in \Delta_0} | a(c_0 ; \sigma_0) \rangle^{(Z)}_{\sigma_0},  \\ 
&| c_1 \rangle := \bigotimes_{\sigma_1 \in \Delta_1} | a(c_1 ; \sigma_1) \rangle^{(Z)}_{\sigma_1}. 
\end{align}

First, consider the time evolution with $H^{\text{TL-Ising}}$, whose Trotterization is written as $T^{\text{TL-Ising}}(t)$.
We take the initial state as {\it any} state defined on the vertices:
\begin{align} \label{eq:1d-ungauged}
| \psi^{(0)}_{\text{ungauged}}\rangle\!\rangle = \sum_{c_0 \in C_0} C(c_0) |c_0 \rangle\!\rangle. 
\end{align}
In particular, we do not impose a $\mathbb{Z}_2$ symmetry for this wave function.

We will load a gauged state on edges and vertices on the dual lattice. 
We emphasize that an edge in the dual lattice is identical to a vertex in the primal lattice, but we treat them as separate degrees of freedom.
We initiate the wave function as 
\begin{align}
|0\rangle^{\otimes |\Delta_0|} |0\rangle^{\otimes |\Delta_1|} = | 0_0, 0_1 \rangle  .
\end{align}
Note that this state satisfies the Gauss law constrained generated by \eqref{eq:gauss-law-mg}. 

\begin{figure*}[htbp]
\includegraphics[width=0.9\linewidth]{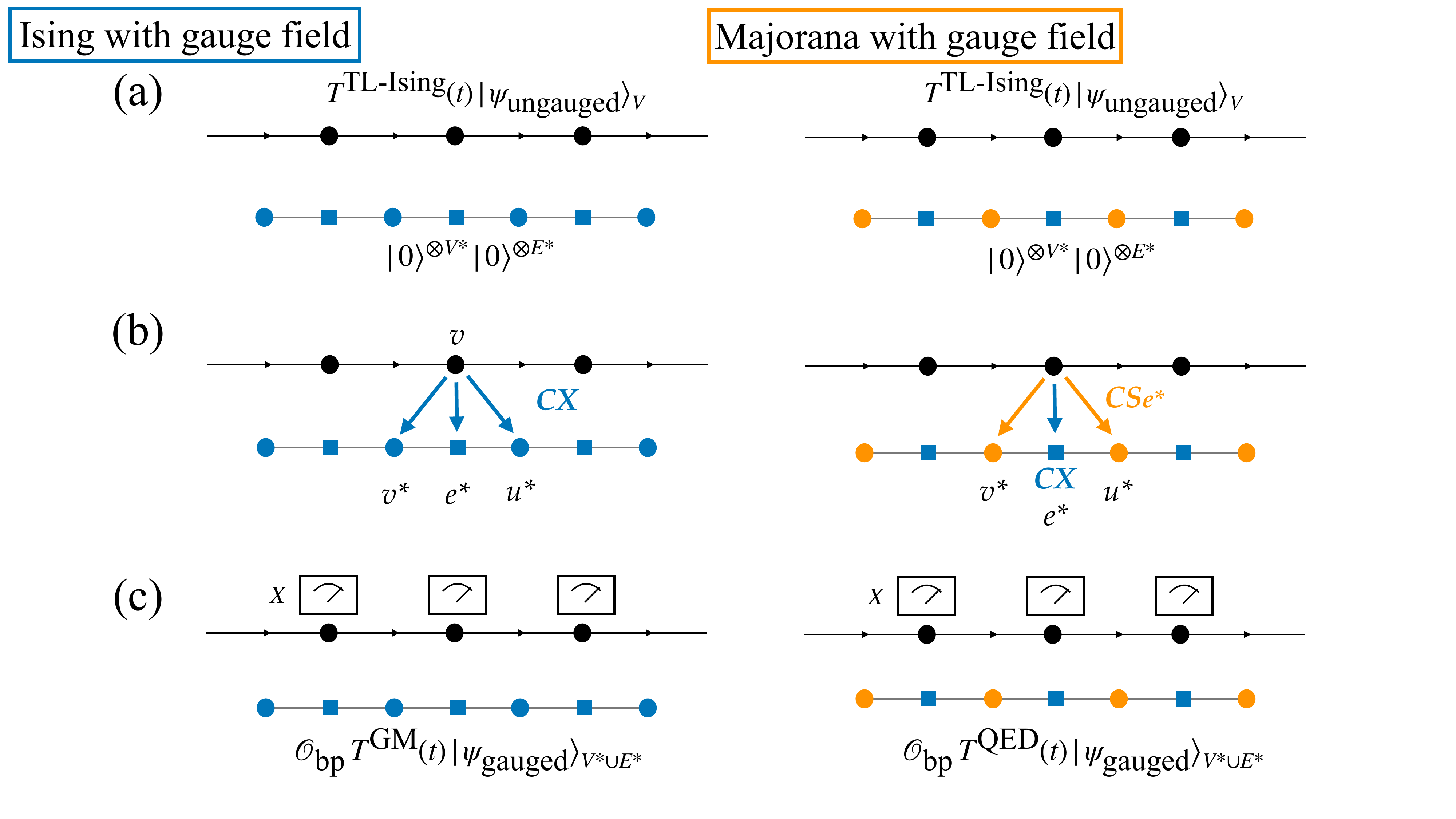}
\caption{
The procedure of obtaining the time evolution with the gauged Ising (left)/Majorana fermion (right) model. 
(a) The ungauged wave function placed on vertices (black dots) is evolved under the Hamiltonian of the transverse and longitudinal Ising model. Separate degrees of freedom are prepared on the dual chain as a product state. 
On the left, to obtain the gauged Ising model, the product state is placed on both dual vertices ($v^*$: blue circle) and dual edges ($e^*$: blue square).
On the right, to obtain the gauged Majorana fermion model ($\mathbb{Z}_2$ QED), the fermionic vacuum state is placed on dual vertices ($v^*$: orange circle) while the bosonic product state is placed on dual edges ($e^*$: blue square).
(b) We apply the entanglers. For the gauged Ising model, we apply $CX$ gates, according to the arrows that point from controlling qubits to target qubits. For the $\mathbb{Z}_2$ QED, we apply $CX$ and $CS$ gates, where the edges are oriented to the right as a convention. (c) We measure the original degrees of freedom (black dots), resulting in a time-evolved gauged wave function on the dual chain (blue and orange dots) up to byproduct operators. Circles are the matter field while squares represent the gauge field.}
\label{fig:gauged-1d-chain-procedure}
\end{figure*}

We consider an entangler 
\begin{align}
\mathcal{U}^{\text{GM}} = \prod_{\sigma_0 \in \Delta_0} 
\Big( \underbrace{CX_{\sigma_0,\sigma_0} }_{\substack{c~:~ {\rm undualized} \\ t~:~ {\rm dualized ~~}} } 
\prod_{\sigma_1 \in \Delta_1} [CX_{\sigma_0,\sigma_1} ]^{a(\partial^* \sigma_0; \sigma_1)} \Big), 
\end{align}
where the first $CX$ gate is controlled by the undualized qubits (also labeled as $c$) and applies $X$ on the dualized degrees of freedom (also labeled as $t$; note that both $c$ and $t$ are on the same 
0-cell $\sigma_0$); see Fig.~\ref{fig:gauged-1d-chain-procedure}. 
The generalized Kramers-Wannier map is now defined as
\begin{align} \label{eq:kw-map-gm}
\widehat{\text{KW}}^{\text{GM}} = \langle\!\langle \widetilde{s_0} | \,\mathcal{U}^{\text{GM}} \, |0_0, 0_1 \rangle. 
\end{align}
We denote the Trotterized time evolution with the Hamiltonian $H^{\text{GM}}$ by $T^{\text{GM}}(t)$.

We show that the time evolution of the gauged Ising model can be obtained by the Kramers-Wannier map KW$^{\text{GM}}$; see Fig.~\ref{fig:gauged-1d-chain-procedure}.
\begin{tcolorbox}
[width=\linewidth, sharp corners=all, colback=white!95!black]
\vspace{-11pt}
\begin{align} \label{kw-time-evolution-gm}
&\mathcal{O}^{\text{GM}}_{\text{bp}} (s_0)\cdot 
T^{\text{GM}}(t) | \psi^{(0,1)}_{\text{gauged}} \rangle \nonumber \\
&= 
\widehat{\text{KW}}^{\text{GM}} \cdot T^{\text{TL-Ising}}(t) | \psi^{(0)}_{\text{ungauged}} \rangle\!\rangle. 
\end{align}
\end{tcolorbox}
\noindent Here the gauged wave function is 
\begin{align}
| \psi^{(0,1)}_{\text{gauged}} \rangle
= \sum_{c_0 \in C_0} C(c_0) |c_0, \partial^* c_0 \rangle, 
\end{align}
and it satisfies the Gauss law constraint.

The byproduct operator for this dualization is given by
\begin{align} \label{eq:bp-gm}
\mathcal{O}^{\text{GM}}_{\text{bp}}(s_0)
=
\prod_{ \sigma_0 \in \Delta_0 } Z(\sigma_0)^{s({\sigma_0})}. 
\end{align}
Here, the exponent $s(\sigma_0)$ is associated with the undualized degrees of freedom, but the operator $Z(\sigma_0)$ acts on the dual degrees of freedom.
We emphasize that it is no longer a string operator; rather, it acts on the 0-chain $s_0$ defined in eq.~\eqref{eq:Z2-measurement-chain} --- the sum of 0-cells that correspond to $s(\sigma_0)=1$. 
Correction can be done directly by applying $\mathcal{O}^{\text{GM}}_{\text{bp}}$.
We present the proof in Appendix~\ref{sec:gauged-ising-proof}.
It turns out this is a generic feature in the gauge theories coupled to matter fields, as we will see in other examples.

\subsection{(1+1)d $\mathbb{Z}_2$ QED with spinless fermion}
\label{sec:QED}

In this subsection, we discuss a Jordan-Wigner transformation of time evolution. 
It is an alternative Jordan-Wigner transformation that leads to a gauged Majorana fermion model, analogous to the case discussed for the Ising model  covariantly coupled to gauge fields; see \cite{2018arXiv180907757R}.
We also refer to~\cite{borla2021gauging} for  the study of the phase diagram of  the gauged Majorana fermion model.

Let $(\chi_{v^*}, \chi'_{v^*})$ ($v^* \in \Delta^*_0$) be a pair of Majorana fermion operators per (dual) site. 
They are related to the fermion operators $c_{v^*}$ and $c^{\dagger}_{v^*}$ as $c = (\chi + i \chi')/2$, $c^{\dagger} = (\chi - i \chi')/2$. 
They satisfy $\{c^\dagger_u,c_v\} = \delta_{u,v}$, $\{\chi_u, \chi_v\} = 2 \delta_{u,v}$, $\{\chi_u , \chi'_v\}=0$.
We write $\partial^* e^* = v^*_+ - v^*_-$.
We define the following two bosonic fermion-bilinear operators,
\begin{align}
S_{e^*} &= -i \chi'_{v^*_-} \chi_{v^*_+}, \\ 
P_{v^*} &= i \chi'_{v^*} \chi_{v^*}  = 1 - 2 c^\dagger_{v^*} c_{v^*}. 
\end{align}
The operator $P_{v^*}$ is a fermion number operator at site $v^*$ and its eigenvector can be written as $| p \rangle_{v^*} = (c^\dagger_{v^*} )^p| 0 \rangle_{v^*}$ with its eigenvalue $(-1)^p$ for $p \in \{0,1\}$. 
We note that locally $(P, \chi, \chi')$ forms the same algebraic relation as $(Z,X,Y)$. 
Thus the fermionic basis can also be written as $|p \rangle_{v^*} = (\chi_{v^*})^p |0\rangle_{v^*}$.

As before, the alternative Jordan-Wigner transformation is given by~\cite{2018arXiv180907757R} 
\begin{align}
X_v &\mapsto S_v X_v = S_{e^*} X_{e^*} \quad \text{for} \quad v=1,...,N-1, \\
X_N &\mapsto q S_N X_N = q \, S_{e_N^*} X_{e_N^*}  \ \quad q \in \{+1,-1\}, \\
Z_v & \mapsto Z_v = Z_{e^*},
\end{align}
with the associated Gauss law constraint being 
\begin{align}
G_{v^*} = Z_{e^*} P_{v^*} Z_{e'^*} = 1  \quad (e^*, e'^* \supset v^* ). 
\end{align}
Every operator on the right-hand side of the map commutes with the operator $G_{v^*}$; note that whenever $S_{e^*} X_{e^*}$ and $G_{v^*}$ have an overlap, $X_{e^*}$ and $Z_{e^*}$ anti-commute, but  $S_{e^*}$ and $P_{v^*}$ also anti-commute.

As noted in \cite{2018arXiv180907757R}, the choice of the parameter $q$ is important for consistency on a periodic chain.
Namely, for the parity even sector, we have to use the anti-periodic boundary condition for the fermion ($q=-1$), when the original spin theory is periodic.
Essentially, this is because if we transport a fermion over the circle by multiplying $S_{v^*}$, we should get a sign $-1$ by commuting with an odd number of fermions.
On the other hand, for the parity odd sector 
we have to use the periodic boundary condition for the fermion ($q=+1$).
A similar result was also obtained for the ordinary Jordan-Wigner transformation of the XY chain in Ref.~\cite{2010arXiv1012.4114W}.

Under this set of maps, the Ising interaction is mapped via $Z_v Z_{u} \mapsto P_{v^*}$, where $v^*=\langle u,v\rangle$. 
The Hamiltonian \eqref{eq:hamiltonian-tl-ising}, 
\begin{align}
H^{\text{TL-Ising}} = - \sum_{ \langle v, v'\rangle \in E} Z_v Z_{v'} - g \sum_{v \in V} X_v - h \sum_{v \in V} Z_v,  
\end{align}
is thus mapped to a Majorana fermion model covariantly coupled to gauge fields (QED):
\begin{align} \label{eq:hamiltonian-qed}
H^{\text{QED}}& = 
- \sum_{v^* \in V^*} P_{v^*} 
- g \sum_{e^* \in E^*} X_{e^*} S_{e^*} (-1)^{\delta_{e^*,N}}\nonumber \\
& \qquad -h \sum_{e^* \in E^*} Z_{e^*}.
\end{align}
We denote the Trotterized time evolution of respective models as $T^{\text{TL-Ising}}(t)$ and $T^{\text{QED}}(t)$.

Our set-up for the Jordan-Wigner transformation that will be applied to the state after  time evolution $T^{\text{TL-Ising}}(t)$ is as follows.
The ungauged wave function is defined in the same way as \eqref{eq:1d-ungauged}.
We initialize the ancillary state as $|0\rangle^{\otimes E^*} |0\rangle^{\otimes V^*}$. 
This state trivially satisfies the Gauss law constraint. 
We consider the entangler defined by
\begin{align} \label{eq:JW-entangler}
\mathcal{U}^{\text{JW}}
= \prod_{v \in V} CS_{v,e^*} CX_{v,e^*} ,
\end{align}
where $CX$ is the usual controlled-X operator and $CS_{v,e^*}$ is a controlled-hopping operator given by
\begin{align}
    CS_{v,e^*}= |0\rangle_v \langle 0 | \otimes I_{v^*,u^*} + |1\rangle_v \langle 1 | \otimes S_{e^*} (-1)^{\delta_{e^*,N}}. 
\end{align}
The Jordan-Wigner map is thus defined as
\begin{align} \label{eq:jw-map}
\widehat{\text{JW}}  = \langle \{\tilde{s}_v\} |_V \,\mathcal{U}^{\text{JW}} |0\rangle^{\otimes E^*} |0\rangle^{\otimes V^*}, 
\end{align}
and we claim that 
\begin{tcolorbox}
[width=\linewidth, sharp corners=all, colback=white!95!black]
\vspace{-11pt}
\begin{align} \label{eq:jordan-wigner-time-evolution}
&\mathcal{O}^{\text{QED}}_{\text{bp}}(s_0) \cdot 
T^{\text{QED}}(t) | \psi_{\text{gauged}} \rangle_{E^* \cup V^* } \nonumber \\
&= 
\widehat{\text{JW}} \cdot T^{\text{TL-Ising}}(t) | \psi_{\text{ungauged}} \rangle_{V}  \ . 
\end{align}
\end{tcolorbox}
\noindent Here, $\mathcal{O}^{\text{QED}}_{\text{bp}}(s_0)$ takes the same form as $\mathcal{O}^{\text{GM}}_{\text{bp}}(s_0)$.

A brief demonstration of this duality is as follows. 
The transformation $Z_v \mapsto Z_{e^*}$ in the exponent of $T^{\text{TL-Ising}}(t)$ is an immediate consequence of the $CX_{v,e^*}$ in the entangler, which is exactly the same as in $\widehat{\text{KW}}^{\text{GM}}$. 
The transverse-field term is conjugated by the entangler as $X_v \mapsto X_v X_{e^*} S_{e^*}$ (up to a sign at $v=N$). 
Each term in the expansion of the exponential  commutes with the Gauss law generator $G_{v^*}$.
We note that for any basis that satisfies the Gauss law constraint, the duality map $Z_v Z_{v'} \mapsto P_{v^*}$ holds. 
Therefore, the first term and the third term in the Hamiltonian~\eqref{eq:hamiltonian-tl-ising} are correctly transformed to the corresponding terms in \eqref{eq:hamiltonian-qed}. 
Now, by measurements, the term $X_v X_{e^*} S_{e^*}$ in the exponent is projected to $(-1)^{s_v } X_{e^*} S_{e^*}$. 
The inner product between $\langle \{s_v\}|_V$ and $|\{a_v\}\rangle_V$ gives us a phase, which is again expressed as the Pauli $Z$ operator $Z^{s_v}_{e^*}$. 
Moving these operators to the front gives us the equation~\eqref{eq:jordan-wigner-time-evolution}.

\subsection{(2+1)d Ising theory coupled to gauge fields from a star-plaquette model}
\label{sec:FS}

\begin{figure}
\includegraphics[width=0.9\linewidth]{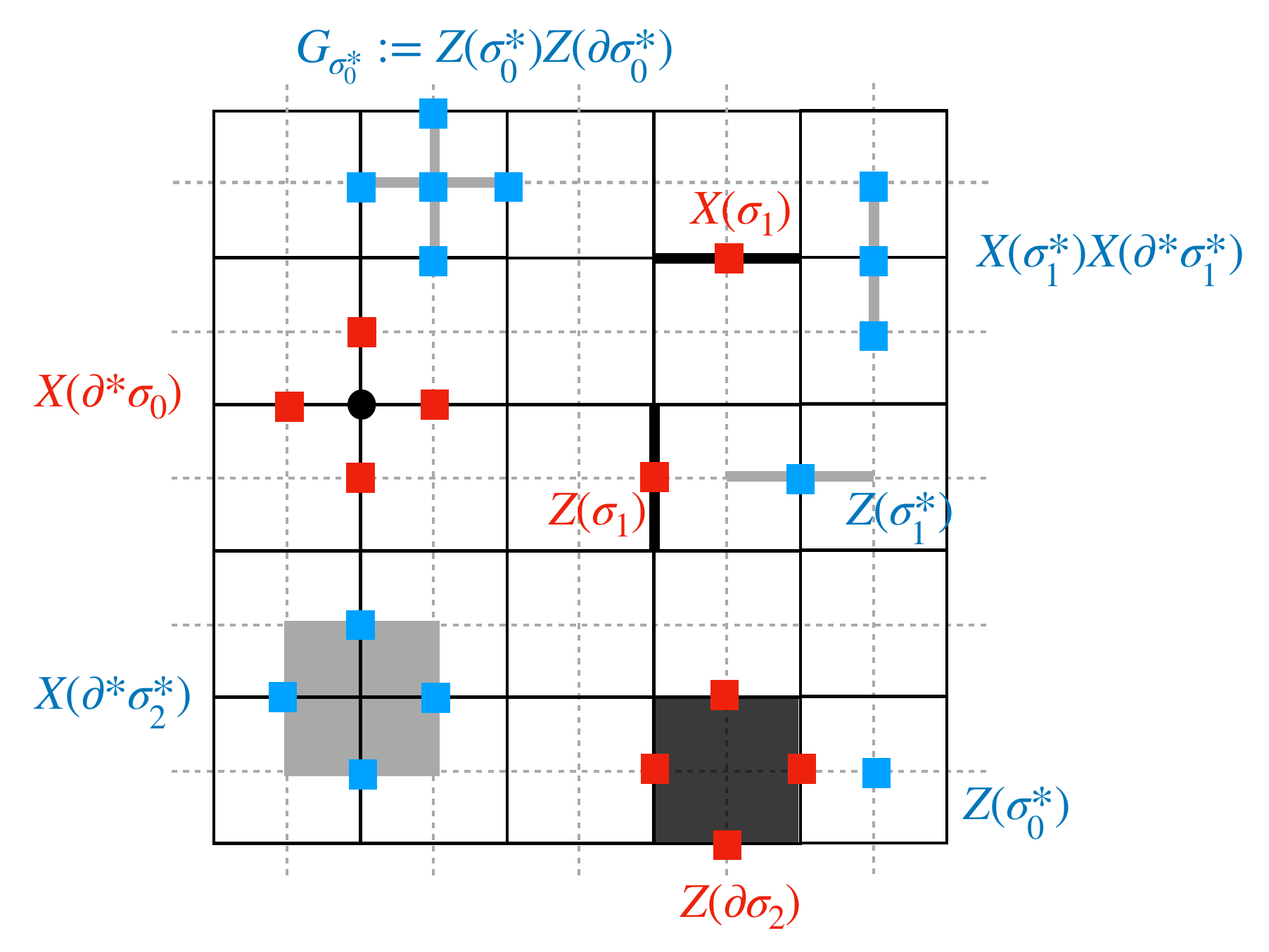}
\caption{The red boxes represent the operators that appear in the Hamiltonian $H^{\text{SP}}$ and the blue ones in the Hamiltonian $H^{\text{FS}}$ and the generator of the gauge transformation.}
\label{fig:2dlattice-FS}
\end{figure}

Consider a model defined on edges in a two-dimensional square lattice, which we call a star-plaquette model:
\begin{align}
H^{\text{SP}}=& - \mu \sum_{\sigma_0 \in \Delta_0} X(\partial^* \sigma_0) - \frac{1}{\mu} \sum_{\sigma_1 \in \Delta_1} Z_{\sigma_1} \nonumber \\
&- \lambda \sum_{\sigma_2 \in \Delta_2} Z(\partial \sigma_2) - \frac{1}{\lambda} \sum_{\sigma_1 \in \Delta_1} X_{\sigma_1}.
\end{align}
We illustrate terms in this Hamiltonian in Fig.~\ref{fig:2dlattice-FS}.
Here the term $X(\partial^* \sigma_0)$ is a product of $X$ over edges surrounding the vertex $\sigma_0$. 
The term $Z(\partial \sigma_2)$ is a product of $Z$ over edges surrounding the plaquette $\sigma_2$.
We consider the following model with matter fields on dual vertices and gauge fields on dual edges; see Fig.~\ref{fig:2dlattice-FS}:
\begin{align} \label{eq:FS-Hamiltonian}
H^{\text{FS}} =& - \mu \sum_{\sigma^*_2\in \Delta^*_2} X(\partial^* \sigma^*_2) - \frac{1}{\mu} \sum_{\sigma^*_1 \in \Delta^*_1} Z_{\sigma^*_1} \nonumber \\
&
- \lambda \sum_{\sigma^*_0 \in \Delta^*_0} Z_{\sigma^*_0}
-  \frac{1}{\lambda}   \sum_{\sigma^*_1 \in \Delta^*_1}
X_{\sigma^*_1} X(\partial^* \sigma^*_1). 
\end{align}
The superscript FS denotes the reference to  the work by Fradkin and Shenker~\cite{PhysRevD.19.3682}, who elaborated the phase diagram of this model. 
This includes deconfinement, confinement, and Higgs phases.
It is invariant under the gauge transformation generated by 
\begin{align}
G_{\sigma^*_0} = Z_{\sigma^*_0}  Z(\partial \sigma^*_0). 
\end{align}
The former model is obtained by a gauge-fixing of the latter, eliminating the matter degrees of freedom~\cite{2010PhRvB..82h5114T, PhysRevD.17.2637}.
We mention that the model eq.~\eqref{eq:FS-Hamiltonian} has been re-investigated recently in Ref.~\cite{verresen2022higgs}, where the Higgs phase was identified as an SPT phase. 
The quantum simulation of the model on Rydberg atom arrays with the protection of gauge invariance was discussed in Ref.~\cite{homeier2022quantum}. 
%We also mention applications of the model in the context of quantum liquid crystals \cite{beekman2017dual}. [If the second sentence sounds unnaturally placed, we can include this in Introduction; probably in the list of other citations in ``...while the low-energy properties of lattice gauge theories have been useful guideposts in condensed matter physics [10-12]"] }

\subsubsection{Results}

We aim to obtain the Trotterized time evolution of the model with the Hamiltonian $H^{\text{FS}}$ from that with $H^{\text{SP}}$. 
We denote the former as $T^{\text{FS}}(t)$ and the latter as $T^{\text{SP}}(t)$.
We show that an application of entanglers followed by measurements of primal edge qubits, which we call the Fradkin-Shenker map in this paper, implements a map from the star-plaquette model to the Fradkin-Shenker model.
For a set of measurement outcomes $s({\sigma_1}) \in \{0,1\}$, we define the associated 1-chain as
\begin{align} \label{eq:measurement-1-chain}
{s_1} = \sum_{{\sigma_1} \in {\Delta_1}} s({\sigma_1}) {\sigma_1}. 
\end{align}
To distinguish the undualized and dualized degrees of freedom, we again write the degrees of freedom on primal 1-cells with the double bracket $| \ \rangle\!\rangle$.
As before, we use the following bases for the wave functions: 
\begin{align} 
&| {c_1} \rangle\!\rangle := \bigotimes_{{\sigma_1} \in \Delta_1} | a(c_1 ; \sigma_1) \rangle\!\rangle^{(Z)}_{{\sigma_1}},   \\
&| \widetilde{ {c_1}} \rangle\!\rangle := \bigotimes_{{\sigma_1} \in \Delta_1} | a(c_1 ; \sigma_1) \rangle\!\rangle^{(X)}_{{\sigma_1}}, \\
&| c_2 \rangle := \bigotimes_{\sigma_2 \in \Delta_2} | a(c_2 ; \sigma_2) \rangle^{(Z)}_{\sigma_2}, \\
&| c_1 \rangle := \bigotimes_{\sigma_1 \in \Delta_1} | a(c_1 ; \sigma_1) \rangle^{(Z)}_{\sigma_1}. 
\end{align}

\begin{figure}
\includegraphics[width=0.5\linewidth]{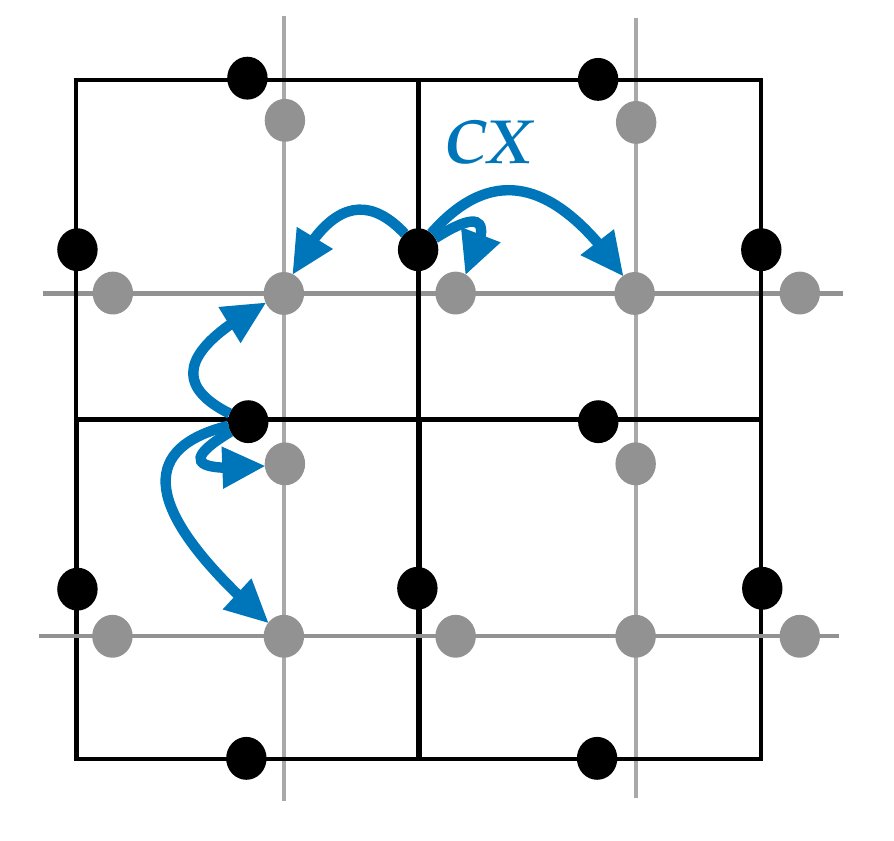}
\caption{The entangler used to implement the Fradkin-Shenker map.}
\label{fig:FS-entangler}
\end{figure}

We write the Fradkin-Shenker map as
\begin{align} \label{eq:fs-map}
\widehat{\text{FS}} &= 
\langle\!\langle  \widetilde{{s_1}}| \,\mathcal{U}^{\text{FS}}\, |0_1,0_2\rangle  \ , \\
\mathcal{U}^{\text{FS}}&= \prod_{{\sigma_1}\in {\Delta_1}} 
\Big( \underbrace{ CX_{\sigma_1,\sigma_1} }_{\substack{c~:~{\rm undualized} \\ t~:~{\rm dualized}~~ } } \prod_{\sigma_2 \in \Delta_2} [CX_{{\sigma_1},\sigma_2}]^{a(\partial^*{\sigma_1};\sigma_2) } \Big),  \label{eq:UFS}
\end{align}
where the first $CX$ gate is controlled by undualized 1-cells and applies $X$ on dualized 1-cells as depicted in Fig.~\ref{fig:FS-entangler}.
We take the ungauged wave function as {\it any} wave function defined for edge qubits:
\begin{align}
|\psi^{(1)}_{\text{ungauged}} \rangle\!\rangle
= \sum_{c_1 \in C_1} C(c_1) | {c_1} \rangle\!\rangle. 
\end{align}
On the other hand, the gauged wave function is 
\begin{align}
|\psi^{(1,2)}_{\text{gauged}} \rangle
&= \sum_{c_1 \in C_1} C(c_1) | c_1,\partial^* c_1 \rangle. 
\end{align}

Our key result here  is summarized as follows:
\begin{tcolorbox}
[width=\linewidth, sharp corners=all, colback=white!95!black]
\vspace{-11pt}
\begin{align} \label{eq:result-FS-map}
&\mathcal{O}^{\text{FS}}_{\text{bp}} (s_1)\cdot  T^{\text{FS}}(t) | \psi^{(1,2)}_{\text{gauged}} \rangle \nonumber \\ 
&= \widehat{\text{FS}} \cdot T^{\text{SP}}(t) | \psi^{(1)}_{\text{ungauged}} \rangle\!\rangle. 
\end{align}
\end{tcolorbox}
\noindent The byproduct operator for this dualization is given by
\begin{align} \label{eq:bp-fs}
\mathcal{O}^{\text{FS}}_{\text{bp}}(s_1)
:=
\prod_{ \sigma_1 \in \Delta_1 } Z(\sigma_1)^{s({\sigma_1})}
\end{align}
with $s_1=\sum_{\sigma_1 \in \Delta_1} s(\sigma_1) \sigma_1$ as defined in eq.~\eqref{eq:measurement-1-chain}.

\subsubsection{Demonstration} 

\begin{figure*}
\includegraphics[width=0.8\linewidth]{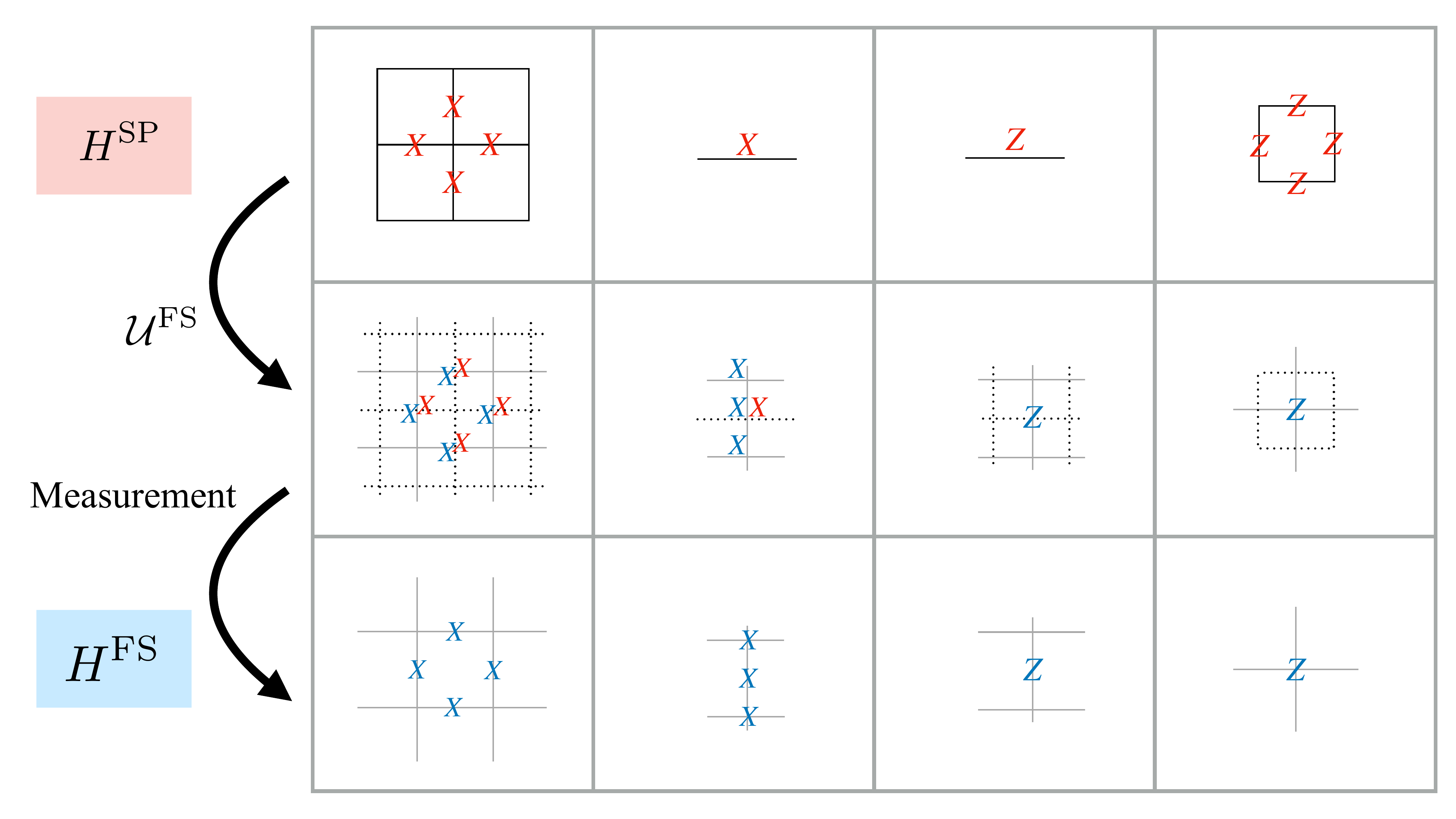}
\caption{The left two columns depict the conjugation of the $X$ operators in eq.~\eqref{eq:FS-X-conjugation} and \eqref{eq:FS-X-conjugation-2} via the entangler $\mathcal{U}_{\text{FS}}$ (top two rows), and the operator resulting after the measurement (bottom two rows).
The right two columns show the operator map for the $Z$ operators; see eq.~\eqref{eq:FS-Z-map_1} and \eqref{eq:FS-Z-map_2}.}
\label{fig:fradkin-shenker-map}
\end{figure*}

The following is the demonstration of the above equality.
To distinguish operators acting on the undualized and dualized degrees of freedom, we write those acting on the undualized qubits with bold symbols such as $\pmb{X}$. 
(In the figures, operators for undualized qubits are indicated by red, dualized ones by blue.)
We propagate the entangler of the Fradkin-Shenker map to the ungauged wave function, and we obtain time evolution terms with the $\pmb{X}$ on primal edges conjugated.
\begin{align} \label{eq:FS-X-conjugation}
    \pmb{X}({\partial^* \sigma_0}) & \mapsto \pmb{X}({\partial^* \sigma_0}) 
    X(\partial^* \sigma_0), \\
    \pmb{X}({\sigma_1}) & \mapsto \pmb{X}({\sigma_1}) X(\sigma_1) X(\partial^* \sigma_1) ,\label{eq:FS-X-conjugation-2}
\end{align}
see the top two rows in the left two columns in Fig.~\ref{fig:fradkin-shenker-map}.
Then, the pre-measurement wave function is the following:
\begin{align}
|\psi_{\text{pre}}\rangle = 
&
\Big(
\mathcal{U}^{\text{FS}} T^{SP}(t) \mathcal{U}^{\text{FS}\dagger}
\Big) \times \sum_{{c_1} \in {C_1}} C({c_1}) |{c_1} \rangle\!\rangle | c_1 ,\partial^* c_1\rangle. 
\end{align}
In the time evolution unitary, the $Z$ operators on primal edges are replaced with those on dual edges/vertices.
We write $|{c_1} \rangle\!\rangle | c_1, \partial^* c_1\rangle = |{c_1};c_1,\partial^* c_1 \rangle$. 
Then, we have
\begin{align}
\label{eq:FS-Z-map_1}
\pmb{Z}({\sigma_1}) |{c_1};c_1,\partial^* c_1 \rangle &
= Z(\sigma_1 ) |{c_1};c_1,\partial^* c_1 \rangle, \\ 
\pmb{Z}({\partial \sigma_2}) |{c_1};c_1,\partial^* c_1 \rangle &= Z(\sigma_2 ) |{c_1};c_1,\partial^* c_1 \rangle,  \label{eq:FS-Z-map_2}
\end{align}
the latter of which is due to $\#(\partial \sigma_2 \cap c_1) = \#(\sigma_2\cap \partial^* c_1)$; see the right two columns in Fig.~\ref{fig:fradkin-shenker-map}.
Note that the action of the operators depicted in the left two columns (the middle row) of Fig.~\ref{fig:fradkin-shenker-map} preserves such replacements,
namely,
\begin{align}
\pmb{Z}({\sigma_1}) \mathbb{X} |{c_1};c_1,\partial^* c_1 \rangle &= Z(\sigma_1 ) \mathbb{X} |{c_1};c_1,\partial^* c_1 \rangle, \\ 
\pmb{Z}({\partial \sigma_2}) \mathbb{X} |{c_1};c_1,\partial^* c_1 \rangle &= Z(\sigma_2 ) \mathbb{X} |{c_1};c_1,\partial^* c_1 \rangle, 
\end{align}
where the operator $\mathbb{X}$ is given as follows,
\begin{align}
\mathbb{X} =&
\prod_{\sigma_0 \in \Delta_0} 
\Big( \pmb{X}({\partial^* \sigma_0})  X(\partial^* \sigma_0) \Big)^{\Lambda(\sigma_0)} \nonumber \\
&\times \prod_{\sigma_1 \in \Delta_1} \Big(  
\pmb{X}({\sigma_1}) X(\sigma_1) X(\partial^* \sigma_1) \Big)^{\Lambda(\sigma_1)}, 
\end{align}
with $\Lambda(\sigma_0),\Lambda(\sigma_1) \in \{0,1\}$. 
Therefore the transformations $\pmb{Z}(\partial \sigma_2) \mapsto Z(\sigma_2)$ and $\pmb{Z}( \sigma_1) \mapsto Z(\sigma_1)$ can be done consistently within the time evolution unitary.
Hence, we have
\begin{widetext}
\begin{align}
|\psi_{\text{pre}} \rangle 
&= \Big(
\prod_{\sigma_0 \in \Delta_0} e^{i \Delta t  \mu \pmb{X}(\partial^* \sigma_0) {X}(\partial^* \sigma_0) } 
 \prod_{\sigma_1 \in \Delta_1} e^{i \frac{ \Delta t }{\mu} Z(\sigma_1)}
\prod_{\sigma_2 \in \Delta_2} e^{i \Delta t \lambda Z( \sigma_2)}  \prod_{\sigma_1 \in \Delta_1} e^{i  \frac{\Delta t}{\lambda} \pmb{X}(\sigma_1) X(\sigma_1) X(\partial^* \sigma_1) } 
\Big)^k  \nonumber \\
& \qquad  \times \sum_{c_1 \in C_1} C(c_1) |c_1 \rangle\!\rangle |c_1 ,\partial^* c_1 \rangle. 
\end{align}

By measurements of primal edge degrees of freedom, the $\pmb{X}$ operators on the primal edges in the time evolution unitaries become $\pmb{X}(\sigma_1)=(-1)^{s({\sigma_1})}$, and also we obtain a phase $\langle\!\langle \widetilde{ s_1} | c_1 \rangle\!\rangle = 2^{-|\Delta_1|/2}(-1)^{\# ({s_1} \cap {c_1})}$.
This phase can be equally written with the byproduct operator $\mathcal{O}^{\text{FS}}_{\text{bp}}(s_1)$.
Moving this operator through the time evolution unitary to the leftmost position flips the signs in the exponent in the time evolution unitary, precisely canceling the unwanted factors $(-1)^{s({\sigma_1})}$.  
We then arrive at 
\begin{align}
|\psi_{\text{post}} \rangle 
&= 2^{-|\Delta_1|/2} \mathcal{O}^{\text{FS}}_{\text{bp}}(s_1) \cdot \Big(
\prod_{\sigma_0 \in \Delta_0} e^{i \Delta t  \mu  {X}(\partial^* \sigma_0) } 
 \prod_{\sigma_1 \in \Delta_1} e^{i \frac{ \Delta t }{\mu} Z(\sigma_1)}
\prod_{\sigma_2 \in \Delta_2} e^{i \Delta t \lambda Z( \sigma_2)}  \prod_{\sigma_1 \in \Delta_1} e^{i  \frac{\Delta t}{\lambda} X(\sigma_1) X(\partial^* \sigma_1) } 
\Big)^k  | \psi_{\text{gauged}} \rangle.  
\end{align}
\end{widetext}
The operators in the exponents of the unitaries have been mapped as
\begin{align}
    \pmb{X}({\partial^* \sigma_0}) & \mapsto   X(\partial^* \sigma_0) = X(\partial^* \sigma^*_2),  \\
    \pmb{X}({\sigma_1}) & \mapsto  X(\sigma_1) X(\partial^* \sigma_1) = X(\sigma^*_1)X(\partial^* \sigma^*_1), 
\end{align}
and
\begin{align}
    \pmb{Z}({\sigma_1}) & \mapsto Z(\sigma_1)  = Z(\sigma^*_1), \\
    \pmb{Z}({\partial \sigma_2}) & \mapsto Z(\sigma_2) = Z(\sigma^*_0),  
\end{align}
and thus we have  demonstrated the claim eq.~\eqref{eq:result-FS-map}.

\section{Conclusions and Discussion}
\label{sec:conclusions-discussion}

We showed that the time evolutions in various models can be transformed into those in corresponding gauge theories via a deterministic procedure involving constant-depth entangling gates, local measurements, and corrections.  
With the presence of noises, when the number of non-trivial outcomes satisfies a certain condition ({\it e.g.,} when it is even, in the case with $\mathbb{Z}_2$), the procedure succeeds, and the resulting wave function is gauge symmetric under the assumption that the duality transformation itself is noise-free.
We also generalized the gauging method with measurement to gauge theories coupled to matters, and we are not aware of any such prior formulation in the literature.

\begin{figure}
    \includegraphics[width=\linewidth]{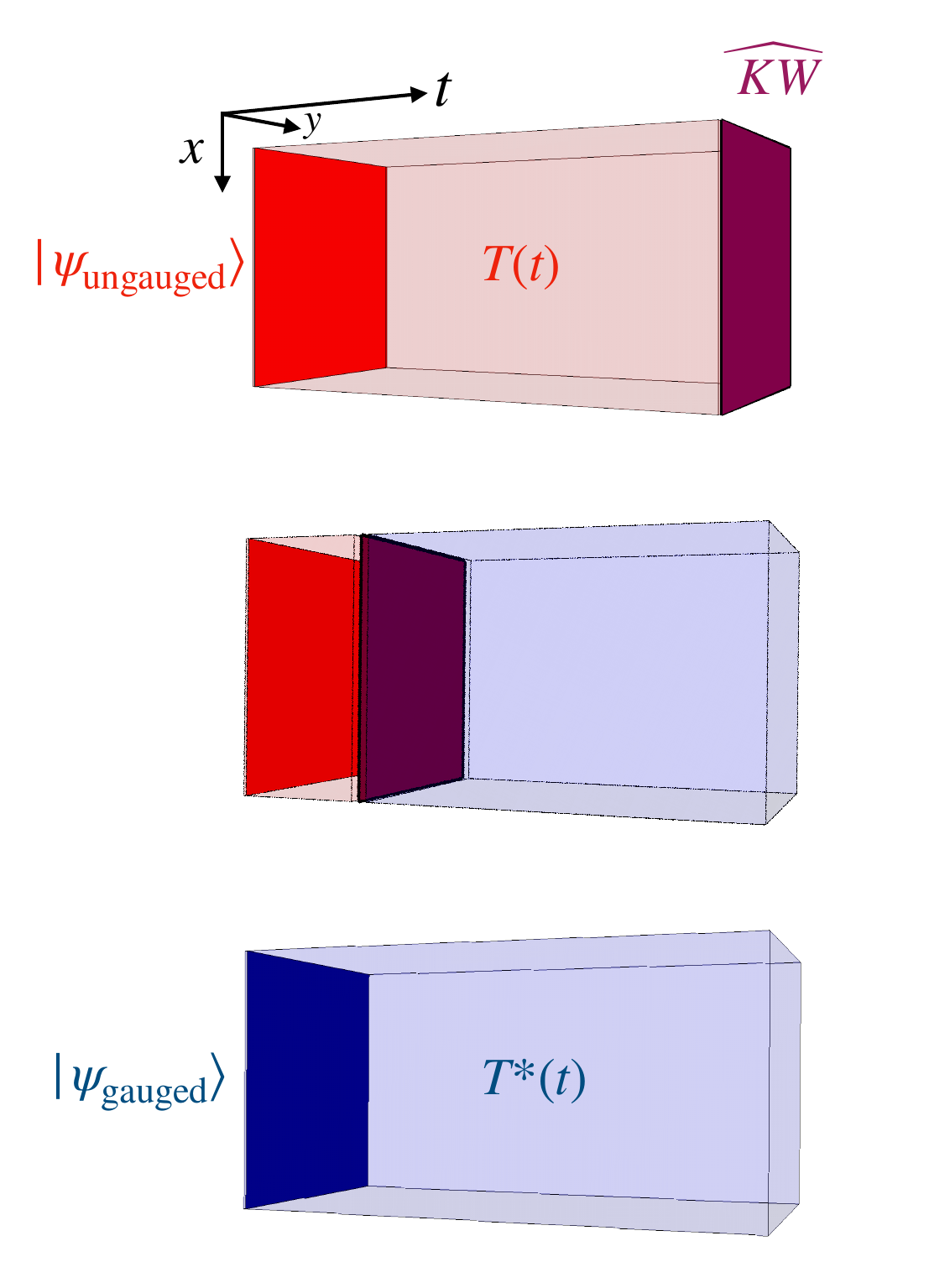}
    \caption{An interpretation of the relation between the (2+1)d TFI and the gauge theory in terms of the corresponding 3d Euclidean lattice theories \cite{Kogut}. A discussion on the mobility of the Kramers-Wannier duality defect/operator is given in Refs.~\cite{aasen2016topological, aasen2020topological,cao2023subsystem}, for example.} 
    \label{fig:KW-interface}
\end{figure}

Our procedure could be interpreted in terms of the corresponding Euclidean lattice theories~\cite{Kogut} as illustrated in Fig.~\ref{fig:KW-interface}.
Namely, we imagine a Kramers-Wannier duality interface that separates two theories, say (2+1)d transverse-field Ising model and the Ising gauge theory. 
The interface turns out to be able to freely move in the spacetime.
Fusing it with the past boundary, which defines the initial ungauged wave function, gives a new boundary that corresponds to the gauged initial wave function.

In two dimensions, a spin model with the $\mathbb{Z}_2$ global symmetry is mapped to a gauge theory restricted by a topological condition.
In our language, it is related to the ambiguity in choosing the paths for the byproduct operator, which gives rise to constraints on non-contractible loop operators.
On the other hand, spin models without global symmetry may be mapped to gauged spin models via a suitable generalization of the Kramers-Wannier map (and to $\mathbb{Z}_2$ QED via a generalized Jordan-Wigner transformation) without any addition of topological sectors~\cite{2018arXiv180907757R}.
Consistently, for gauge theories coupled matter fields, our procedure does not produce string operators, and the correction of the phase factors has nothing to do with the topology of the background manifold.

We showed that the time evolution under a symmetric Hamiltonian describing an SPT phase is mapped to that under a twisted gauge theory.
An appropriate generalization of the Kramers-Wannier map to higher spacetime dimensions would give us a time evolution describing a corresponding twisted higher-form gauge theory.
Another interesting direction of generalization is to non-Abelian gauge theories.
For solvable groups, it has been shown that appropriate generalizations of the Kramers-Wannier map can transform the short-range entangled states to that of non-Abelian (twisted) quantum double models \cite{2003AnPhy.303....2K,2013PhRvB..87l5114H} \cite{2021arXiv211201519T,Hierarchy, Bravyi2022a}.
It would be interesting to explicitly formulate a procedure to obtain the time evolution of non-Abelian lattice gauge theories by measurements.

Our method enables us a shortcut to the quantum simulation in the long-range entanglement regime, {\it e.g.,} a quenched dynamics.
Some current quantum devices may already furnish the basic demands of our procedure, such as locally addressed mid-circuit measurements and sufficient coherence.
The recent result of demonstrating the gauging method on real devices on trapped ions~\cite{iqbal2023topological,foss2023experimental, iqbal2023creation} encourages us to take further steps and implement the duality transformation of the time evolution in gauge theories.
The Rydberg atom arrays~\cite{2000PhRvL..85.2208J,2001PhRvL..87c7901L,2009NatPh...5..110U,2009NatPh...5..115G,2010PhRvL.104a0503I,2010PhRvL.104a0502W,2010NatPh...6..382W,2019Sci...365..775D,verresen2021prediction, omran2019generation, semeghini2021probing, homeier2022quantum} may also be suitable for processing the gauging procedures~\cite{SchrodingersCat}.
Other quantum devices which support the mid-circuit measurements are also interesting to be considered for implementing the idea in this work.

\section*{Acknowledgement}
The authors thank Wenhan Guo, Yabo Li, Mikhail Litvinov, Aswin Parayil Mana for their discussions. In particular, HS thanks Takuya Okuda for discussions on related topics. This work was supported by the National Science Foundation under Award No. PHY 2310614. T.-C.W. acknowledges the support of Stony Brook University’s Center for Distributed Quantum Processing.

\appendix

\section{Dualizaing to $\mathbb{Z}_2$ QED: expanded form}

We explain our result in Section~IV.A with an expanded notation.

The undualized model is defined on the vertices in the one-dimensional lattice, and the Hamiltonian is given by
\begin{align}
H^{\text{TL-Ising}}  
=& - \sum_{ e \in E} \prod_{v \subset e} Z_v - \sum_{v \in V} ( g X_v + h Z_v) . 
\end{align}
We consider a generalized Kramers-Wannier duality by introducing a topological gauge field on the edges in the dual lattice denoted by $e^*$ (which is dual to a primal vertex; $e^* \simeq v$), and imposing the Gauss law constraint on the degrees of freedom in the dual lattice.
The duality map is given by 
\begin{align}
X_v &\mapsto X_{e^*} \prod_{v^* \subset e^*} X_{v^*},  \\
Z_v & \mapsto Z_{e^*},
\end{align}
and the Gauss law constraint is given by 
\begin{align} 
G_{v^*} := Z_{v^*} \prod_{e^* \supset v^* } Z_{e^*} =1.  
\end{align}
We also have another transformation $\prod_{v \subset e} Z_v \mapsto Z_{v^*}$, which is obtained by substituting the second line of the transformations to the Gauss law constraint.
The resulting theory is a gauge-matter theory:
\begin{align}
H^{\text{GM}}= 
- \sum_{v^* \in V^*} Z_{v^*} 
-  \sum_{e^* \in E^*} \Big(g X_{e^*} \prod_{v^* \subset e^*} X_{v^*}+ h  Z_{e^*} \Big)\,. 
\end{align}
The Hamiltonian is invariant under the gauge transformation, $[H^{\text{GM}},G_{v^*}]=0$.
The first term is an ordinary matter term.
The second term is a matter kinetic term covariantly coupled to the gauge field. 
And the last term can be seen as an electric term in the gauge sector.

To distinguish the undualized and dualized degrees of freedom, we write the degrees of freedom on primal 0-cells with the double bracket $| \ \rangle\!\rangle$. The original wave function can be expanded with the basis
\begin{align} 
\bigotimes_{v \in V} | a_v \rangle\!\rangle_v  
\end{align}
and that in the dualized theory can be expanded with
\begin{align}
\bigotimes_{v \in V }| a_v \rangle_v   \text{ and }\bigotimes_{e \in E }| a_e  \rangle_e \, .
\end{align}

First, consider the time evolution with $H^{\text{TL-Ising}}$, whose Trotterization is written as $T^{\text{TL-Ising}}(t)$.
We take the initial state as {\it any} state defined on the vertices:
\begin{align} 
| \psi^{(0)}_{\text{ungauged}}\rangle\!\rangle = \sum_{ \{a_v\} \in \{0,1\}^{\otimes V}} C(\{a_v\}) \bigotimes_{v \in V} | a_v \rangle\!\rangle_v  . 
\end{align}
In particular, we do not impose a $\mathbb{Z}_2$ symmetry for this wave function.

We will load a gauged state on edges and vertices on the dual lattice. 
We emphasize that an edge in the dual lattice is identical to a vertex in the primal lattice, but we treat them as separate degrees of freedom.
We initiate the wave function as 
\begin{align}
|0\rangle^{\otimes V} |0\rangle^{\otimes E}  .
\end{align}
Note that this state satisfies the Gauss law constraint. 

We consider an entangler 
\begin{align}
\mathcal{U}^{\text{GM}} = \prod_{v \in V} 
\Big( \underbrace{CX_{v,v} }_{\substack{c~:~ {\rm undualized} \\ t~:~ {\rm dualized ~~}} } 
\prod_{\substack{e \supset v \\ e \in E }} CX_{v,e}  \Big), 
\end{align}
where the first $CX$ gate is controlled by the undualized qubits (also labeled as $c$) and applies $X$ on the dualized degrees of freedom (also labeled as $t$; note that both $c$ and $t$ are on the same vertex but separate degrees of freedom). 
The generalized Kramers-Wannier map is now defined as
\begin{align}
\widehat{\text{KW}}^{\text{GM}} = 
\bigotimes_{v \in V} \langle\!\langle s_v |^{(X)}_v \,\mathcal{U}^{\text{GM}} \, 
\bigotimes_{v \in V}|0\rangle^{(Z)}_v\bigotimes_{e \in E} |0 \rangle^{(Z)}_e , 
\end{align}
where the notation $(Z)$ and $(X)$ indicates that the basis is the eigenvector of the operator, and $s_v \in \{0,1\}$ is the measurement outcome.
We denote the Trotterized time evolution with the Hamiltonian $H^{\text{GM}}$ by $T^{\text{GM}}(t)$.

We claim that the time evolution of the gauged Ising model can be obtained by the Kramers-Wannier map KW$^{\text{GM}}$.
\begin{tcolorbox}
[width=\linewidth, sharp corners=all, colback=white!95!black]
\vspace{-11pt}
\begin{align} 
&\mathcal{O}^{\text{GM}}_{\text{bp}} (\{s_v\})\cdot 
T^{\text{GM}}(t) | \psi^{(0,1)}_{\text{gauged}} \rangle \nonumber \\
&= 
\widehat{\text{KW}}^{\text{GM}} \cdot T^{\text{TL-Ising}}(t) | \psi^{(0)}_{\text{ungauged}} \rangle\!\rangle. 
\end{align}
\end{tcolorbox}
\noindent Here the gauged wave function is 
\begin{align}
| \psi^{(0,1)}_{\text{gauged}} \rangle
= 
\sum_{ \{a_v\} \in \{0,1\}^{\otimes V}} C(\{a_v\}) \bigotimes_{v\in V} |a_v\rangle_v
\bigotimes_{e\in E}\Big| \sum_{v \subset e} a_v \Big\rangle_e, 
\end{align}
and it satisfies the Gauss law constraint, $Z_{e} \prod_{v \subset e } Z_{v} =1$.

The byproduct operator for this dualization is given by
\begin{align} 
\mathcal{O}^{\text{GM}}_{\text{bp}}(\{s_v\})
=
\prod_{ v \in V } (Z_v)^{s_v}. 
\end{align}
Here, the exponent $s_v$ is associated with measurement outcomes of un-dualized degrees of freedom, but the operator $Z_v$ acts on the dual degrees of freedom.
We emphasize that it is no longer a string operator. 
Correction can be done directly by applying $\mathcal{O}^{\text{GM}}_{\text{bp}}$ after measurements.

\section{Replacing phase operators}
\label{sec:replacing}

Here we show the replacement $Z(\partial \sigma_1) \rightarrow Z(\sigma_1)$ in the Trotter unitary, which we omitted from the main text.
We note that 
\begin{align}
Z(c_i) X(c'_i) = (-1)^{\#(c_i \cap c'_i)} X(c'_i) Z(c_i), 
\end{align}
by appropriately regarding one of the chains as its dual.
For $\sigma_1 \in \Delta_1$ and $\sigma_0 \in \Delta_0$,
\begin{align}
&Z(\partial \sigma_1)
X_{\sigma_0} X(\partial^* \sigma_0)  \nonumber\\
&\qquad = (-1)^{\#( \partial \sigma_1 \cap \sigma_0 )}
X_{\sigma_0} X(\partial^* \sigma_0)
Z(\partial \sigma_1), 
\end{align}
and 
\begin{align}
 Z(\sigma_1)
X_{\sigma_0} X(\partial^* \sigma_0) 
= (-1)^{\#( \sigma_1 \cap \partial^* \sigma_0 )}
X_{\sigma_0} X(\partial^* \sigma_0) 
Z(\sigma_1).  
\end{align}
Due to the Poincare duality, we have $(-1)^{\#( \partial \sigma_1 \cap \sigma_0 )}=(-1)^{\#( \sigma_1 \cap \partial^* \sigma_0 )}$. 
Using the relation~\eqref{eq:Z-replacement}, we find
\begin{align}
&Z(\partial \sigma_1)
\Big(\prod_{\sigma_0 \in \Delta_0} (  X_{\sigma_0} X(\partial^* \sigma_0) )^{\Lambda(\sigma_0)} \Big) | c_0, \partial^* c_0 \rangle  \nonumber \\ 
& =
 Z(\sigma_1)
 \Big(\prod_{\sigma_0 \in \Delta_0} (  X_{\sigma_0} X(\partial^* \sigma_0) )^{\Lambda(\sigma_0)} \Big) | c_0, \partial^* c_0 \rangle,
\end{align}
with $\Lambda(\sigma_0) \in \{0,1\}$.

\section{Delegated proof of Kramers-Wannier transformation of time evolutions}

\subsection{Twisted gauge theory in (2+1)d}

The Kramers-Wannier transformation of the tTFI is mostly identical to that of TFI.
It is calculated as follows.
We first apply the entangler and obtain
\begin{align}
&\Big(\prod CX_{v,e}\Big) T_{\text{tTFI}} (t) |\psi_{\text{ungauged}} \rangle_V |0\rangle_E \nonumber \\
& = 
\Big(
\prod_{u \in V}
e^{i \Delta t \mathcal{O}_u \prod_{e \supset u} X_e}
\prod_{\langle u , u' \rangle \in E}
e^{ig \Delta t  Z_u Z_{u'} }
\Big)^k \nonumber \\
& \qquad \sum_{a_v = 0,1} C(\{a_v\})|\{a_v + a_{v'}\} \rangle_E | \{a_v\} \rangle_V. 
\end{align}
As before, $Z_u Z_{u'}$ can be replaced by $Z_{\langle u,u'\rangle}$ because the operator $\mathcal{O}_u \prod_{e \supset u} X_e$ preserves such structure.
Furthermore, we can replace the operator $\mathcal{O}_u \prod_{e \supset u} X_e$ by $X_u \tilde{\mathcal{O}}_u$.

Now we take the inner product between $\langle \{\tilde{s}_v\}|_V$ and $| \{a_v\} \rangle_V$. 
We find the resultant wave function for the edge degrees of freedom is equal to
\begin{align}
&\Big( 
\prod_{u \in V}
\exp \Big(i \Delta t (-1)^{s_u} \tilde{\mathcal{O}}_u \Big)
\prod_{e\in E}
e^{ig \Delta t  Z_e }
\Big)^k  \nonumber \\ 
&\times \mathcal{O}_{\text{bp}} (\rho_1) | \psi_{\text{gauged}}\rangle_E. 
\end{align}
Using the commutation relation $(-1)^{s_u} \tilde{\mathcal{O}}_u \mathcal{O}_{\text{bp}}(\rho_1) =\mathcal{O}_{\text{bp}} (\rho_1)\tilde{\mathcal{O}}_u$, we obtain 
\begin{align}
\mathcal{O}_{\text{bp}}(\rho_1) \cdot  \Big( 
\prod_{u \in V}
\exp \Big(i \Delta t  \tilde{\mathcal{O}}_u \Big)
\prod_{e \rangle \in E}
e^{ig \Delta t  Z_e }
\Big)^k  | \psi_{\text{gauged}}\rangle_E  . 
\end{align}

\subsection{Gauged Ising model in (1+1)d}
\label{sec:gauged-ising-proof}

To distinguish operators acting on the undualized and dualized degrees of freedom, we write those acting on the undualized qubits with bold symbols such as $\pmb{X}$.
We propagate the entangler $\mathcal{U}^{\text{GM}}$ to the ungauged wave function, and we obtain time evolution terms with the $\pmb{X}$ on primal vertices conjugated,
\begin{align}
    \pmb{X}({\sigma_0}) & \mapsto \pmb{X}({\sigma_0})  X(\sigma_0)
    X(\partial^* \sigma_0). 
\end{align}
Then, the pre-measurement wave function is the following:
\begin{align}
|\psi_{\text{pre}} \rangle 
=
& \Big( \mathcal{U}^{\text{GM}} T^{\text{TL-Ising}}(t) \mathcal{U}^{\text{GM}\dagger}\Big) \nonumber \\
&\sum_{{c_0} \in {C_0}} C({c_0}) |{c_0} \rangle\!\rangle | c_0 , \partial^* c_0\rangle_{\Delta_1}. 
\end{align}
With this wave function, the $Z$ operators on primal vertices are replaced with those on dual edges/vertices.
We write $|{c_0} \rangle\!\rangle | c_0, \partial^* c_0\rangle = |{c_0};c_0,\partial^* c_0 \rangle$ and then we have
\begin{align}
\pmb{Z}({\sigma_0}) |{c_0};c_0,\partial^* c_0 \rangle &
= Z(\sigma_0 ) |{c_0};c_0,\partial^* c_0 \rangle, \\ 
\pmb{Z}({\partial \sigma_1}) |{c_0};c_0,\partial^* c_0 \rangle &= Z(\sigma_1 ) |{c_0};c_0,\partial^* c_0 \rangle, 
\end{align}
the latter of which is due to $\#(\partial \sigma_1 \cap c_0) = \#(\sigma_1\cap \partial^* c_0)$. 
Note that for arbitrary $\sigma_0 \in \Delta_0$, we have
\begin{align}
\pmb{Z}({\sigma_0}) \mathbb{X} |{c_0};c_0,\partial^* c_0 \rangle &
= Z(\sigma_0 ) \mathbb{X}  |{c_0};c_0,\partial^* c_0 \rangle, \\ 
\pmb{Z}({\partial \sigma_1}) \mathbb{X} |{c_0};c_0,\partial^* c_0 \rangle &= Z(\sigma_1 ) \mathbb{X} |{c_0};c_0,\partial^* c_0 \rangle,
\end{align}
with 
\begin{align}
\mathbb{X} = \prod_{\sigma_0 \in \Delta_0} \Big( \pmb{X}(\sigma_0) X(\sigma_0) X(\partial^* \sigma_0) \Big)^{\Lambda(\sigma_0)}
\end{align}
with $\Lambda(\sigma_0) \in \{0,1\}$.
So the replacement of $Z$ operators holds even with the action of the conjugated $\pmb{X}$ operators in the time evolution unitary.
The pre-measurement wave function is thus
\begin{widetext}
\begin{align}
|\psi_{\text{pre}} \rangle
= \Big(
\prod_{\sigma_1} e^{ i \Delta t Z(\sigma_1) }
\prod_{\sigma_0} e^{ i \Delta t h Z(\sigma_0) }
\prod_{\sigma_0} e^{ i \Delta t g \pmb{X}(\sigma_0) X(\sigma_0) X(\partial^* \sigma_0) }
\Big)^k 
\sum_{{c_0} \in {C_0}} C({c_0}) |{c_0} \rangle\!\rangle | c_0, \partial^* c_0\rangle_{\Delta_1}. 
\end{align}

By measurements of the undualized degrees of freedom, the $\pmb{X}$ operators on the primal vertices in the time evolution unitaries become $(-1)^{s({\sigma_0})}$. We also obtain a phase $ \langle\!\langle \widetilde{s_0}|c_0 \rangle\!\rangle = 2^{-|\Delta_0|/2}(-1)^{\# ({s} \cap {c_0})}$:
\begin{align}
|\psi_{\text{post}} \rangle
= \Big(
\prod_{\sigma_1} e^{ i \Delta t Z(\sigma_1) }
\prod_{\sigma_0} e^{ i \Delta t h Z(\sigma_0) }
\prod_{\sigma_0} e^{ i \Delta t g (-1)^{s(\sigma_0)} X(\partial^* \sigma_0) X(\sigma_0) }
\Big)^k 
2^{-|\Delta_0|/2}
\sum_{{c_0} \in {C_0}} C({c_0}) (-1)^{\#(s_0 \cap c_0)} | c_0, \partial^* c_0\rangle_{\Delta_1}. 
\end{align}
This phase $(-1)^{\#(s_0 \cap c_0)}$ can be equally written as 
\begin{align} 
\mathcal{O}^{\text{GM}}_{\text{bp}}(s_0)
=
\prod_{ \sigma_0 \in \Delta_0 } Z({\sigma_0})^{s({\sigma_0})}
\end{align}
acting on $|\psi_{\text{gauged}} \rangle$. 
Moving this operator through the time evolution unitary to the left flips the signs in the exponent in the time evolution unitary, precisely canceling the unwanted factors $(-1)^{s({\sigma_0})}$.  
Thus we have mapped the operators in the exponents of the unitaries as
\begin{align}
&\pmb{Z}({\sigma_0})  \mapsto Z(\sigma_0)  =  Z(\sigma^*_1),  \\
& \pmb{Z}({\partial \sigma_1}) \mapsto Z(\sigma_1)  =  Z(\sigma^*_0),  \\ 
&\pmb{X}(\sigma_0) \mapsto X(\sigma_0) X(\partial^* \sigma_0) = X(\sigma^*_1) X(\partial^* \sigma_1). 
\end{align}
More concretely, we have shown
\begin{align}
|\psi_{\text{post}} \rangle
&= 
2^{-|\Delta_0|/2}
\mathcal{O}^{\text{GM}}_{\text{bp}}(s_0) \cdot 
\Big(
\prod_{\sigma_1} e^{ i \Delta t Z(\sigma_1) }
\prod_{\sigma_0} e^{ i \Delta t h Z(\sigma_0) }
\prod_{\sigma_0} e^{ i \Delta t g X(\sigma_0) X(\partial^* \sigma_0) }
\Big)^k |\psi_{\text{gauged}} \rangle  \nonumber \\
&=  
2^{-|\Delta_0|/2}\mathcal{O}^{\text{GM}}_{\text{bp}} (s_0)\cdot T^{\text{GM}}(t ) |\psi_{\text{gauged}} \rangle. 
\end{align}
\end{widetext}

\section{2d gauged Jordan-Wigner transformation}

The authors of Ref.~\cite{borla2022quantum} considered a duality map between a Pauli spin model on the 2d square lattice, which involves up to six-body interaction terms, and a lattice gauge theory coupled to spinless fermions. 
Here, we consider the mapping in~Ref.~\cite{borla2022quantum} and implement it using local unitary operators and measurement. We also remark that authors of Ref.~\cite{zohar2018eliminating} constructed a unitary transforation to eliminate matter degrees of freedom from lattice gauge theories with fermionic matter fields.

The Hamiltonian of the latter is defined on a lattice with oriented edges (with $\partial e = v_+ - v_-$) pointing towards $+x$ and $-y$ directions, and is expressed in terms of Majorana fermions as
\begin{align}
H_{\rm MGT} 
=& t \sum_{e} \big( i \chi'_{v_-} \widetilde{Z}_e \chi_{v_+} - i \chi_{v_-} \widetilde{Z}_e \chi'_{v_+}  \big)
+ \frac{\mu}{2} \sum_v P_v \nonumber \\
&- J \sum_{p} \prod_{e \subset p } \widetilde{Z}_e 
-h \sum_{e} \widetilde{X}_e \, .
\end{align}
The Majorana fermions above are related to the complex fermion as $c=(\chi + i\chi')/2$, $c^\dagger=(\chi - i\chi')/2$, so that $\{\chi, \chi\}=\{\chi', \chi'\}=2$ and $\{\chi, \chi'\}=0$.
The operator $P_v$ is the local fermion parity operator $P_v:=i \chi'_v \chi_v$.
We also make use of a fermionic bilinear operator $S_e = - i \chi'_{v_-} \chi_{v_+}$. 

We generalize the construction of the 2d Jordan-Wigner transformation enabled by entangler and measurement in Ref.~\cite{2021arXiv211201519T} to incorporate gauge fields in the Majorana fermion model.

We begin with a simpler setup with a Hilbert space $\mathcal{H}_V \otimes \mathcal{H}_E$ of qubits defined on vertices and edges of the 2d (periodic) square lattice.
Let $CS_{e}$ be an operator such that it applies $S_{e}$ to two Majorana fermions controlled by the qubit on $e$. 
We set
\begin{equation}
\mathcal{U}_{CS} := \prod_{e} CS_{e},
\end{equation}
with the following ordering.
Within a horizontal layer, the operators $CS_{e}$ commute with each other, and we let them appear in the product simultaneously.
Such a product is ordered so that as we go down in the $-y$ direction, we go to the left within the product.
It was noted that 
\begin{align}
\mathcal{U}_{CS}  P_{v} \mathcal{U}_{CS}^{-1} 
 = 
\begin{array}{ccccc}
&& Z_{e_u} &&\\
&&|&&\\
Z_{e_\ell}&\!\!\text{\bf ---}\!\!&P_v&\!\!\text{\bf ---}\!\!&Z_{e_r}\\
&&|&&\\
&&Z_{e_d}&& 
\end{array} \ , 
\end{align}
\begin{align}
\mathcal{U}_{CS}  
\left(
\begin{array}{c}
| \\
X_e \\
|
\end{array}
\right)
\mathcal{U}_{CS}^{-1}
= 
\begin{array}{ccccc}
&& i \chi'_{v_-} &&\\
&&|&&\\
&&X_e&&\\
&&|&&\\
Z_{e'}&\!\!\text{\bf ---}\!\!&\chi_{v_+}&& 
\end{array} \ , 
\end{align}
\begin{align}
\mathcal{U}_{CS}  
\left(
\begin{array}{ccc}
\!\text{\bf ---}\!\!\!&X_e& \!\!\!\text{\bf ---}\!
\end{array}
\right)
\mathcal{U}_{CS}^{-1}
= 
\begin{array}{ccccc}
\chi'_{v_-}&\!\!\text{\bf ---}\!\!&X_e&\!\!\text{\bf ---}\!\!&i \chi_{v_+}\\
|&&&&\\
Z_{e'}&&&& 
\end{array} \ .
\end{align}
As discussed in Refs.~\cite{2021arXiv211201519T}, the local entanglers $CS_e$ do not commute (i.e., the one in the vertical direction and the one in the horizontal direction that overlap at a vertex), and different orderings of the product in entanglers give different duality transformations, where dual Pauli spin models have different spatial anisotropy. 
Here, we have chosen a particular ordering following Ref.~\cite{sukeno2023mbqs}.

Generalizing the entangler above, we consider a Hilbert space $\mathcal{H}_V \otimes \mathcal{H}_E \otimes \mathcal{H}_{\widetilde{E} }$ and we define (note a similar use of notation in eq.~(\ref{eq:UFS})),
\begin{align}
\mathcal{U}_{\rm MGT} := \mathcal{U}_{CS} \times  \prod_{e} \underbrace{  C\widetilde{Z}_{e,e} }_{ \substack{ c \in E \\ t \in \widetilde{E} } },  
\end{align}
where the wide tilde denotes the Pauli operator that acts on qubits on edges different from the ones in $\mathcal{U}_{CS}$. 
{(Namely, for every edge, we have a copy of qubits --- tilded and untilded --- and the controlled-Z gate acts on them.)}
We get 
\begin{align}
\mathcal{U}_{\rm MGT}   P_{v} \,\mathcal{U}_{\rm MGT} ^{-1} 
 = 
\begin{array}{ccccc}
&& Z_{e_u} &&\\
&&|&&\\
Z_{e_\ell}&\!\!\text{\bf ---}\!\!&P_v&\!\!\text{\bf ---}\!\!&Z_{e_r}\\
&&|&&\\
&&Z_{e_d}&& 
\end{array}
 =\mathcal{U}^{-1}_{\rm MGT} P_{v} \,\mathcal{U}_{\rm MGT}, 
\end{align}
\begin{align}
\mathcal{U}_{\rm MGT}   
\left(
\begin{array}{c}
| \\
X_e \\
|
\end{array}
\right)
\mathcal{U}_{\rm MGT} ^{-1}
=\! 
\begin{array}{ccccc}
&& i \chi'_{v_-} &&\\
&&|&&\\
&&X_e \widetilde{Z}_e&&\\
&&|&&\\
Z_{e'}&\!\!\text{\bf ---}\!\!&\chi_{v_+}&& 
\end{array} \ , 
\end{align}
\begin{align}
\mathcal{U}_{\rm MGT}  
\left(
\begin{array}{ccc}
\!\text{\bf ---}\!\!\!&X_e & \!\!\!\text{\bf ---}\!
\end{array}
\right)
\mathcal{U}_{\rm MGT}^{-1}
\!= 
\begin{array}{ccccc}
\chi'_{v_-}&\!\!\text{\bf ---}\!\!&X_e \widetilde{Z}_e&\!\!\text{\bf ---}\!\!&i \chi_{v_+}\\
|&&&&\\
Z_{e'}&&&& 
\end{array} \ .
\end{align}

Let $|0\rangle^V$denote the fermionic vacuum state. 
We define a duality operator $\widehat{{\rm D}}:\mathcal{H}_E \rightarrow \mathcal{H}_{\widetilde{E} }\otimes \mathcal{H}_V$,
\begin{align}
\widehat{{\rm D}} := \langle +|^E \mathcal{U}_{\rm MGT} |+\rangle^{\widetilde{E}} |0\rangle^V \, . 
\end{align}
We immediately obtain the following equalities:
\begin{align}
\widehat{{\rm D}} \left(
\begin{array}{cccc}
&&&I \\
&&&| \\
&&&X_e \\
&&& | \\
& Z_{e'} & \!\! \text{\bf ---}\!\! & I 
\end{array}
\right) 
= 
\left(
\begin{array}{cccc}
&&& i\chi'_{v_-} \\
&&&| \\
&&&\widetilde{Z}_e \\
&&& | \\
& I & \!\! \text{\bf ---}\!\! & \chi_{v_+} 
\end{array}
\right) 
\widehat{{\rm D}} \ , 
\end{align}

\begin{align}
\widehat{{\rm D}} \left(
\begin{array}{ccccc}
I&\!\!\text{\bf ---}\!\!&X_e&\!\!\text{\bf ---}\!\!&I  \\ 
|&&&& \\ 
Z_{e'} &&&& 
\end{array}
\right) 
= 
\left(
\begin{array}{ccccc}
\chi'_{v_-}&\!\!\text{\bf ---}\!\!&\widetilde{Z}_e&\!\!\text{\bf ---}\!\!&i\chi_{v_+}  \\ 
|&&&& \\ 
I &&&& 
\end{array}
\right) 
\widehat{{\rm D}} \ , 
\end{align}

\begin{align}
\widehat{{\rm D}} 
\left( 
\begin{array}{ccccc}
&& Z_{e_u} &&\\
&&|&&\\
Z_{e_\ell}&\!\!\text{\bf ---}\!\!&I&\!\!\text{\bf ---}\!\!&Z_{e_r}\\
&&|&&\\
&&Z_{e_d}&& 
\end{array}
\right)
=
\left( 
\begin{array}{ccccc}
&& I &&\\
&&|&&\\
I&\!\!\text{\bf ---}\!\!&P_v&\!\!\text{\bf ---}\!\!&I\\
&&|&&\\
&&I&& 
\end{array}
\right)
\widehat{{\rm D}}  \ , 
\end{align}

\begin{align}
\widehat{{\rm D}} 
\left(
\begin{array}{ccc}
\text{\bf ---}\!\! & Z_e &  \!\! \text{\bf ---}
\end{array}
\right)
= 
\left(
\begin{array}{ccc}
\text{\bf ---}\!\! & \widetilde{X}_e &  \!\! \text{\bf ---}
\end{array}
\right)
\widehat{{\rm D}} \, 
\end{align}

\begin{align} \label{eq:Z-to-X}
\widehat{{\rm D}} 
\left(
\begin{array}{c}
|\\
Z_e \\
|
\end{array}
\right)
= 
\left(
\begin{array}{c}
|\\
\widetilde{X}_e \\
|
\end{array}
\right)
\widehat{{\rm D}}  \, .
\end{align}

Combining some of the duality relations above, some algebras, and $\chi' P = i \chi$ and $\chi P = - i \chi'$, we also find the dual Pauli terms for the other Majorana fermion terms coupled to the gauge field:
\begin{align}
\widehat{{\rm D}} 
\left(
\begin{array}{ccccc}
&&Z&& \\
&&|&& \\
Z&\!\!\text{\bf---}\!\!&I&\!\!\text{\bf---}\!\!&Z \\
&&|&& \\ 
&&X_e&&\\
&&|&& \\
I&\!\!\text{\bf---}\!\!&I&\!\!\text{\bf---}\!\!&Z \\
&&|&& \\
&&Z&&
\end{array}
\right)
=
\left(
\begin{array}{ccc}
&& i \chi_{v_-} \\
&&| \\
&&\widetilde{Z}_e \\
&& | \\
I & \!\!\text{\bf---}\!\! & \chi'_{v_+}
\end{array}
\right)
\widehat{{\rm D}} 
\end{align}
and 
\begin{align}
&\widehat{{\rm D}} 
\left(
\begin{array}{ccccccccc}
&&Z&&&&Z&& \\
&&|&&& &|&& \\
Z &\!\!\text{\bf---}\!\!&I&\!\!\text{\bf---}\!\!&X_e &\!\!\text{\bf---}\!\!&I&\!\!\text{\bf---}\!\!&Z  \\
&&|&&& &|&& \\
&&I&&& &Z&& 
\end{array}
\right) \nonumber \\
&\quad = 
\left(
\begin{array}{ccccc}
\chi_{v_-}&\!\!\text{\bf ---}\!\!&\widetilde{Z}_e&\!\!\text{\bf ---}\!\!&i\chi'_{v_+}  \\ 
|&&&& \\ 
I &&&& 
\end{array}
\right) 
\widehat{{\rm D}} \ .
\end{align}
Similarly, combining the minimal coupling terms yields the dualization of the plaquette operator:
\begin{align}
\widehat{{\rm D}} 
\left(
\begin{array}{ccccccc}
&&&&Z&& \\ 
&&&&|&& \\ 
I&\!\!\text{\bf ---}\!\!&XZ&\!\!\text{\bf ---}\!\!&I&\!\!\text{\bf ---}\!\!&Z \\ 
|&&&&|&& \\ 
X&&p&&XZ&& \\ 
|&& &&|&& \\ 
I&\!\!\text{\bf ---}\!\!&X&\!\!\text{\bf ---}\!\!&I&& 
\end{array} 
\right) 
= 
\left(
\begin{array}{ccccc}
I &\!\!\text{\bf ---}\!\! & \widetilde{Z} &\!\!\text{\bf ---}\!\! & I \\
| & & & & | \\
\widetilde{Z} & & p & & \widetilde{Z} \\ 
| & & & & | \\ 
I &\!\!\text{\bf ---}\!\! & \widetilde{Z} &\!\!\text{\bf ---}\!\! & I \\
\end{array}
\right) 
\widehat{{\rm D}} \ .
\end{align}

In order to summarize, let us write the L-shaped Pauli operators as $L_e(=X_eZ_{e'})$ with the appropriate anisotropic assignment of $e'$.  
Let $W_v = Z_{e_u} Z_{e_\ell} Z_{e_r} Z_{e_d}$ be the plaquette operator associated with the vertex $v$ (which is dual to the dual plaquette). 
The six-body terms can be expressed as a product $L_e W_{v_+} W_{v_-}$, where $\partial e = v_+ - v_-$. 
Then, the duality transformation implemented by $\widehat{{\rm D}}$ is given by
\begin{align} \label{eq:MGT-dualization}
\widehat{{\rm D}} H_{\rm dual-MGT} = H_{\rm MGT} \widehat{{\rm D}}
\end{align}
with 
\begin{align}
H_{\rm dual-MGT} =& 
t \sum_{e} (L_e - L_e W_{v_+} W_{v_-}) 
+ \frac{\mu}{2} \sum_{v} W_v \nonumber \\
&- J \sum_{p} W_{{\rm ne}(p)} \prod_{e \subset p} X_e 
- h \sum_e Z_e \ ,
\end{align}
where ${\rm nw}(p)$ is the vertex at the northeast corner of the plaquette $p$. 
This dual model was obtained in Ref.~\cite{borla2022quantum}.
 
The equation~\eqref{eq:MGT-dualization} implies 
\begin{align}
\widehat{{\rm D}} 
\Big( e^{-it H_{\rm dual-MGT} }|\psi\rangle_{E}\Big) 
&= e^{- it H_{\rm MGT} } \widehat{{\rm D}}|\psi\rangle_{E}   \nonumber \\
&= e^{- it H_{\rm MGT} } |\psi_{\rm gauged}\rangle_{\tilde{E} \cup V} \,  ,
\end{align}
where $|\psi_{\rm gauged}\rangle_{\tilde{E} \cup V} = \widehat{{\rm D}}|\psi\rangle_{E}  $ is the gauged initial wave function.
Thus one can obtain the time evolution with the lattice gauge theory with spinless Majorana fermions from that with the dual model $H_{\rm dual-MGT}$.

The operator $\widehat{{\rm D}}$ can be realized by 
(1) introducing ancillas as a product state of $|+\rangle$ on edges and $|0\rangle$ (the fermion vacuum) on vertices, (2) applying the entangler $\mathcal{U}_{\rm MGT}$, and (3) measuring the (original) edge degrees of freedom in the $X$ basis. 
In the third step, the measurement outcomes might be the $|-\rangle$ state, which differs from the bra state in $\widehat{{\rm D}}$. 
The difference can be accounted for just as in the case with the Fradkin-Shenker model, but here we present a concise argument to show that the correction is possible. 
Since $|-\rangle_e = Z_e |+\rangle_e$ and the $Z_e$ operator commutes with $\mathcal{U}_{\rm MGT}$, the minus measurement outcome can be expressed as the $Z_e$ operator acting on the Hilbert space $\mathcal{H}_E$ of the model to be dualized. 
Due to the duality map~\eqref{eq:Z-to-X}, the operator $Z_e$ is mapped to the $\widetilde{X}_e$ operator that acts on the Hilbert space $\mathcal{H}_{\widetilde{E}}$. 
Each $\widetilde{X}_e$ operator converted from each $Z_e$ operator can be thus corrected based on the information gathered from the measurement outcomes after the third step. 
In sum, one can realize a clean duality operator $\widehat{{\rm D}}$ deterministically.

\bibliography{references}

\end{document}